\documentclass[twocolumn,trackchanges]{aastex631}
\usepackage{amsmath}
\usepackage{epsf}
\usepackage{color}
\usepackage{color}
\usepackage{enumitem}
\usepackage{mathtools}
\usepackage[mathcal]{eucal}
 \let\mathscr\relax
\usepackage[scr]{rsfso}

\shorttitle{The Complete CEERS $z >$ 9 Sample}
\shortauthors{Finkelstein et al.}

\newcommand{\sol}{$_{\odot}$}

\def\arcs{\hbox{$^{\prime\prime}$}}

\begin{document}
\title{The Complete CEERS Early Universe Galaxy Sample:\\ A Surprisingly Slow Evolution of the Space Density of Bright Galaxies at $z \sim$ 8.5--14.5}

\suppressAffiliations

\author[0000-0001-8519-1130]{Steven L. Finkelstein}
\affiliation{Department of Astronomy, The University of Texas at Austin, Austin, TX, USA}
\email{stevenf@astro.as.utexas.edu}

\author[0000-0002-9393-6507]{Gene C. K. Leung}
\affiliation{Department of Astronomy, The University of Texas at Austin}

\author[0000-0002-9921-9218]{Micaela B. Bagley}
\affiliation{Department of Astronomy, The University of Texas at Austin, Austin, TX, USA}

\author[0000-0001-5414-5131]{Mark Dickinson}
\affiliation{NSF's National Optical-Infrared Astronomy Research Laboratory, 950 N. Cherry Ave., Tucson, AZ 85719, USA}

\author[0000-0001-7113-2738]{Henry C. Ferguson}
\affiliation{Space Telescope Science Institute, Baltimore, MD, USA}

\author[0000-0001-7503-8482]{Casey Papovich}
\affiliation{Department of Physics and Astronomy, Texas A\&M University, College Station, TX, 77843-4242 USA}
\affiliation{George P.\ and Cynthia Woods Mitchell Institute for Fundamental Physics and Astronomy, Texas A\&M University, College Station, TX, 77843-4242 USA}

\author[0000-0003-3596-8794]{Hollis B. Akins}
\affiliation{The University of Texas at Austin, 2515 Speedway Blvd Stop C1400, Austin, TX 78712, USA}

\author[0000-0002-7959-8783]{Pablo Arrabal Haro}
\affiliation{NSF's National Optical-Infrared Astronomy Research Laboratory, 950 N. Cherry Ave., Tucson, AZ 85719, USA}

\author[0000-0003-2842-9434]{Romeel Dav\'e}
\affiliation{Institute for Astronomy, University of Edinburgh, Blackford Hill, Edinburgh, EH9 3HJ UK}
\affiliation{Department of Physics and Astronomy, University of the Western Cape, Robert Sobukwe Rd, Bellville, Cape Town 7535, South Africa}

\author[0000-0003-4174-0374]{Avishai Dekel}
\affil{Racah Institute of Physics, The Hebrew University of Jerusalem, Jerusalem 91904, Israel}

\author[0000-0001-9187-3605]{Jeyhan S. Kartaltepe}
\affiliation{Laboratory for Multiwavelength Astrophysics, School of Physics and Astronomy, Rochester Institute of Technology, 84 Lomb Memorial Drive, Rochester, NY 14623, USA}

\author[0000-0002-8360-3880]{Dale D. Kocevski}
\affiliation{Department of Physics and Astronomy, Colby College, Waterville, ME 04901, USA}

\author[0000-0002-6610-2048]{Anton M. Koekemoer}
\affiliation{Space Telescope Science Institute, 3700 San Martin Dr., Baltimore, MD 21218, USA}

\author[0000-0003-3382-5941]{Nor Pirzkal}
\affiliation{ESA/AURA Space Telescope Science Institute}

\author[0000-0002-6748-6821]{Rachel S. Somerville}
\affiliation{Center for Computational Astrophysics, Flatiron Institute, 162 5th Avenue, New York, NY, 10010, USA}

\author[0000-0003-3466-035X]{{L. Y. Aaron} {Yung}}
\altaffiliation{NASA Postdoctoral Fellow}
\affiliation{Astrophysics Science Division, NASA Goddard Space Flight Center, 8800 Greenbelt Rd, Greenbelt, MD 20771, USA}
\affiliation{Space Telescope Science Institute, 3700 San Martin Dr., Baltimore, MD 21218, USA}

\author[0000-0001-5758-1000]{Ricardo O. Amor\'{i}n}
\affiliation{ARAID Foundation. Centro de Estudios de F\'{\i}sica del Cosmos de Arag\'{o}n (CEFCA), Unidad Asociada al CSIC, Plaza San Juan 1, E--44001 Teruel, Spain}
\affiliation{Departamento de Astronom\'{i}a, Universidad de La Serena, Av. Juan Cisternas 1200 Norte, La Serena 1720236, Chile}

\author[0000-0001-8534-7502]{Bren E. Backhaus}
\affiliation{Department of Physics, 196 Auditorium Road, Unit 3046, University of Connecticut, Storrs, CT 06269, USA}

\author[0000-0002-2517-6446]{Peter Behroozi}
\affiliation{Department of Astronomy and Steward Observatory, University of Arizona, Tucson, AZ 85721, USA}
\affiliation{Division of Science, National Astronomical Observatory of Japan, 2-21-1 Osawa, Mitaka, Tokyo 181-8588, Japan}

\author[0000-0003-0492-4924]{Laura Bisigello}
\affiliation{Dipartimento di Fisica e Astronomia "G.Galilei", Universit\'a di Padova, Via Marzolo 8, I-35131 Padova, Italy}
\affiliation{INAF--Osservatorio Astronomico di Padova, Vicolo dell'Osservatorio 5, I-35122, Padova, Italy}

\author[0000-0003-0212-2979]{Volker Bromm}
\affiliation{Department of Astronomy, The University of Texas at Austin, Austin, TX, USA}

\author[0000-0002-0930-6466]{Caitlin M. Casey}
\affiliation{Department of Astronomy, The University of Texas at Austin, Austin, TX, USA}

\author[0000-0003-2332-5505]{\'Oscar A. Ch\'avez Ortiz}
\affiliation{Department of Astronomy, The University of Texas at Austin, Austin, TX, USA}

\author[0000-0001-8551-071X]{Yingjie Cheng}
\affiliation{University of Massachusetts Amherst, 710 North Pleasant Street, Amherst, MA 01003-9305, USA}

\author[0000-0003-4922-0613]{Katherine Chworowsky}\altaffiliation{NSF Graduate Fellow}
\affiliation{Department of Astronomy, The University of Texas at Austin, Austin, TX, USA}

\author[0000-0001-7151-009X]{Nikko J. Cleri}
\affiliation{Department of Physics and Astronomy, Texas A\&M University, College Station, TX, 77843-4242 USA}
\affiliation{George P.\ and Cynthia Woods Mitchell Institute for Fundamental Physics and Astronomy, Texas A\&M University, College Station, TX, 77843-4242 USA}

\author[0000-0003-1371-6019]{M. C. Cooper}
\affiliation{Department of Physics \& Astronomy, University of California, Irvine, 4129 Reines Hall, Irvine, CA 92697, USA}

%\author[0000-0002-3892-0190]{Asantha R. Cooray}
%\affiliation{Department of Physics \& Astronomy, University of California, Irvine, 4129 Reines Hall, Irvine, CA 92697, USA}

\author[0000-0001-8047-8351]{Kelcey Davis}
\affiliation{Department of Physics, 196 Auditorium Road, Unit 3046, University of Connecticut, Storrs, CT 06269, USA}

\author[0000-0002-6219-5558]{Alexander de la Vega}
\affiliation{Department of Physics and Astronomy, University of California, 900 University Ave, Riverside, CA 92521, USA}

\author[0000-0002-7631-647X]{David Elbaz}
\affiliation{Universit{\'e} Paris-Saclay, Universit{\'e} Paris Cit{\'e}, CEA, CNRS, AIM, 91191, Gif-sur-Yvette, France}

\author[0000-0002-3560-8599]{Maximilien Franco}
\affiliation{Department of Astronomy, The University of Texas at Austin, Austin, TX, USA}

\author[0000-0003-3820-2823]{Adriano Fontana}
\affiliation{INAF - Osservatorio Astronomico di Roma, via di Frascati 33, 00078 Monte Porzio Catone, Italy}

\author[0000-0001-7201-5066]{Seiji Fujimoto}
\affiliation{Cosmic Dawn Center (DAWN), Jagtvej 128, DK2200 Copenhagen N, Denmark}
\affiliation{Niels Bohr Institute, University of Copenhagen, Lyngbyvej 2, DK2100 Copenhagen \O, Denmark}

\author[0000-0002-7831-8751]{Mauro Giavalisco}
\affiliation{University of Massachusetts Amherst, 710 North Pleasant Street, Amherst, MA 01003-9305, USA}

\author[0000-0001-9440-8872]{Norman A. Grogin}
\affiliation{Space Telescope Science Institute, Baltimore, MD, USA}

\author[0000-0002-4884-6756]{Benne W. Holwerda}
\affil{Physics \& Astronomy Department, University of Louisville, 40292 KY, Louisville, USA}

\author[0000-0002-1416-8483]{Marc Huertas-Company}
\affil{Instituto de Astrof\'isica de Canarias, La Laguna, Tenerife, Spain}
\affil{Universidad de la Laguna, La Laguna, Tenerife, Spain}
\affil{Universit\'e Paris-Cit\'e, LERMA - Observatoire de Paris, PSL, Paris, France}

\author[0000-0002-3301-3321]{Michaela Hirschmann}
\affiliation{Institute of Physics, Laboratory of Galaxy Evolution, Ecole Polytechnique Fédérale de Lausanne (EPFL), Observatoire de Sauverny, 1290 Versoix, Switzerland}

\author[0000-0001-9298-3523]{Kartheik G. Iyer}
\affiliation{Dunlap Institute for Astronomy \& Astrophysics, University of Toronto, Toronto, ON M5S 3H4, Canada}

\author[0000-0002-1590-0568]{Shardha Jogee}
\affiliation{Department of Astronomy, The University of Texas at Austin, Austin, TX, USA}

\author[0000-0003-1187-4240]{Intae Jung}
\affiliation{Space Telescope Science Institute, Baltimore, MD, 21218, USA}

\author[0000-0003-2366-8858]{Rebecca L. Larson}
\affiliation{NSF Graduate Fellow}
\affiliation{The University of Texas at Austin, Department of Astronomy, Austin, TX, United States}

\author[0000-0003-1581-7825]{Ray A. Lucas}
\affiliation{Space Telescope Science Institute, 3700 San Martin Drive, Baltimore, MD 21218, USA}

\author[0000-0001-5846-4404]{Bahram Mobasher}
\affiliation{Department of Physics and Astronomy, University of California, 900 University Ave, Riverside, CA 92521, USA}

\author[0000-0003-4965-0402]{Alexa M.\ Morales}
\affiliation{Department of Astronomy, The University of Texas at Austin, 2515 Speedway, Austin, TX, 78712, USA}

\author[0000-0002-4404-0456]{Caroline V. Morley}
\affiliation{Department of Astronomy, University of Texas at Austin, 2515 Speedway, Austin, TX 78712, USA}

\author[0000-0003-1622-1302]{Sagnick Mukherjee}
\affiliation{Department of Astronomy and Astrophysics, University of California, Santa Cruz, CA 95064, USA}

\author[0000-0003-4528-5639]{Pablo G. P\'erez-Gonz\'alez}
\affiliation{Centro de Astrobiolog\'{\i}a (CAB), CSIC-INTA, Ctra. de Ajalvir km 4, Torrej\'on de Ardoz, E-28850, Madrid, Spain}

\author[0000-0002-5269-6527]{Swara Ravindranath}
\affiliation{Astrophysics Science Division, NASA Goddard Space Flight Center, 8800 Greenbelt Road, Greenbelt, MD 20771, USA}
\affiliation{Center for Research and Exploration in Space Science and Technology II, Department of Physics, Catholic University of America, 620 Michigan Ave N.E., Washington DC 20064, USA}

\author[0000-0002-9415-2296]{Giulia Rodighiero}
\affiliation{Department of Physics and Astronomy, Università degli Studi di Padova, Vicolo dell’Osservatorio 3, I-35122, Padova, Italy}
\affiliation{INAF - Osservatorio Astronomico di Padova, Vicolo dell’Osservatorio 5, I-35122, Padova, Italy}

\author[0000-0003-4225-6314]{Melanie J. Rowland}
\affil{University of Texas at Austin, Department of Astronomy, 2515 Speedway C1400, Austin, TX 78712, USA}

\author[0000-0002-8224-4505]{Sandro Tacchella}
\affiliation{Kavli Institute for Cosmology, University of Cambridge, Madingley Road, Cambridge, CB3 0HA, UK}\affiliation{Cavendish Laboratory, University of Cambridge, 19 JJ Thomson Avenue, Cambridge, CB3 0HE, UK}

\author[0000-0003-1282-7454]{Anthony J. Taylor}
\affiliation{Department of Astronomy, The University of Texas at Austin, Austin, TX, USA}

\author[0000-0002-1410-0470]{Jonathan R. Trump}
\affiliation{Department of Physics, 196 Auditorium Road, Unit 3046, University of Connecticut, Storrs, CT 06269, USA}

\author[0000-0003-3903-6935]{Stephen M.~Wilkins} %
\affiliation{Astronomy Centre, University of Sussex, Falmer, Brighton BN1 9QH, UK}
\affiliation{Institute of Space Sciences and Astronomy, University of Malta, Msida MSD 2080, Malta}

\begin{abstract}
We present a sample of 88 candidate $z \sim$ 8.5--14.5 galaxies selected from the completed NIRCam imaging from the Cosmic Evolution Early Release Science (CEERS) survey.  These data cover $\sim$90 arcmin$^2$ (10 NIRCam pointings) in six broad-band and one medium-band imaging filter.  With this sample we confirm at higher confidence early {\it JWST} conclusions that bright galaxies in this epoch are more abundant than predicted by most theoretical models.  We construct the rest-frame ultraviolet luminosity functions at $z \sim$ 9, 11 and 14, and show that the space density of  bright ($M_{UV} = -$20) galaxies changes only modestly from $z \sim$ 14 to $z \sim$ 9, compared to a steeper increase from $z \sim$ 8 to $z \sim$ 4. While our candidates are photometrically selected, spectroscopic followup has now confirmed 13 of them, with only one significant interloper, implying that the fidelity of this sample is high.  Successfully explaining the evidence for a flatter evolution in the number densities of UV-bright $z >$ 10 galaxies may thus require changes to the dominant physical processes regulating star formation. While our results indicate that significant variations of dust attenuation with redshift are unlikely to be the dominant factor at these high redshifts, they are consistent with predictions from models which naturally have enhanced star-formation efficiency and/or stochasticity.  An evolving stellar initial mass function could also bring model predictions into better agreement with our results.   Deep spectroscopic followup of a large sample of early galaxies can distinguish between these competing scenarios.
\end{abstract}


\section{Introduction}\label{sec:intro}

The first 500 million years of cosmic time ($z \gtrsim$ 10), when the first stars and galaxies formed, began to grow, and kick-started the process of reionization, was largely hidden from view until recently.  The depth achievable with {\it JWST} near-infrared imaging, along with the capabilities to deeply probe beyond $\sim$1.6 $\mu$m for the first time, were expected to revolutionize our understanding of this early epoch.  As soon as the first data from \textit{JWST} were released in July 2022, this renaissance in our understanding unfolded immediately.

Prior to {\it JWST}, the high-redshift community had debated about the evolution of the rest-frame ultraviolet (UV) luminosity function (and by extension, the cosmic star-formation rate density) at $z >$ 8.  While there was good agreement that these quantities evolved smoothly downward from $z =$ 4 to $z =$ 8 \citep[e.g.][]{bouwens15,finkelstein15}, results differed at $z >$ 8, with some advocating for continued evolution with the same declining slope from lower redshifts \citep[e.g.][]{mcleod16,finkelstein16}, while others claimed evidence of an accelerated decline towards higher redshifts \citep[e.g.][]{oesch18,bouwens19}.  Early {\it JWST} surveys, including the Cosmic Evolution Early Release Science Survey (CEERS; PID 1345, PI Finkelstein) and GLASS (PID 1324, PI Treu) surveys, were designed in part to determine which of these evolutionary possibilities was correct.

In defiance of expectations, several studies immediately reported the presence of bright ($m \lesssim$ 27.5) galaxies at $z >$ 10 from the CEERS and GLASS surveys \citep[e.g.,][]{castellano22,naidu22,finkelstein22c,donnan23a}.  These galaxies were both brighter and at higher redshifts than expected from these early surveys, which were neither extremely wide nor deep.  While some early results changed due to the uncertain characterization of the NIRCam photometric zeropoint \citep{boyer22}, within a few months more robust samples of galaxies were in place \citep[e.g.][]{finkelstein23,harikane23,mcleod23}.

These first studies with larger ($\sim$ 20 object) samples agreed that the abundance of $z \gtrsim$ 10 galaxies was in excess of both theoretical and empirical predictions, with explanations ranging from an evolving initial mass function (IMF), changes in star-formation efficiency, changes in dust attenuation, contribution from active galactic nuclei, rampant sample contamination \citep[e.g.][]{finkelstein23, harikane23,ferrara23,dekel23,mason22}, to even changes to the underlying cosmology \citep[e.g.][]{mbk22,liu22}.  Such explanations are compelling, yet these early datasets spanned small dynamic ranges in UV luminosity, with few galaxies yet included.

Here we report the results from a search for $z \geq$ 8.5 galaxies over the completed CEERS dataset.  This follows on the work of \citet[][hereafter F23]{finkelstein23}\defcitealias{finkelstein23}{F23} 
% \citep[hereafter F23]{finkelstein23}\defcitealias{finkelstein23}{F23} 
who did a similar search, finding 26 galaxies over the first four CEERS pointings.  Importantly, here we combine our results with those from \citet{leung23} who used a near-identical analysis procedure to identify a sample of 38 galaxies at similarly high redshift from the deep NGDEEP (the Next Generation Deep Extragalactic Exploratory Public survey; \citealt{bagley23}) NIRCam imaging, extending our dynamic range 1.5 magnitudes fainter.

In \S 2 we describe our imaging dataset, and give a detailed explanation of our photometry procedure, which has evolved from F23 to increase our color and flux accuracy.  In \S 3 we outline our photometric redshift sample selection procedure, and discuss the available spectroscopy for our candidate galaxies.  Our results are presented in \S 4, where in \S 4.2 we compare the observed surface density of galaxies to pre-launch predictions, and in \S 4.3 we calculate the rest-UV luminosity functions at $z \sim$ 9, 11 and 14.  We discuss these results in the context of a variety of more recent simulation predictions in \S 5, and present our conclusions in \S 6.   We assume the latest {\it Planck} flat $\Lambda$CDM cosmology with H$_{0}=\ $67.36 km s$^{-1}$ Mpc$^{-1}$, $\Omega_m=\ $0.3153 and $\Omega_{\Lambda}=\ $0.6847 \citep{planck20}.  All magnitudes are in the absolute bolometric system \citep[AB;][]{oke83}.

\section{Data}\label{sec:data}

The CEERS NIRCam imaging survey consists of 10 NIRCam pointings in the CANDELS Extended Groth Strip (EGS) field, done in parallel with prime MIRI and NIRSpec observations.  These data were taken in two epochs.  On June 21-22, 2022 four NIRCam pointings were obtained (known as pointings NIRCam1, 2, 3 and 6; results presented in F23).  The remaining six pointings (NIRCam4, 5, 7, 8, 9 and 10) were completed on Dec 20-24, 2022.  All 10 pointings include the same filter coverage of F115W, F150W, and F200W in the short-wavelength channel, and F277W, F356W, F410M, and F444W in the long-wavelength channel.  In this analysis we make use of the official CEERS publicly released mosaics (DR0.5 for the June data, and DR0.6 for the December data). These images are available on the CEERS website (\url{https://ceers.github.io/releases.html}) and on MAST as High Level
Science Products via \dataset[10.17909/z7p0-8481]{\doi{10.17909/z7p0-8481}}.

\subsection{Data Reduction}
We reduce the NIRCam imaging following the procedures outlined in \citet{bagley22b}, which we briefly summarize here. We use the \textit{JWST} Calibration Pipeline \citep{jwstpipeline} with custom modifications and additional steps needed to remove features such as snowballs, wisps, and $1/f$ noise. Our reduction process is the same for all images, though we use different Pipeline and Calibration Reference Data System (CRDS) versions for the two epochs of CEERS NIRCam imaging: Pipeline version 1.7.2 and CRDS context 0989 for the Epoch 1 images (CEERS DR0.5, obtained in June, 2022, and including pointings NIRCam1, 2, 3 and 6), and Pipeline version 1.8.5 and CRDS context 1023 for the Epoch 2 images (CEERS DR0.6, obtained in December, 2022, and including pointings NIRCam4, 5, 7, 8, 9 and 10). There were no major changes between these pipeline$+$CRDS versions which we would expect to affect the photometry.

For our astrometric alignment and analysis we make use of the archival {\it HST} imaging from the All-wavelength Extended Groth Strip International Survey \citep[AEGIS,][]{davis07}, the Cosmic Assembly Deep Extragalactic Legacy Survey \citep[CANDELS, ][]{grogin11,koekemoer11}, and the 3D-HST \citep{momcheva16} surveys. The entire CEERS field is covered by F606W, F814W, F125W, F140W, and F160W; portions are covered by F105W (we note that the F140W imaging is very shallow [800 sec], thus while we include it in our catalog, we do not include these data in any figures below).  
We use the CEERS v1.9\footnote{\url{ceers.github.io/releases.html\#hdr1}} {\it HST} EGS mosaics, which are created from these datasets, aligned to {\it Gaia} DR3, and on a 30mas pixel scale. 
We use a modified version of the \texttt{TweakReg} routine to align the images, using the \textit{HST} F160W mosaic as the astrometric reference in all pointings except NIRCam3 and 9. In these two pointings, a low-level guide star tracking issue in the \textit{HST} imaging caused a sub-PSF shift across a portion of the F160W mosaic, and so we use the NIRCam F277W (pointing 3) and F356W (pointing 9) mosaics as the absolute references. In all pointings, the RMS of the relative, NIRCam-to-NIRCam astrometry is $\sim$5--10 mas.

We create mosaics on a 30mas pixel scale in all filters, and extract smaller cutouts of the \textit{HST} mosaics to match the footprints of the drizzled NIRCam images, providing pixel-aligned imaging in up to 13 filters per field from $\sim0.5-5$\micron. Finally, we perform a global, two-dimensional background subtraction on the mosaics to remove any residual background variations. This method first performs a tiered source detection to identify progressively smaller sources in each filter. Then the source masks in each filter, including all available \textit{HST} images, are combined into a single merged mask such that pixels with source flux identified in any filter are excluded when measuring the background. See \citet{bagley22b} and \citetalias{finkelstein23} for more information on the background method and details on the method's performance in CEERS images.

%Following F23, in our analysis we make use of the archival {\it HST} imaging from the All-wavelength Extended Groth Strip International Survey (AEGIS, \citep{davis07}), the Cosmic Assembly Deep Extragalactic Legacy Survey \citep[CANDELS, ][]{grogin11,koekemoer11}, and the 3D-HST \citep{momcheva16} surveys.  The entire CEERS field is covered by F606W, F814W, F125W, (shallow 800 s) F140W, and F160W; portions are covered by F105W.  In each of the {\it HST} mosaics, we create smaller cutouts to match the footprints of the drizzled NIRCam mosaics, providing pixel-aligned imaging in up to 13 filters per field from $\sim0.5-5$\micron.  We make use of the CEERS v1.9\footnote{https://ceers.github.io/releases.html\#hdr1} {\it HST} EGS mosaics, which are aligned to {\it Gaia} DR3, and on the same 30 mas pixel scale as our NIRCam images.

\subsection{Photometry}

We perform photometry on all 10 CEERS fields using \textsc{Source
  Extractor} (hereafter \texttt{\textsc{SE}}, \citealt{bertin96}).  
 Photometry is performed on each of the 10 pointings independently (the high-redshift galaxy sample is screened for duplicates in the small overlapping areas, as discussed below).  The photometry process here is similar to F23, with some key
differences designed to improve the photometric validity, as described below. 
  Our fiducial photometry is measured in elliptical Kron apertures, using a
Kron factor $=$1.1 and a minimum radius $=$ 1.6, following F23.  These
small Kron apertures result in optimal signal-to-noise.  We derive
accurate colors in these apertures by matching the image PSFs between different filters, and
calculate accurate total fluxes via a two-step simulation-based 
aperture correction process.  

\subsubsection{PSF Matching}

We  create empirical point-spread functions (PSFs) in each filter
by stacking stars.  We select stars by identifying the stellar locus
in a plot of half-light radius versus magnitude in a preliminary
\texttt{\textsc{SE}} run in each filter.  Each star is then inspected to ensure it appears to be a non-saturated point source in a non-crowded region.
We make a single PSF per filter by stacking stars across all 10
pointings (as all observations used the same dither pattern).  
For each star, we extract a 101x101 pixel box, upsample by a factor of
10, measure the centroid, and shift the star to be centered in this
upsampled image.  We then downsample back to the native resolution,
rotate the star by a random position angle (to account for situations
when the position angle of the observations was not identical), and
normalize the star's peak flux to unity.  The final PSF is made by
median-combining the individual stars.  The final PSFs have a
centroiding accuracy of $\sim$0.05--0.1 pixels.

We then use these PSFs to derive kernels to match the PSFs in images
with PSF FWHMs smaller than that of F277W to the F277W PSF.  This
includes the NIRCam F115W, F150W and F200W images, and the {\it
  HST}/ACS F606W and F814W images.  Kernels were created with the
\textsc{pypher} Python routine\footnote{https://pypher.readthedocs.io}
\citep{boucaud16}.  The NIRCam F356W, F410M and F444W filters, along
with all four {\it HST}/WFC3 filters, F105W, F125W, F140W, and F160W, have PSF FWHMs larger than that
of the NIRCam F277W filter.  To correct for missing flux when
performing aperture photometry on these larger PSF images, we first convolve the F277W image to match the PSF of a given
larger-PSF image, and then derive source-specific correction factors
as the ratio of the F277W flux prior to convolution to that after
convolution (where the larger PSF images will have less flux in the
aperture).  These correction factors are then applied to photometry
measured on the larger PSF images, to account for the missing flux.  We tested our PSF-matching process by measuring curves-of-growth of
the PSF stars in the images, finding that the median enclosed flux at
an aperture diameter of 0.3\arcs\ was within 5\%\ (and often less) of
the F277W value for all filters.

In F23 we did not employ these correction factors as we matched all
NIRCam bands to the F444W PSF.  However, by matching here to the
smaller-PSF F277W image we better optimize signal-to-noise by not smoothing to the largest PSF.  Additionally, as the
{\it HST}/WFC3 images have larger PSFs than even F444W, in F23 we
devised correction factors by comparing the {\it HST} fluxes to
previous {\it HST} catalogs.  Our updated method here is independent
  of other catalogs, and is thus more self consistent.  We confirm that our fluxes here are consistent to within 5\% on average (and often lower) with the fluxes from F23.

\subsubsection{Catalog Creation}

We use the inverse-variance-weighted sum of the non-PSF-matched F277W
and F356W images as our detection image, to better detect faint
sources.  Using this detection image, we run \texttt{\textsc{SE}}
cycling through the seven NIRCam images and six {\it HST} images as
the measurement image.  The key \texttt{\textsc{SE}} parameters were:
DETECT\_THRESH$=$1.4, DETECT\_MINAREA$=$5 pixels, and a top-hat
convolution kernel with a width of 4 pixels, similar to F23.
We force \texttt{\textsc{SE}} to skip the background subtraction step
as this was previously removed (\S 2.1).  We use MAP\_RMS for the
source weighting.  As the pipeline-produced ERR images include Poisson
noise, they are not appropriate for source detection.  We thus convert
the weight map associated with the detection image into an effective
RMS map by taking 1/sqrt(WHT), and assign this to the detection image.
For the measurement image, we use the pipeline ERR image.

\begin{deluxetable*}{cccccccc}
\vspace{2mm}
%\tabletypesize{\small}
\tablecaption{NIRCam Imaging Summary}
\tablewidth{\textwidth}
\tablehead{\multicolumn{1}{c}{Field} & \multicolumn{1}{c}{F115W} & \multicolumn{1}{c}{F150W} & \multicolumn{1}{c}{F200W} & \multicolumn{1}{c}{F277W} & \multicolumn{1}{c}{F356W} & \multicolumn{1}{c}{F410M} & \multicolumn{1}{c}{F444W}}
\startdata
CEERS1&29.06&28.91&29.13&29.13&29.13&28.32&28.58\\
CEERS2&29.04&28.92&29.12&29.13&29.14&28.32&28.56\\
CEERS3&29.18&29.01&29.16&29.16&29.15&28.37&28.57\\
CEERS4&29.16&29.02&29.18&29.15&29.13&28.31&28.51\\
CEERS5&29.37&29.02&29.16&29.15&29.43&28.32&28.50\\
CEERS6&29.19&29.00&29.15&29.17&29.15&28.38&28.60\\
CEERS7&29.43&29.04&29.17&29.14&29.41&28.31&28.50\\
CEERS8&29.43&29.02&29.14&29.16&29.42&28.32&28.52\\
CEERS9&29.55&28.96&29.16&29.15&29.37&28.31&28.76\\
CEERS10&29.20&29.00&29.13&29.04&29.12&28.34&28.50\\
\hline
Median&29.20&29.01&29.16&29.15&29.15&28.32&28.56\\
\hline
PSF~FWHM&\phantom{0}0.068$^{\prime\prime}$&\phantom{0}0.072$^{\prime\prime}$&\phantom{0}0.080$^{\prime\prime}$&\phantom{0}0.128$^{\prime\prime}$&\phantom{0}0.143$^{\prime\prime}$&\phantom{0}0.156$^{\prime\prime}$&\phantom{0}0.163$^{\prime\prime}$\\
Flux~Enclosed&0.776&0.775&0.742&0.625&0.574&0.537&0.513\\
\enddata
\tablecomments{The depths given represent 5$\sigma$ limiting magnitudes, measured in d=0.2\arcs\ diameter circular apertures and corrected to total fluxes assuming a point-source.  The PSF FWHM and fraction of the flux enclosed in a 0.2\arcs\ diameter circular aperture are given below the horizontal line.}
\label{tab:tab1}
\vspace{-8mm}
\end{deluxetable*}

We estimate an aperture correction to the total flux for these small
apertures by performing a second run of \texttt{\textsc{SE}} on the
F277W image with the Kron parameters set to the default ``MAG\_AUTO"
parameters of (2.5, 3.5), deriving an aperture correction as the ratio
between the flux in this larger aperture to that in the smaller
aperture for each object in the F277W catalog, which we then applied multiplicatively to
the fluxes and uncertainties for all filters (thereby correcting fluxes to an estimated total, but not changing colors or signal-to-noise values).  

As several previous
studies have noted that the default Kron parameters we use for this
aperture correction can miss light in the wings of the PSF
\citep[e.g.,][]{bouwens15,finkelstein22}, we estimate residual
aperture corrections using source-injection simulations, adding 3000
mock sources to our real images in each field. 
We add sources from $m =$ 22--28.5 mag (to ensure a robust photometric
measurement), with a log-normal half-light radius distribution peaking
at $\sim$1.5 pixels ($\sim$0.2 kpc at $z =$ 10; compact but modestly
resolved, comparable to high-redshift sources), with a log-normal S\'ersic parameter distribution,
peaking at 1.2.  These mock sources were generated with
\textsc{galfit} \citep{peng02} and added at random positions to the
F277W and F356W images.  In \textsc{galfit},
the total flux is calculated such that half the flux is within r$_e$.
We combined the two images to create a
detection image, ran \texttt{\textsc{SE}} in the same way as on
our real data to generate a F277W catalog, and estimated residual
aperture corrections as the ratio of input-to-recovered fluxes for
recovered sources.  While in F23 we derived a single factor, here we
note that the residual correction needed is magnitude dependent.  We
thus fit a linear function to the flux ratios as a function of
magnitude over 24 $< m <$ 28, finding a correction ranging from
$\sim$2\% at $m <$ 22, to $\sim$20\% at $m =$ 28.  We performed this
linear fitting in each field, applying these residual aperture
corrections to every source (placing a bound on the corrections
applied to be from 1.0 -- 1.2).  We note that in the magnitude range
of 25 $< m <$ 26 used by F23, we see a similar correction factor
(1.08) as they derived.

\subsubsection{Flux Uncertainties}

We derive flux uncertainties empirically based on the number of pixels in an aperture.  We fit for the noise as a function of aperture size by measuring the fluxes in circular apertures with 30 different diameters, ranging from 0.1\arcs\ (3.33 pixels) to 3\arcs\ (100 pixels).  We place non-overlapping apertures randomly, avoiding pixels with zero values in the error image, and positive values in the segmentation map.  We do this in two iterations, placing 3000 apertures with diameters $<$1.5\arcs, and 500 apertures with larger diameters.   We measure fluxes at these positions in all aperture sizes, calculating the 1$\sigma$ noise in each aperture size by measuring the median absolute deviation of the measured flux values (multiplying by 1.48 to convert to a Gaussian-like standard deviation).  Finally, we fit a curve to the noise in a given aperture as a function of pixels in that aperture, using this equation \citep{gawiser06a}:
\begin{equation}
\sigma_{N} = \sigma_1 (\alpha N^{\beta} + \gamma N^{\delta})
\end{equation}
where $\sigma_N$ is the noise in an aperture containing N pixels, and $\sigma_1$ is the pixel-to-pixel noise measured in each image as the sigma-clipped standard deviation of all non-object pixels (see Figure 3 in \citealt{finkelstein22} for an example of this process).  Compared to F23, here we include the second term in the parentheses as we find a better fit to the data.  We fit the four free parameters with an IDL implementation of \textsc{emcee} (see \citealt{finkelstein19} for details), taking the median of the posterior as our fiducial values.  

We use these functional form fits for each filter to calculate the photometric uncertainties for each object, using both the number of pixels in its Kron aperture (Area $=$ $\pi ab$, where the semi-major [minor] axis $a$ [$b$] is equal to the A[B]\_IMAGE keywords multiplied by the KRON\_RADIUS value), as well as the area value for a given circular aperture.  We scale these values by the ratio of the error image value at the central position of a given source to the median error value of the whole map, thereby allowing the noise to be representative of the noise level around a given galaxy.  We refer to these measurements as our ``global'' noise measurements, which we use as our fiducial value.  Finally, to account for variable image noise not captured by the error image value at the central pixel, for each object in our catalog we also calculate a ``local" noise measurement.  This local noise was calculated in apertures with 0.2\arcs, 0.3\arcs, 0.4\arcs\ and 0.5\arcs-diameters, measured as 1.48 times the median absolute deviation of the flux distribution in the 200 closest apertures from the above process.

\begin{deluxetable}{cccc}
\vspace{2mm}
%\tabletypesize{\small}
\tablecaption{HST Imaging Summary}
\tablewidth{\textwidth}
\tablehead{\multicolumn{1}{c}{Filter} & \multicolumn{1}{c}{5$\sigma$ Limiting} & \multicolumn{1}{c}{FWHM} & \multicolumn{1}{c}{PSF Enclosed}\\
\multicolumn{1}{c}{} & \multicolumn{1}{c}{magnitude} & \multicolumn{1}{c}{} & \multicolumn{1}{c}{(d=0.2\arcs)}}
\startdata
ACS F606W&28.73&0.118$^{\prime\prime}$&0.696\\
ACS F814W&28.50&0.124$^{\prime\prime}$&0.625\\
WFC3 F105W&27.28&0.235$^{\prime\prime}$&0.348\\
WFC3 F125W&27.32&0.244$^{\prime\prime}$&0.327\\
WFC3 F140W&26.66&0.247$^{\prime\prime}$&0.317\\
WFC3 F160W&27.38&0.254$^{\prime\prime}$&0.303\\
\enddata
\tablecomments{Similar to Table 1, for the {\it HST} imaging used.  All depths given represent 5$\sigma$ limiting magnitudes, measured in d=0.2\arcs\ diameter circular apertures and corrected to total fluxes assuming a point-source.}
\label{tab:tab2}
\vspace{-8mm}
\end{deluxetable}

\subsubsection{Multi-band Catalog}

For each object in the catalog we use astropy.wcs \textsc{wcs\_pix2world} to derive celestial coordinates from the \texttt{\textsc{SE}} x, y positions (\texttt{\textsc{SE}} cannot presently parse the world-coordinate system in the {\it JWST} data model image headers).  We calculate physical fluxes by applying a photometric zeropoint to convert the image from MJy sr$^{-1}$ to erg s$^{-1}$ cm$^{-2}$ Hz$^{-1}$, and apply both aperture corrections derived above to all flux and flux error estimates.  We correct for Galactic extinction using an E(B-V) of 0.006 for the EGS field and a \citet{cardelli89} Milky Way attenuation curve.

We create a multi-band catalog from the individual-filter catalogs created by \texttt{\textsc{SE}}, including our fiducial Kron apertures and fluxes measured in circular apertures with diameters ranging from 0.05\arcs\ to 2.0\arcs.  For the latter we include fluxes measured from both the PSF-matched and native-resolution images; the latter are used below as a measure of detection significance.  These circular apertures are corrected for Galactic attenuation, but not corrected to total, as we will use them solely for detection significance.  For all flux measurements we calculate the noise per source following the above methods.  We flag any sources that had either a zero or NaN in any error column, replacing their flux error with 10$^{12}$ nJy (several orders of magnitude larger than any real source error) such that these flux measurements do not impact any analysis.  The final catalog contains only objects with valid measurements in the six broad-band filters, excluding the short-wavelength chips gaps, covering a total area of 88.1 arcmin$^2$.

In Table 1, we include an estimate of the limiting 5$\sigma$ magnitude for our catalog.  To calculate this, we use the noise functions described above to derive the flux density uncertainty in an aperture of diameter 0.2\arcs.  We then measure the enclosed flux at this radius from the stacked PSF.  We then divide the flux uncertainty by the enclosed fraction of flux to estimate the total noise for a point source.  Finally, we multiply this value by five, and convert to an AB magnitude.  This final multi-band catalog was known as ''v0.51" internally to the CEERS team, and has been used in a variety of analyses \citep[e.g.][]{arrabalharo23a,arrabalharo23b,vega23,larson23a,huertascompany23,ronayne23}.
%Both the enclosed flux values and the limiting magnitudes are listed in Table 1.  While the depths were broadly as expected based on the pre-launch exposure-time calculator, the F115W image (with double the exposure time), was expected to be 0.3 mag deeper.  The very low background at that wavelength has led to those images being more read-noise dominated than expected, thus this image has a depth comparable to the bulk of the NIRCam filters.  The primary impact is that photometric redshifts will be slightly more uncertain at $m >$ 29 than originally planned.

\section{Sample Selection}\label{sec:sample}

\subsection{Photometric Redshifts}

The final photometric catalog contains 101,808 sources across the entire CEERS field (86.8\% of which have signal-to-noise greater than three in F277W).  We create a new numerical identifier as an ascending integer starting at 1 in CEERS1, adding in each field sequentially.  We measure photometric redshifts for all sources in our 13-band photometric catalog using \textsc{EAZY} \citep{brammer08}.  We perform three iterations of \textsc{EAZY}: (1) \emph{Fiducial}: using our fiducial Kron, aperture corrected photometry with a maximum redshift of 20, (2) \emph{Low-z}: The same as the fiducial run, but with the maximum redshift set to seven (allowing visualization of the best-fitting low-redshift model), and (3) \emph{circular}: Replacing the Kron fluxes with the flux measured in $d=$0.2\arcs\ diameter apertures, with a maximum redshift of 20 (see \S 3.2.2).

\texttt{\textsc{EAZY}} fits non-negative linear combinations of user-supplied templates to derive probability distribution functions
(PDFs) for the redshift, based on the quality of fit of the various
template combinations to the observed photometry for a given source.  We use the same customized template list as F23, including the 12 FSPS \citep{conroy10} templates in the recommended ``tweak\_fsps\_QSF\_12\_v3" set, supplemented with six templates created by \citet{larson23b} to span the blue colors expected for early galaxies,.  We assume a flat prior in luminosity, and include a
systematic error of 5\% of the observed flux values, and fit to our measured total flux and flux error values.  

\subsection{Selection Criteria}\label{sec:selectioncriteria}

Here we describe the selection criteria we use to identify candidate $z >$ 8.5 galaxies.  Following our previous work \citep{finkelstein10,finkelstein15,finkelstein22,finkelstein22c,finkelstein23}, we use a combination of flux detection significance values and quantities derived from the full photometric redshift PDF, denoted $\mathcal{P}(z)$, to select our galaxy sample.  We also make use of the peak $\mathcal{P}(z)$ redshift, denoted $z_{best}$.  All signal-to-noise ratios (SNRs), unless stated otherwise, are measured in 0.2\arcs\ diameter apertures in the native resolution (non-PSF-matched) images.  We note that while we primarily use the global empirical noise measurement, we also make use of the local noise measurements with slightly relaxed criteria as described below.  

We also make use of SNR criteria in filters blue-ward of the Ly$\alpha$ break (``dropout'' filters, which should contain no significant flux).  To identify these filters, for each object we first zero out the $\mathcal{P}(z)$ at $z <$ 7.5 such that any low-redshift solution does not impact the dropout filter choice (followed by renormalizing the $\mathcal{P}[z]$).  We then consider a filter to be a dropout filter if the wavelength corresponding to the red side of the filter transmission's FWHM is less than the observed Ly$\alpha$ wavelength at the 16th percentile of the renormalized $\mathcal{P}(z)$.  In this way we make use of the full $\mathcal{P}(z)$ when calculating which bands should be considered a dropout filter.  Here we list our selection criteria, separated into two categories: 
\vspace{1mm}

\noindent Detection Significance Criteria:
\begin{itemize}
    \item Detection Signal-to-Noise: SNR $>$ 5.5 ($>$5.0) in at least two of the F150W, F200W, F277W, F356W, or F444W filters, using the global (local) empirical noise measurements. 
    \item Dropout Signal-to-Noise: SNR $<$ 2.0 (3.0) in all bands fully blue-ward of the Ly$\alpha$ break, and no more than one filter with SNR $>$ 1.5 (2.0), using the global (local) empirical noise measurements.  For this we consider the ACS F606W and F814W filters, and the NIRCam F115W, F150W, and F200W filters (we note that some candidates have $\sim$1--3$\sigma$ flux measurements in WFC3 bands nominally below the Ly$\alpha$ break; however we consider these spurious as such objects show significant non-detections in the deeper short-wavelength NIRCam bands).
    \item Error-map values at the central pixel of the object $<$1000 in the F115W, F150W, F200W, and F277W filters.  This ensures valid flux measurements in a minimum set of filters to robustly select $z >$ 9 galaxies (and explicitly excludes the NIRCam short-wavelength chip gaps).
    \item Total F277W magnitude $\leq$ 29.2 as a conservative estimate to limit low-significance sources.
\end{itemize}

\noindent Photometric Redshift Criteria:
\begin{itemize}
    \item $\int\mathcal{P}$($z$ $>$ 7) $\geq$ 0.7, requiring at least 70\% of the integrated $\mathcal{P}(z)$ at $z >$ 7.
    \item $z_{best}$ $>$ 8.5.
    \item $\mathcal{S}_z \geq$ 9, where $\mathcal{S}_z$ is calculated as the unit redshift where the integral in a $z \pm 0.5$ bin is the maximum compared to all other unit redshift bins.
    \item Total $\chi^2$ $\leq$ 60 (for the 13 available bands).
    \item $\Delta \chi^2$ $>$ 4, calculated as the difference between the lowest $\chi^2$ value at $z <$ 7 to the best-fitting $\chi^2$ value.
\end{itemize}

These criteria are very similar to those in F23, with the key differences being the selection of dropout bands, the more conservative ${\rm SNR} < 2$ dropout criteria (compared to ${\rm SNR} < 3$ in F23), and the addition of the local noise measurements.  

The combination of this sample selection criteria yielded an initial sample of 185 $z >$ 8.5 galaxy candidates across the full CEERS field.  We visually inspected all objects (examining cutout images in all filters, as well as the photometric SED and the $\mathcal{P}(z)$ curves) to identify any spurious sources, or potential cases where the \texttt{\textsc{SE}} photometry could be questionable (e.g., the presence of a bright neighbor).  We removed 91 objects through this screening process.  Cutout images for all removed objects as well as a table of their positions are shown in Appendix B.

Objects were removed for a variety of reasons, but the vast majority (78/91) were sources of an obvious spurious nature, including 10 sources identified as diffraction spikes, 45 sources associated with image edges, and 23 sources identified as bad pixels.  The latter had a characteristic observational signature of a compact (few pixel), boxy morphology in the long-wavelength channel images only.  We note that a few of our removed bad-pixel sources appear slightly less boxy than the bulk.  It is possible these are real astrophysical objects at $z \gtrsim$ 16 (where the Ly$\alpha$ break would be redward of F200W).  However, as they are visible in the three long-wavelength channels only it is much more plausible (if indeed they are not bad pixels) that they are faint objects associated with the known $z \sim$ 4.9 overdensity in this field \citep{arrabalharo23a}, or low-redshift dusty galaxies \citep[e.g.][]{bisigello23}, rather than true $z \sim$ 16 galaxies.  Improved outlier pixel rejection in future iterations of our data reduction could reduce the frequency of such objects.  Likewise, use of the context map (``CON") extension of the {\it JWST} data model could remove the need to manually identify image edges.  We do note that in some cases, objects removed due to image edges have legitimate long-wavelength photometry, but are adjacent to a short-wavelength chip gap.

The remaining 13 sources were a combination of objects identified as being associated with nearby bright galaxies, inadvertently split into separate objects by \textsc{SE} (10 objects), and objects where the photometric SED did not appear accurate (three objects); e.g., obvious flux visible in a dropout filter that was not accurately recorded (usually due to crowding).   Of the full sample of 91 sources removed, a small number (five) are plausible to still be true high-redshift galaxies.  While we conservatively keep them out of our sample, we note them specifically in Tables~\ref{tab:appendix-spurious1} and \ref{tab:appendix-spurious2} in Appendix B.  

Finally, we check this sample for duplicates which could arise as the 10 NIRCam CEERS pointings have slightly overlapping edges.  We find one object which appears in our catalog twice, encouragingly satisfying our sample selection with two independent photometric measures.  This object appears in CEERS4 (as ID 42447) and CEERS10 (as ID 93725), with similar $\mathcal{P}(z)$ distributions.  As this object has slightly higher signal-to-noise in the CEERS4 photometric catalog, we keep ID$=$42447, and remove ID$=$93725 from our catalog.  We note that future work making use of a complete mosaic of all 10 NIRCam pointings will make better use of these near-edge regions by combining the images from both pointings.  After removal of the 91 spurious sources and the one duplicate, the sample consisted of 93 candidate galaxies.

\begin{figure}[!t]
\epsscale{1.1}
\plotone{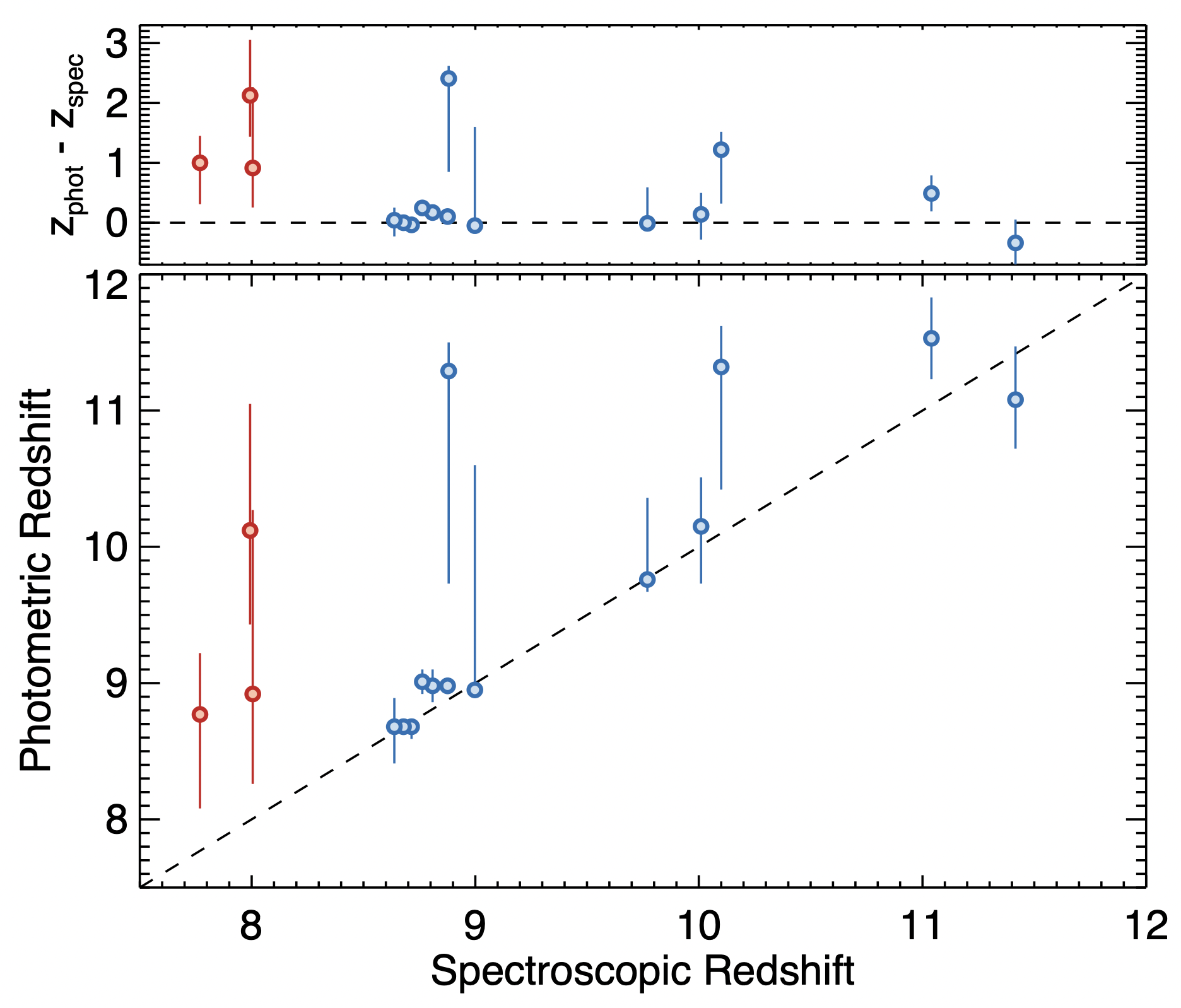}
\caption{A comparison of the photometric redshifts to the measured spectroscopic redshifts for the 17 galaxies in our initial sample with spectroscopic confirmation.  With the exception of the (not shown) $z \sim$ 16 candidate at $z_{spec} =$ 4.9 \citep{arrabalharo23a}, the agreement is generally good.  After removal of three sources (in red) with spectroscopic redshifts below our cut of $z >$ 8.5, we find a  median (mean) $z_{phot} - z_{spec} = $0.1 (0.3), with 8/13 sources having  $|z_{phot} - z_{spec}| <$ 0.2.  As noted by \citet{arrabalharo23b} and \citet{fujimoto23} there does appear to be a mild systematic offset towards higher photometric redshifts.  We also note that there are four sources with $z_{phot} >$ 10 observed with NIRSpec that showed no detectable signatures, which is consistent with $z >$ 10, where all strong lines are shifted out of the NIRSpec range. }
\label{fig:specz}
\end{figure}

\subsubsection{Spectroscopic Redshifts}\label{sec:specz}

Among the more transformative advances of {\it JWST} is NIRSpec's ability to efficiently spectroscopically confirm galaxies via strong [\ion{O}{3}] line emission to $z \sim$ 9.5 and via Ly$\alpha$ spectral breaks to higher redshift.  We make use of followup spectroscopy from both the CEERS spectroscopic program (which in its second epoch placed NIRSpec slits on some sources observed in the first epoch) as well as a Director's Discretionary Time (DDT) program in the CEERS field (PID 2750, PI Arrabal Haro; \citealt{arrabalharo23a}).  \citet{fujimoto23} presented [\ion{O}{3}]-based redshifts for sources in the $\sim$3100 sec CEERS second-epoch medium-resolution grating and prism observations, while \citet{arrabalharo23b} presented both Ly$\alpha$ break and [\ion{O}{3}]-based redshifts for sources in the $\sim$3100 sec CEERS third-epoch prism observations (additional redshifts were also presented in \citealt{larson23a} and \citealt{tang23}).  \citet{arrabalharo23a} presented both Ly$\alpha$ break and line-based ([\ion{O}{2}] and [\ion{O}{3}]) redshifts for sources in the $\sim$5 hr prism DDT observations (one of these sources had its redshift of $z =$ 11.04 first presented in \citealt{harikane23}).

Cross-matching our sample of 93 galaxy candidates to these spectroscopic lists, we find 17 sources which have published spectroscopic redshifts.  A comparison between the photometric redshifts and spectroscopic redshifts is shown in Figure~\ref{fig:specz}.  Not shown in this figure is the single catastrophic redshift failure (defined as $|z_{spec} - z_{phot}|/(1+z_{spec}) > 0.3$), the galaxy (ID$=$13256) originally presented in \citet{donnan23a} as having $z \sim$ 16.5, with similar redshifts proposed by \citealt{harikane23} and \citet{finkelstein23}; a lower redshift of $z_{phot} =$ 4.6 was proposed in \citealt{perezgonzalez23b}.  As discussed in \citet{arrabalharo23a}, this object has a confirmed redshift of $z \sim$ 4.9, and is the result of a very pathological situation where at this specific redshift a red galaxy with extreme line emission can mimic a $z \sim$ 16 galaxy as the H$\alpha$ line falls in all three of the F356W, F410M, and F444W filters, while [\ion{O}{3}] enhances F277W (see also discussion in \citealt{zavala22}).

In addition we remove one object that also has $z_{phot} \sim$ 16 (ID$=$43382, with a 68\% confidence range on the photometric redshift of $z =$ 15.9 -- 19.2).  While fainter than ID$=$13256 (F277W $=$ 28.8 versus 26.5), its spectral signature is almost identical, thus we consider it a likely fainter companion to the $z \sim$ 4.9 overdensity confirmed in \citet{arrabalharo23a}.

We find three additional sources with spectroscopic redshifts below our nominal redshift cut of $z >$ 8.5, which we thus remove.  These are: ID$=$4774, ID$=$4777 and ID$=$23084.  ID$=$4774 has a photometric redshift 68\% confidence limit (CL) of 8.26--10.27, with $z_{spec} =$ 8.01.  ID$=$4777 has a photometric redshift 68\% CL of 9.43--11.05, with $z_{spec} =$ 7.99.  ID$=$23084 has a photometric redshift 68\% CL of 8.08--9.22, with $z_{spec} =$ 7.77.  While the spectroscopic redshift is outside the 68\% confidence range on the photometric redshift, the redshifts are not catastrophically low.  We tabulate the five removed sources, including their celestial coordinates, in Table~\ref{tab:appendix-spectab} in the Appendix.

Beyond these five removed sources, we find generally good agreement between the photometric and spectroscopic redshifts, with a median (mean) $z_{phot} - z_{spec} = $0.1 (0.3), with 8/13 sources having  $|z_{phot} - z_{spec}| <$ 0.2 (the median offset was 0.2 prior to removal of the five sources in the preceding paragraph).  As noted by \citet{arrabalharo23b} there does appear to be a systematic offset towards higher photometric redshifts.  One likely explanation for this is that the shape of the Ly$\alpha$ break in these galaxy spectra is more extended than the sharp break assumed in the IGM attenuation models employed by \textsc{EAZY}.  This could be due to a variety of factors, including stellar population properties not accounted by the typically used templates \citep[e.g.,][]{arrabalharo23b}, the Ly$\alpha$ damping wing from an increasingly neutral IGM \citep[e.g.,][]{curtislake23,arrabalharo23b,umeda23}, and/or extremely dense line-of-sight damped Ly$\alpha$ systems in close proximity to a given galaxy \citep{heintz23,hsiao23}.  Additionally some of the spectroscopic redshifts we compare to come from the Ly$\alpha$ break alone, which has additional uncertainties (see discussion in \citealt{fujimoto23b}).  It is important to note that several additional sources were spectroscopically observed but not detected (indicated as ``Nz" in Tables~\ref{tab:highz} and \ref{tab:lesshighz}); this is modest evidence in favor of $z >$ 9.6, as at lower redshifts [\ion{O}{3}]$+$H$\beta$ should have been detectable.  For the remainder of our analysis, we use the spectroscopic redshift values when they are available, which is the case for 13 of our final sample of 88 candidate galaxies.

\subsubsection{Kron Aperture Corrections}\label{sec:kroncorr}

During the visual inspection step we found that some legitimate high-redshift galaxies had Kron apertures which appeared much larger than the galaxy in question, stretched by nearby galaxies.  Similar to F23, we devise a correction to ensure that flux from neighboring galaxies does not bias the colors nor the total fluxes.  To identify sources where this is needed, we explore the ratio between the area of the Kron aperture and the area of a $d=$0.2\arcs\ circular aperture.  Our galaxy sample shows a log-normal distribution, with a peak aperture ratio of $\sim$2, with a tail to higher values.  There is a notable gap at a ratio of $\sim$10, thus we flag sources with aperture size ratios larger than this as potentially needing a correction.

We find just one source meets this criterion, ID$=$11384 (with an aperture ratio of 14.7; the next highest was 9.3).  Upon inspection of this source, it is very compact, but is in a region of high background with a very bright galaxy $\sim$1.5\arcs\ to the NW, and a modestly bright galaxy $\sim$0.5\arcs\ to the S, resulting in an elongated Kron aperture in the N-S direction.  For this object we thus make use of colors measured in $d=$0.2\arcs\ circular apertures.  To calculate total fluxes we derive an aperture correction as the median of the ratio of the total F277W flux to the flux measured in a $d=$0.2\arcs\ circular aperture for all sources in our full photometry catalog with $d=$0.2\arcs\ fluxes within 20\% of the F277W $=$0.2\arcs\ flux for this object.  We find this correction factor is 2.9 for ID$=$11384, consistent with the values for other sources in our galaxy sample (median of 2.3 $\pm$ 0.8).

For this object we then replace its default fluxes with these new values, and adopt the photometric redshift results from colors measured in the small circular apertures.  This increases the photometric redshift from 10.8--11.4 to 11.2--11.8.  Of note is that this source has a spectroscopic redshift of 11.043 \citep{harikane23,arrabalharo23a}.  While the uncorrected photometric redshift is more consistent, the higher value from our improved photometry matches the observed very shallow Ly$\alpha$ break observed in the prism spectrum of this source (the redshift inferred from this break in the prism spectrum is $z \approx$ 11.4; \citealt{arrabalharo23a}), which can lead to minor photometric redshift overestimates as discussed in the previous subsection.

\subsubsection{Ly$\alpha$ Break}\label{sec:lyabreak}

As a cross-check on our Ly$\alpha$-break criterion, we examine our sample for objects where there is a discrepancy between the primary photometric redshift peak and the SNR in bands nominally below the break; such sources can still satisfy our criterion for inclusion in a high-redshift bin if their $\mathcal{P}(z)$ is bimodal or somewhat broad.

We identify two sources in our nominal $z \sim$ 11 sample (see \S 3.4 for sample definitions) which have $\mathcal{P}(z)$'s that exhibit two high-redshift peaks; a larger peak at $z >$ 9.5, and a significant secondary peak at $z \sim$ 9.  These objects are ID$=$17898 and 42447 (neither have spectroscopic redshifts).  Both objects exhibit SNR $>$ 2 in F115W.   As the lower 16th percentile of their $\mathcal{P}(z)$ was at $z <$ 9.54 (the redshift of Ly$\alpha$ at the red edge of the F115W filter FWHM), this significant F115W flux did not violate our selection criteria.  However, as thus flux is measured as significant (SNR=3.1 and 5.7 for these two objects, respectively), the $z \sim$ 9 peak is more likely to be correct.  For these two objects, we thus applied a prior to their $\mathcal{P}(z)$, setting them to zero at $z >$ 9.54, renormalizing to unity and recomputing the best-fit photometric redshift as the new peak.  We find that the peak photometric redshift for these sources changes from $z =$ 10.45 to $z =$ 9.07 for ID$=$17898 and $z =$ 10.30 to $z =$ 9.13 for ID$=$42447.  We confirm that both sources continue to satisfy our selection criterion of $\mathcal{P}(z>7) >$ 0.7.

We also do a similar analysis for sources in our $z >$ 13 sample.  We find two sources with SNR in F150W $>$ 2.  These objects (ID $=$ 2067 with SNR $=$ 2.57, and ID $=$ 77647 with SNR $=$ 2.45) both exhibit a broad $\mathcal{P}(z)$, extending down to $z \sim$ 11, thus this F150W flux did not violate our selection criteria.  We apply a similar prior, setting $\mathcal{P}(z >$ 12.72; corresponding to the red edge of F150W) to zero.  We find that the peak photometric redshift for these three sources changes from $z =$ 13.69 to 12.70 and $z =$ 13.57 to 12.70 (e.g., the new peak is at the edge of the prior).

\begin{figure}[!t]
\epsscale{1.15}
\plotone{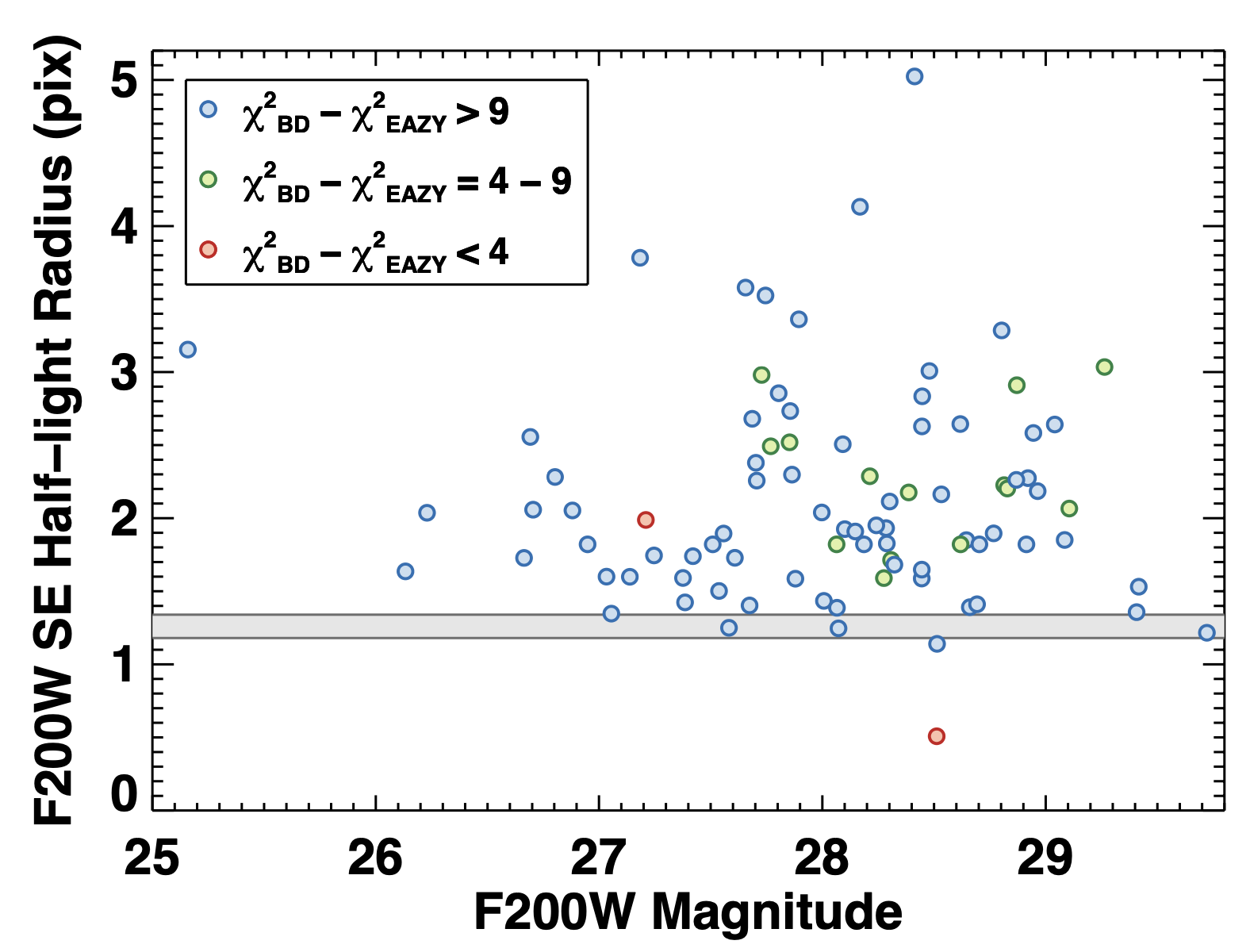}
\caption{The F200W half-light radius (measured from \textsc{SE}) versus F200W apparent magnitude.  The gray bar shows the half-light radius from stars in the image (with the width showing the 68\% spread in the values).  The data points are color-coded by the difference in $\chi^2$ between the best-fitting (sub)stellar model, and the best-fitting \textsc{EAZY} galaxy model.  Most objects are clear resolved, and thus extragalactic in origin.  Only one object is formally compact with a stellar $\chi^2$ comparable to the best-fitting EAZY model, though even for this object we conclude it is likely extragalactic due to its non-compact appearance in the imaging, and the very large ($\sim$4 kpc) implied distance were it stellar.  We conclude that stellar contamination is not significant in our sample.}
\label{fig:rh_stars}
\end{figure}

We acknowledge that while \textsc{EAZY} had knowledge of these observed SNR ratios and still found a preferred peak at higher redshift, the obvious real flux in the images left us confident that applying this prior to the $\mathcal{P}(z)$ will result in more accurate redshift estimates.  These $\mathcal{P}(z)$ priors were applied in the completeness simulations discussed in \S~\ref{sec:completeness}.  In Figures~\ref{fig:z14bioplots} and \ref{fig:z11bioplots} we show the original $\mathcal{P}(z)$ as a faded black curve for these sources.

\subsection{Stellar Contamination}\label{sec:stars}
The colors of high-redshift galaxies, especially between 1--2 $\mu$m, can be degenerate with low-mass stars and substellar objects \citep[e.g.][]{wilkins2014,finkelstein16}.  While the photometric coverage from 1--5$\mu$m should mitigate this confusion, here we explore whether the colors of our candidate galaxies could plausibly be consistent with low-mass stars or brown dwarfs. 
We fit each candidate to a grid of low-temperature, cloudy, chemical equilibrium substellar atmosphere models from Sonora-Diamondback (Morley et al.~2023, in prep). We explore a range of temperatures $T\sim 900$--$2400$ K, surface gravities $g=100$ and $3160$, and metallicities $[{\rm M}/{\rm H}] = 0$ and $-0.5$. 
We convolve the model SEDs with the \textit{HST}+\textit{JWST} filter curves and perform a simple grid-fitting routine, scaling the fluxes of each model to minimize the $\chi^2$. 
We adopt the model with the lowest $\chi^2$ as the best-fitting stellar model. 
We estimate the implied distance by scaling the model fluxes and assuming an intrinsic radius of 1 Jupiter radius. 
We note that we also ran fits with cloud-free models grids extending to particularly cold (Sonora-Bobcat, $T\sim 200$--$1300$ K; \citealt{marleySonora2021}) and low-metallicity (LOWZ, $[{\rm M}/{\rm H}]=-1$; \citealt{meisnerLOWZ2021}) parameter spaces; however, none provided a better fit over the Sonora-Diamondback models. 

\begin{deluxetable*}{ccccccccccc}
\vspace{2mm}
%\tabletypesize{\small}
\tablecaption{Summary of $z >$ 9.7 Candidate Galaxies}
\tablewidth{\textwidth}
\tablehead{\multicolumn{1}{c}{ID} & \multicolumn{1}{c}{RA} & \multicolumn{1}{c}{Dec} & \multicolumn{1}{c}{m$_{F277W}$} & \multicolumn{1}{c}{M$_{1500}$} & \multicolumn{1}{c}{$\mathcal{C}_{FUV}$} & \multicolumn{1}{c}{$\int_7^{20} \mathcal{P}(z)$} & \multicolumn{1}{c}{$\Delta \chi^2$} & \multicolumn{1}{c}{Photometric} & \multicolumn{1}{c}{Spectroscopic}\\
\multicolumn{1}{c}{$ $} & \multicolumn{1}{c}{(J2000)} & \multicolumn{1}{c}{(J2000)} & \multicolumn{1}{c}{(mag)} & \multicolumn{1}{c}{(mag)} & \multicolumn{1}{c}{(mag)} & \multicolumn{1}{c}{$ $} & \multicolumn{1}{c}{$ $} & \multicolumn{1}{c}{Redshift} & \multicolumn{1}{c}{Redshift}}
\startdata
CEERS-90891&214.945818&52.829729&28.3&$-$19.8$^{+0.3}_{-0.1}$&$-$0.12$^{+0.16}_{-0.03}$&0.96&4.8&14.44$^{+0.48}_{-1.89}$&---\\
CEERS-64602&215.074304&52.951154&28.8&$-$19.3$^{+0.3}_{-0.2}$&0.00$^{+0.11}_{-0.15}$&0.99&8.2&14.32$^{+0.75}_{-1.41}$&---\\
CEERS-57438&214.871989&52.845014&28.3&$-$19.6$^{+0.3}_{-0.2}$&0.03$^{+0.23}_{-0.07}$&0.96&4.2&13.99$^{+0.75}_{-1.35}$&---\\
CEERS-77647&215.054220&52.923839&29.0&$-$18.9$^{+0.3}_{-0.0}$&$-$0.19$^{+0.11}_{-0.01}$&1.00&9.4&12.70$^{+-0.09}_{-0.66}$&---\\
CEERS-2067&215.010026&53.013641&27.8&$-$20.1$^{+0.2}_{-0.0}$&0.02$^{+0.09}_{-0.03}$&1.00&8.3&12.70$^{+-0.09}_{-0.72}$&Nz\\
CEERS-36796&214.727248&52.748045&28.2&$-$19.9$^{+0.2}_{-0.2}$&$-$0.11$^{+0.11}_{-0.06}$&0.98&6.5&12.28$^{+1.59}_{-0.24}$&---\\
CEERS-70831&215.100921&52.936270&28.1&$-$19.6$^{+0.3}_{-0.3}$&0.06$^{+0.22}_{-0.08}$&0.93&4.7&12.07$^{+1.98}_{-1.08}$&---\\
CEERS-34925&214.738486&52.765665&28.4&$-$19.0$^{+0.4}_{-0.4}$&0.33$^{+0.36}_{-0.19}$&1.00&10.5&11.89$^{+1.92}_{-1.68}$&---\\
CEERS-34685&214.700083&52.752419&28.4&$-$19.4$^{+0.3}_{-0.2}$&$-$0.06$^{+0.21}_{-0.03}$&0.96&6.1&11.53$^{+0.51}_{-0.72}$&---\\
CEERS-16943&214.943152&52.942442&27.9&$-$20.2$^{+0.1}_{-0.1}$&$-$0.12$^{+0.03}_{-0.02}$&1.00&30.2&11.08$^{+0.39}_{-0.36}$&11.416 $_{-0.005}^{+0.005}$\\
CEERS-26112&214.818999&52.865299&28.3&$-$19.9$^{+0.3}_{-0.1}$&$-$0.09$^{+0.20}_{-0.01}$&0.93&5.0&11.38$^{+0.30}_{-0.81}$&---\\
CEERS-54306&214.858815&52.850712&28.8&$-$19.3$^{+0.1}_{-0.1}$&$-$0.19$^{+0.00}_{-0.00}$&1.00&11.3&11.23$^{+0.36}_{-0.36}$&---\\
CEERS-76686&214.976311&52.873417&28.2&$-$19.2$^{+0.1}_{-0.2}$&0.03$^{+0.07}_{-0.11}$&1.00&13.6&11.11$^{+0.33}_{-0.36}$&---\\
CEERS-87379&214.932064&52.841873&27.3&$-$20.7$^{+0.2}_{-0.1}$&$-$0.19$^{+0.14}_{-0.00}$&1.00&23.7&11.08$^{+0.24}_{-0.48}$&Nz\\
CEERS-85546&214.885963&52.819060&27.7&$-$20.3$^{+0.2}_{-0.1}$&$-$0.15$^{+0.11}_{-0.00}$&1.00&26.3&11.08$^{+0.24}_{-0.60}$&---\\
CEERS-11384&214.906640&52.945504&27.5&$-$20.0$^{+0.1}_{-0.1}$&0.20$^{+0.09}_{-0.09}$&1.00&11.2&11.53$^{+0.30}_{-0.30}$&11.043 $_{-0.003}^{+0.003}$\\
CEERS-77367&214.989018&52.879278&27.9&$-$20.0$^{+0.2}_{-0.2}$&$-$0.07$^{+0.08}_{-0.08}$&0.97&6.7&10.84$^{+0.54}_{-0.51}$&---\\
CEERS-47141&214.910771&52.873928&28.7&$-$18.6$^{+0.4}_{-0.2}$&0.12$^{+0.36}_{-0.02}$&0.97&6.1&10.66$^{+1.50}_{-0.48}$&---\\
CEERS-16984&214.866488&52.887854&28.5&$-$19.2$^{+0.1}_{-0.2}$&$-$0.10$^{+0.11}_{-0.02}$&0.97&7.0&10.63$^{+0.39}_{-0.24}$&---\\
CEERS-54903&214.811119&52.813829&29.0&$-$18.3$^{+0.2}_{-0.3}$&0.05$^{+0.23}_{-0.08}$&0.93&5.7&10.60$^{+0.81}_{-0.36}$&---\\
CEERS-57400&214.869658&52.843646&28.7&$-$19.4$^{+0.2}_{-0.2}$&$-$0.00$^{+0.07}_{-0.15}$&1.00&12.4&10.60$^{+0.60}_{-0.66}$&Nz\\
CEERS-10332&215.044001&52.994302&28.4&$-$19.5$^{+0.4}_{-0.1}$&$-$0.06$^{+0.20}_{-0.01}$&1.00&14.3&10.57$^{+0.18}_{-1.05}$&Nz\\
CEERS-92463&214.975831&52.841961&28.3&$-$18.8$^{+0.4}_{-0.1}$&0.17$^{+0.23}_{-0.02}$&1.00&21.1&10.51$^{+0.69}_{-0.60}$&---\\
CEERS-74300&214.969256&52.882119&28.2&$-$19.3$^{+0.1}_{-0.2}$&$-$0.14$^{+0.06}_{-0.05}$&1.00&12.3&10.51$^{+0.30}_{-0.57}$&---\\
CEERS-101746&214.881212&52.772963&28.4&$-$19.6$^{+0.2}_{-0.1}$&$-$0.19$^{+0.07}_{-0.00}$&1.00&20.4&10.33$^{+0.24}_{-0.72}$&---\\
CEERS-19996&214.922787&52.911529&28.3&$-$19.3$^{+0.2}_{-0.1}$&0.18$^{+0.04}_{-0.17}$&1.00&11.7&11.32$^{+0.30}_{-0.90}$&10.10 $_{-0.26}^{+0.13}$\\
CEERS-98518&214.817113&52.748343&27.4&$-$19.9$^{+0.2}_{-0.1}$&0.21$^{+0.08}_{-0.05}$&1.00&17.4&10.09$^{+0.18}_{-0.39}$&---\\
CEERS-35590&214.732525&52.758090&27.7&$-$20.2$^{+0.1}_{-0.0}$&$-$0.07$^{+0.06}_{-0.05}$&1.00&34.9&10.15$^{+0.36}_{-0.42}$&10.01 $_{-0.19}^{+0.14}$\\
CEERS-99715&214.811852&52.737110&27.1&$-$20.5$^{+0.1}_{-0.0}$&0.06$^{+0.06}_{-0.05}$&1.00&37.7&9.76$^{+0.60}_{-0.09}$&9.77 $_{-0.29}^{+0.37}$\\
CEERS-61941&214.850131&52.808053&28.9&$-$18.6$^{+0.1}_{-0.3}$&0.08$^{+0.07}_{-0.15}$&1.00&12.8&9.76$^{+1.02}_{-0.18}$&---\\
\enddata
\tablecomments{A summary of the key properties for the 30 galaxies in our sample at $z >$ 9.7 (the remaining 55 galaxies with 8.5 $\leq z <$ 9.7 are presented in the appendix). $\mathcal{C}_{FUV}$ is defined in \S 3.4 as the rest-frame far-UV color.  The photometric redshift is ``za'' from \texttt{\textsc{EAZY}}, which is the redshift where the $\chi^2$ is minimized.  The $\int_7^{20} \mathcal{P}(z)$ quantity is the integrated redshift probability density between $z =$ 7 and 20, which was used in the sample selection.  The $\Delta \chi^2$ compares the best-fitting low-redshift (0 $< z <$ 7) model to the best-fitting high-redshift model; a value of $\geq$ 4 was required for selection.  Spectroscopic redshifts come from \citet{arrabalharo23a} and \citet{arrabalharo23b}; we list ``Nz" when an object was spectroscopically observed but no robust redshift was determined.  As discussed in the text, this is modest evidence in favor of $z >$ 9.6, as at lower redshifts [\ion{O}{3}]+$H\beta$ should have been detectable.}
%\vspace{-8mm}
\end{deluxetable*} \label{tab:highz}

In Figure~\ref{fig:rh_stars} we show the results of this analysis.  Here we plot the \textsc{SE} measured F200W half-light radius for our 88 candidate galaxies versus their apparent F200W magnitude, compared to the half-light radius of the F200W PSF as measured from stars in the image.  One can see that the majority of our galaxy sample is clearly resolved, thus non-stellar in origin.  To diagnose the potential stellar nature of the more compact objects, the data points are color-coded by the difference in the goodness-of-fit ($\chi^2$) between the best-fitting (sub)stellar model and the best-fitting EAZY model.  We find that no objects are better fit by the stellar model (e.g., the EAZY $\chi^2$ is always lower).  We do find two sources where the difference between the stellar and EAZY $\chi^2$ is $<$ 4.  One is significantly resolved, but the other object (ID$=$34925) is measured by \textsc{SE} as being very compact (in fact, unphysically smaller than the PSF). However, we conclude this object is much more likely a galaxy.  First, examining the imaging of this object, it does not appear to be obviously point-like; rather, it is faint, barely above our significance thresholds, thus the \textsc{SE} half-light radius is quite uncertain.  Second, due to its faint brightness, its implied distance (were it stellar in origin) would be $\sim$4 kpc, which would be extremely far into the halo, and thus highly unlikely.  We conclude that we find no evidence for stellar contamination in the sources in our sample.

\subsection{Sample Summary}\label{sec:summary}

After removing visually-identified spurious sources, the four sources with $z_{spec} <$ 8.5, and the faint $z \sim$ 16 candidate which is likely at $z \sim$ 4.9, our sample contained 88 candidate $z >$ 8.5 galaxies. For our analysis we divide these candidate galaxies into three sub-samples: $z \sim$ 9, which contains the 58 galaxies with 8.5 $\leq z_{best} \leq$ 9.7; $z \sim$ 11, which contains the 27 galaxies with 9.7 $< z_{best} \leq$ 13; and $z \sim$ 14, which contains the three galaxies with $z_{best} >$ 13.

For all objects in our sample we calculate an observed rest-UV absolute magnitude following \citet{finkelstein15}.  Briefly, we perform a simple round of SED fitting with BC03 \citep{bruzual03} models to derive a best-fitting model spectrum.  We then calculate the bandpass-averaged flux from this spectrum in a top-hat filter curve spanning 1450--1550 \AA\ in the rest-frame, converting to an apparent magnitude and then applying the cosmological distance modulus for a given redshift.  As a part of this process, we run Monte Carlo simulations sampling the photometric redshift $\mathcal{P}(z)$, such that the resulting uncertainty on $M_{1500}$ is inclusive of both the photometric scatter and redshift uncertainty.  We note that in this process we set $\mathcal{P}(z<6) =$ 0, such that any low-redshift solutions (which are small by design) do not bias the magnitude calculation.  During this process we also calculated a rest-far-UV color (discussed in \S 5), dubbed $\mathcal{C}_{FUV}$, calculated as
\begin{equation}
    \mathcal{C}_{FUV} = -2.5~\log_{10}~\left(\frac{f_{1470}}{f_{1850}}\right),
\end{equation}
where $f_{1470}$ and f$_{1850}$ are the bandpass averaged fluxes in top-hat filters spanning 1430--1510 and 1800-1900, respectively (where these windows were designed to probe the color in the far-UV avoiding strong spectral features).  %We also record the stellar mass from these fits for use in our discussion in \S \ref{sec:discussion}.

\begin{figure*}
\epsscale{1.15}
\plotone{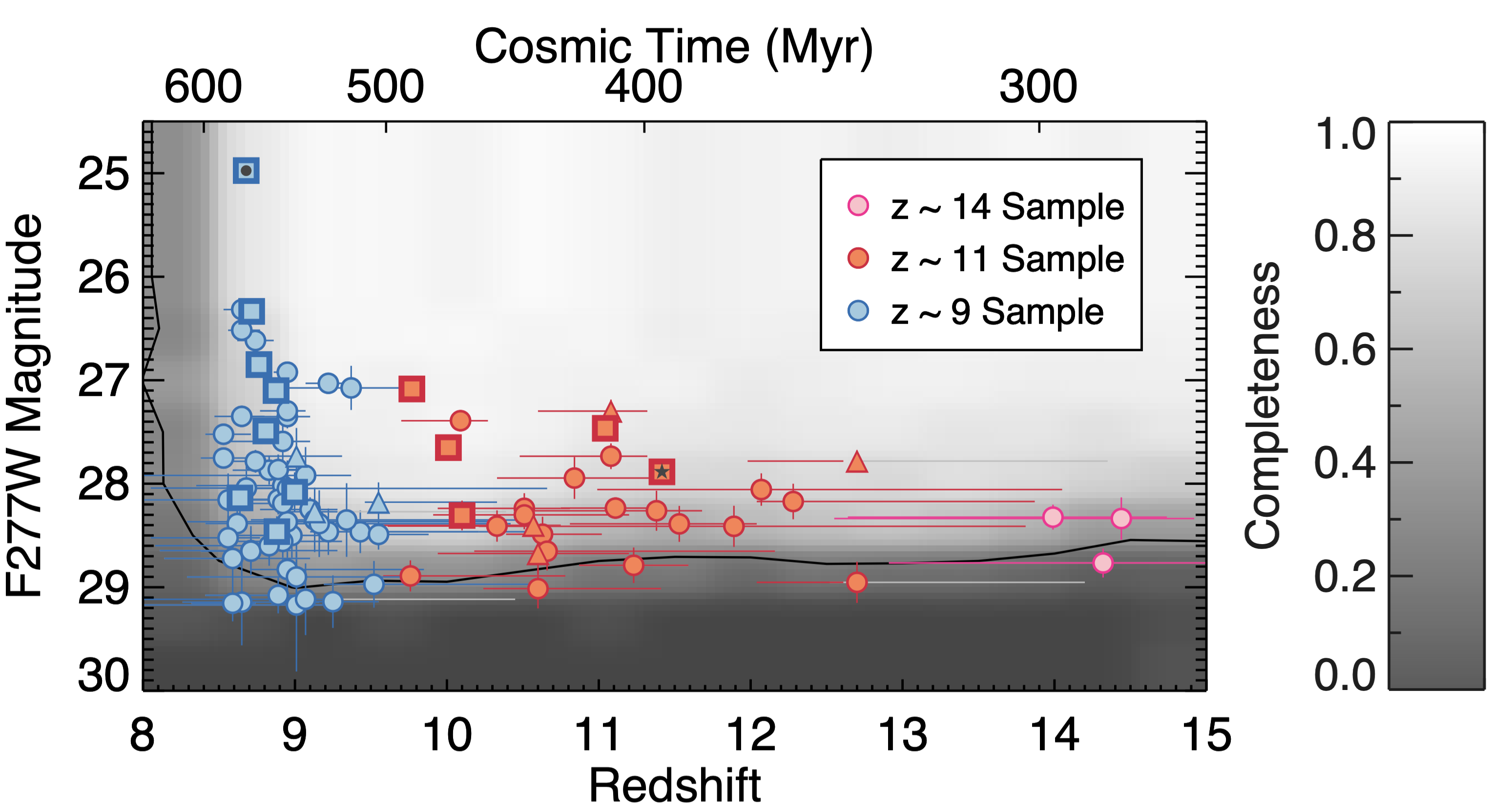}
\caption{The symbols show our sample of 88 $z >$ 8.5 galaxy candidates in a plane of F277W magnitude versus redshift, with the different colors representing the different redshift samples.  Squares denote objects with spectroscopic confirmation, while the circles are plotted at the photometric redshifts.  Triangles denote objects with spectroscopic observations but no confirmation.  The small star denotes Maisie's Galaxy \citep{finkelstein22c}, one of the first {\it JWST} very high-redshift galaxy discoveries, while the small dot denotes CEERS-1019, a confirmed $z =$ 8.7 galaxy which appears to have broad H$\beta$ emission, indicative of an active super-massive black hole \citep{larson23a}.  The background shading shows the completeness (inclusive of both photometric and sample selection completeness) of our sample (for sources with half-light radii of 3.3 pixels), as described in \S 4.1; the black line shows the 20\% completeness contour.}
\label{fig:zmag}
\end{figure*}

\begin{figure*}[!t]
\figurenum{4}
\epsscale{0.57}
\plotone{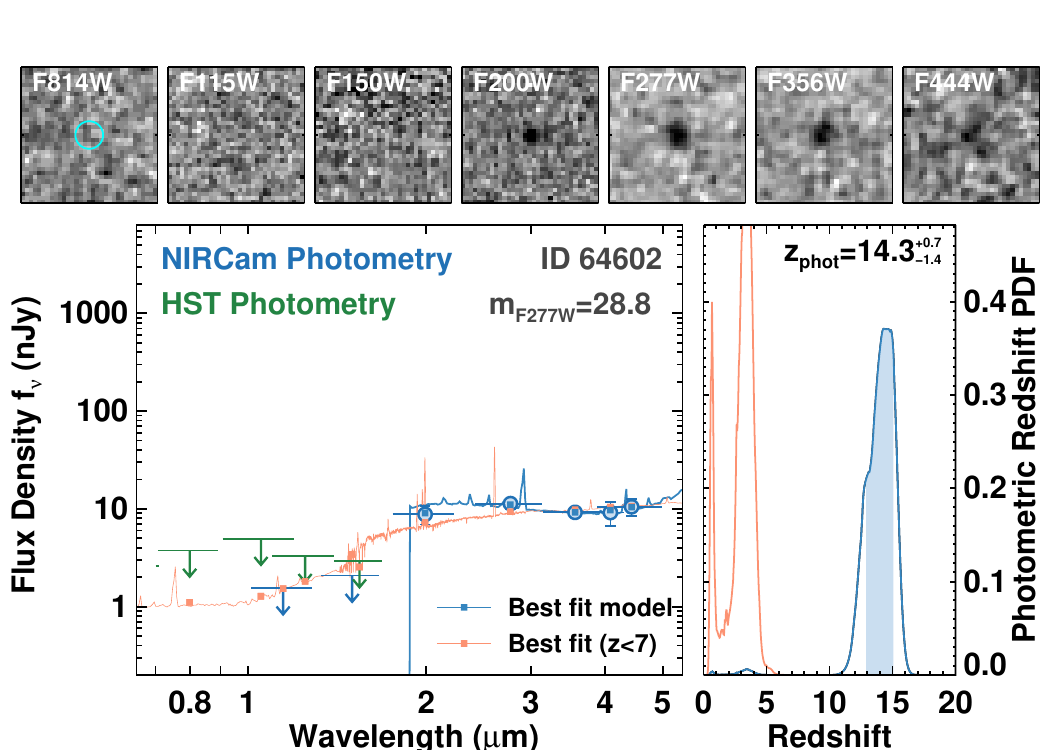}
\plotone{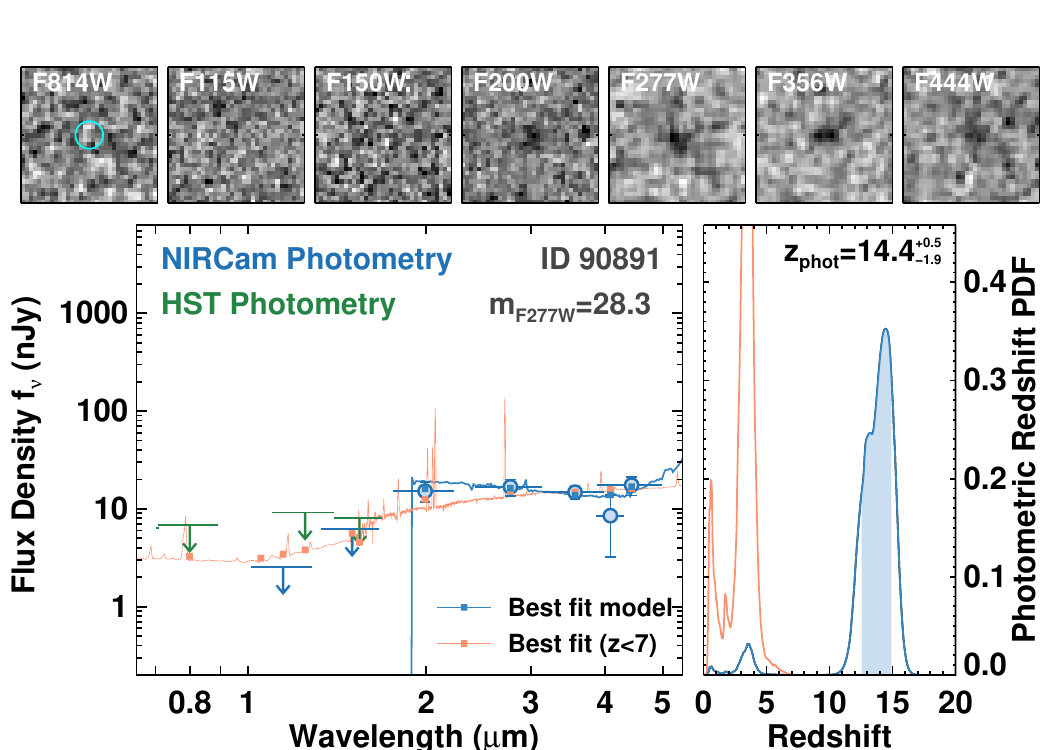}
\plotone{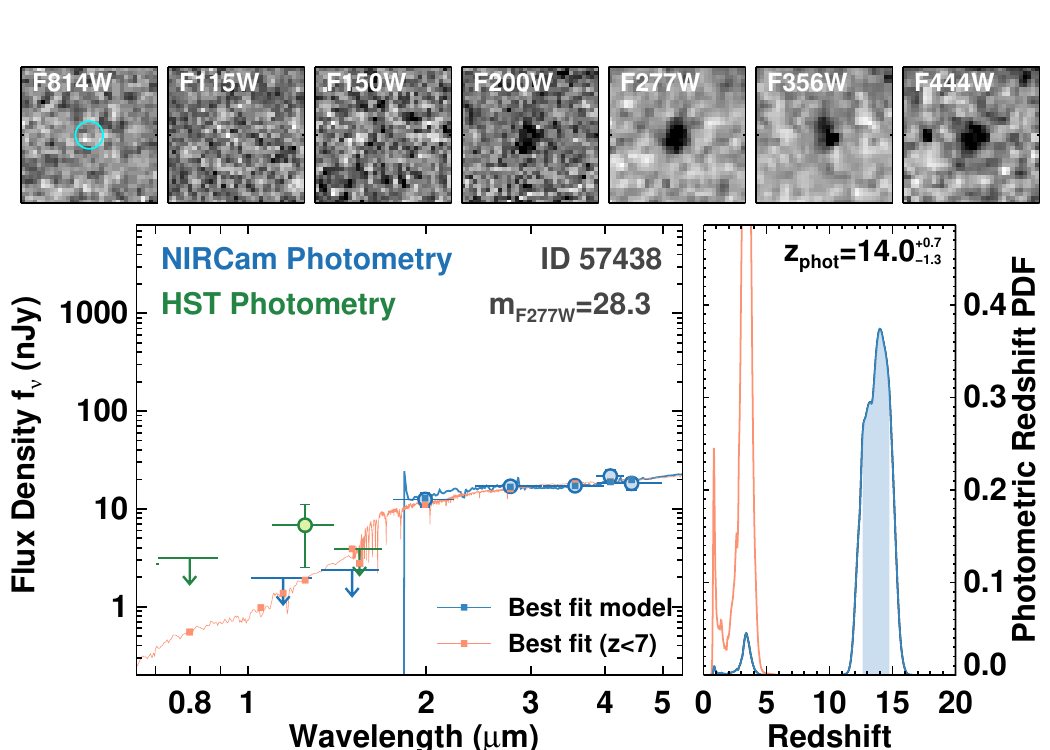}
\caption{Summary figures of the three objects in our sample with $z_{best} >$ 13.  The top row shows 1\arcs\ cutout images in the F814W, F115W, F150W, F200W, F277W, F356W and F444W filters (the F410M and remaining {\it HST} bands are not shown for brevity).  The cyan circle in the first panel shows a 0.2\arcsec\ diameter circle (the size used to measure detection significance).  The bottom-left panel shows the SED, with blue (green) points representing photometry from NIRCam ({\it HST}).  Upper limits shown are 1$\sigma$.  The bottom-right panel shows the photometric redshift distribution in blue (with the corresponding best-fit model in blue in the bottom-left panel).  The light red curves show the photometric redshift results when constrained to $z <$ 7.}
\label{fig:z14bioplots}
\end{figure*}

\begin{figure*}
\figurenum{5}
\epsscale{0.57}
\plotone{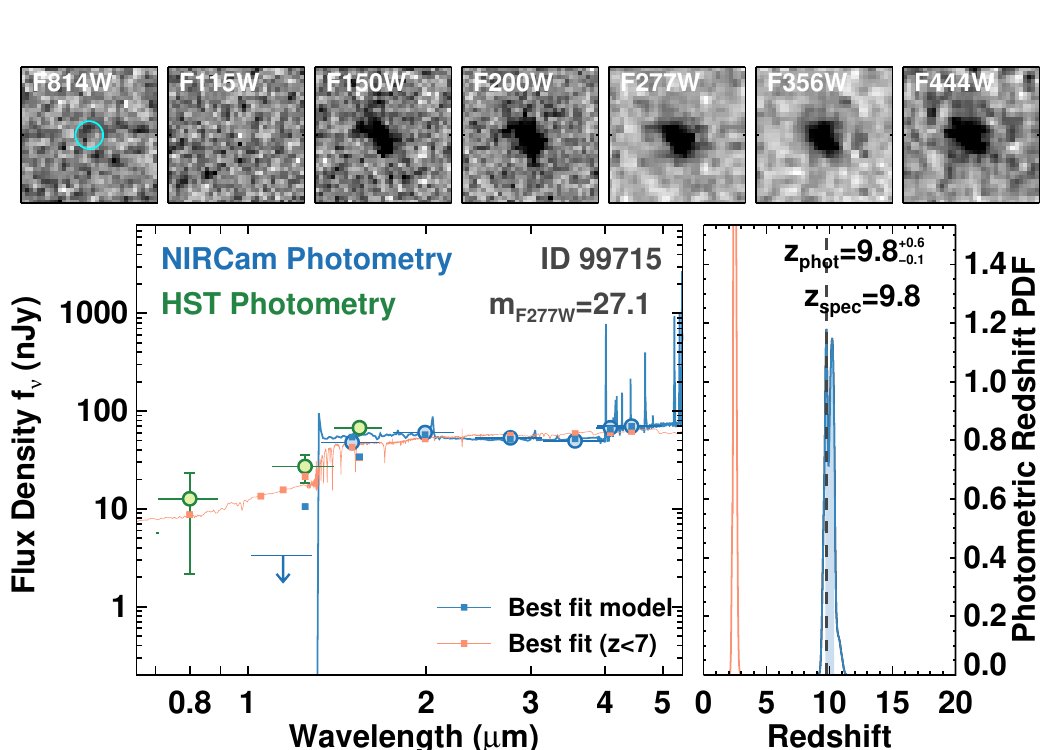}
\plotone{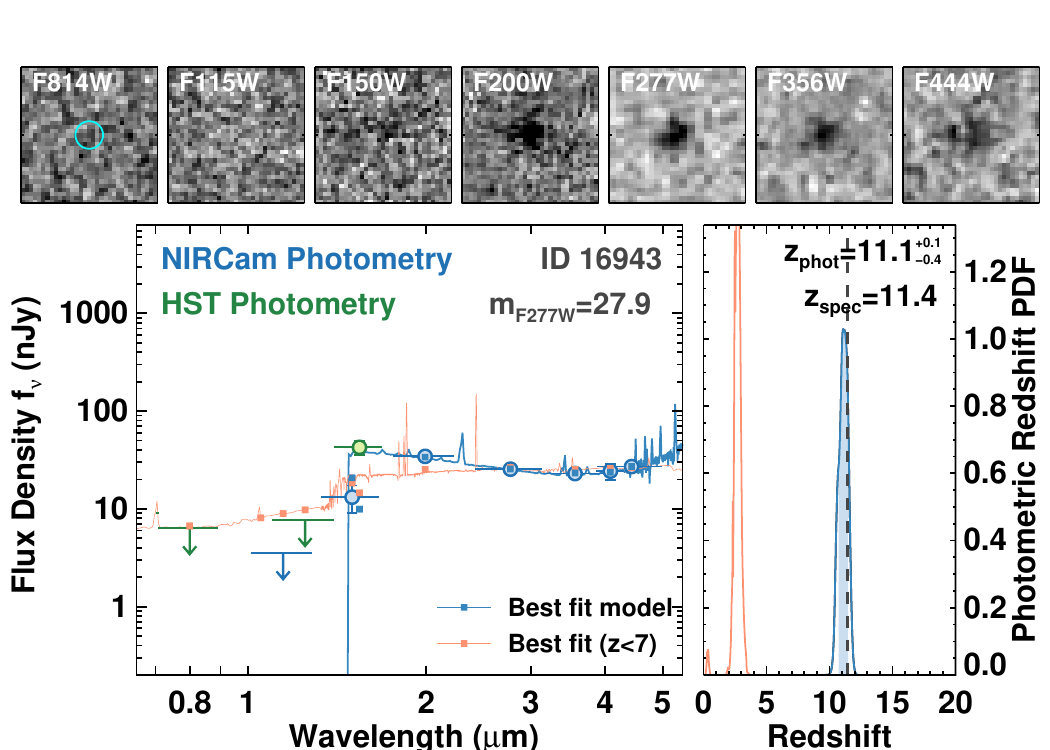}
\caption{Same as Figure 4, for two example sources in the $z \sim$ 11 galaxy candidate sample.  The object on the left has a confirmed spectroscopic measurement via detection of the Ly$\alpha$ break from \citet{arrabalharo23b}, while the object on the right is the spectroscopically confirmed Maisie's Galaxy \citep{finkelstein22c,arrabalharo23a}.  The complete figure set (27 images) is available in the online journal.}
\label{fig:z11bioplots}

\end{figure*}

\begin{figure*}
\figurenum{6}
\epsscale{0.57}
\plotone{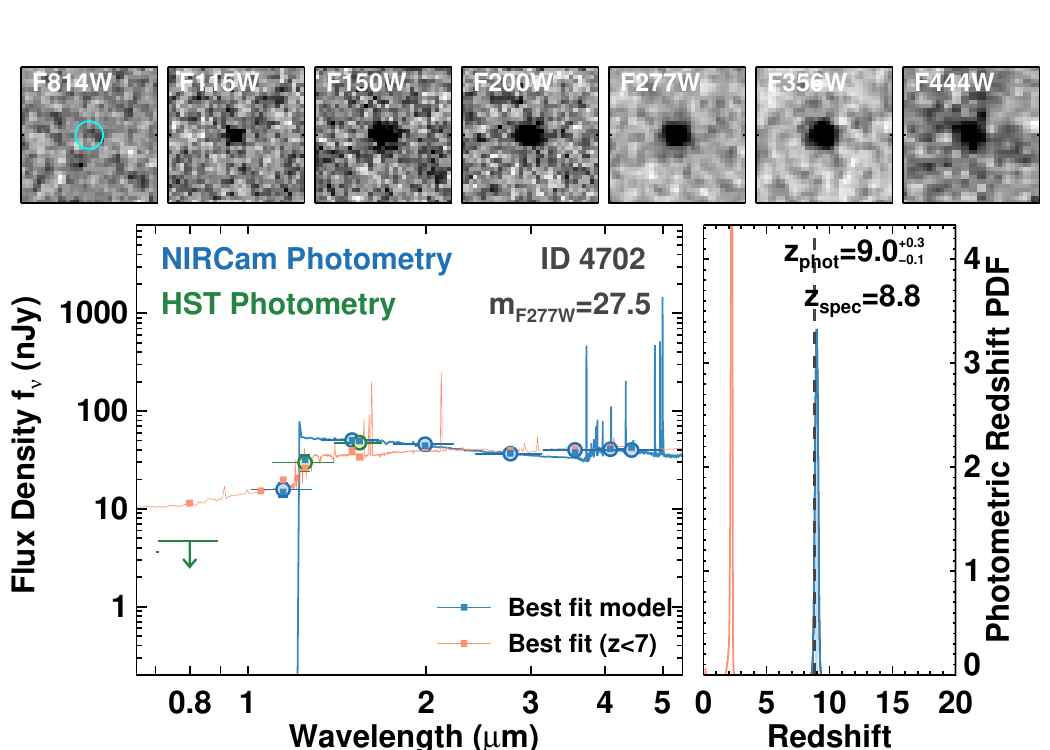}
\plotone{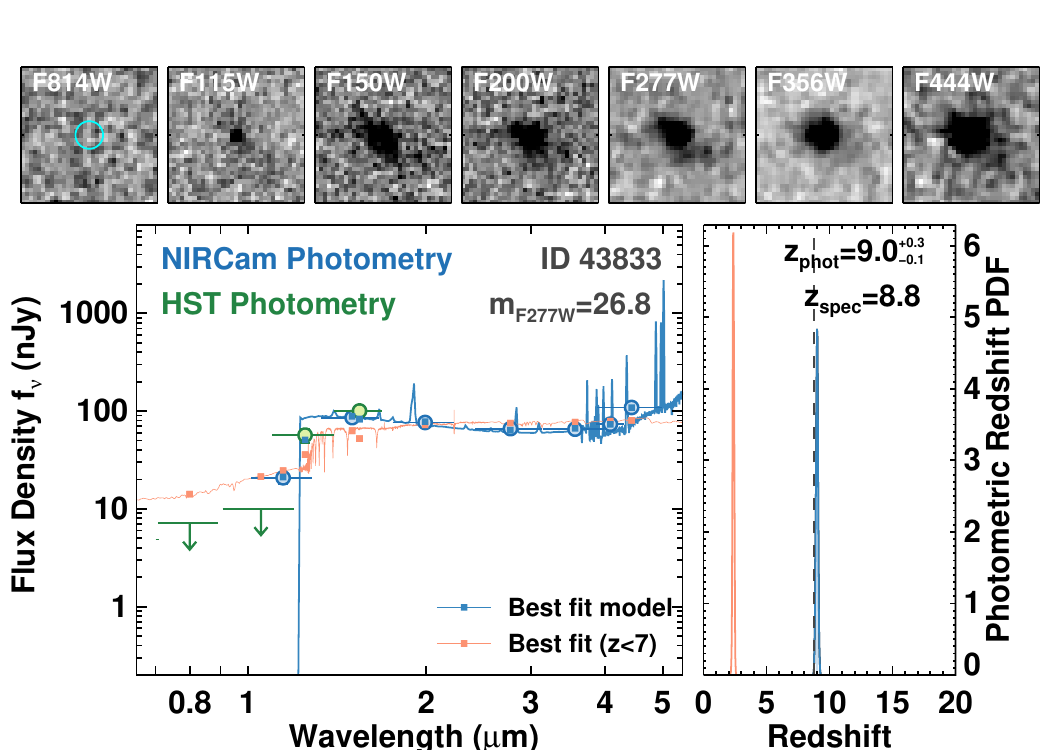}
\caption{Same as Figure 4, for two example sources in the $z \sim$ 9 galaxy candidate sample.  Both example sources have confirmed spectroscopic measurements via [\ion{O}{3}] line emission from \citet{fujimoto23}.  Of note is that both sources were beyond the detection limit of the CANDELS imaging in this field, yet appear very bright and extended in these $\sim$2900 sec {\it JWST}/NIRCam data.  The complete figure set (58 images) is available in the online journal.}
\label{fig:z9bioplots}
\end{figure*}

We show the distribution of our sample in F277W magnitude and photometric redshift in Figure~\ref{fig:zmag}.  Figures~\ref{fig:z14bioplots}, \ref{fig:z11bioplots}, \ref{fig:z9bioplots} contain summary figures of sources in the $z \sim$ 14, 11 and 9 samples, respectively.  The latter two are figure sets, with two example objects shown, and with all objects viewable in the electronic version of the manuscript.  We tabulate key properties of our sample in Table~\ref{tab:highz} for $z >$ 9.7, and Table~\ref{tab:lesshighz} in the Appendix for $z <$ 9.7.

\subsubsection{Comparison to \citealt{mcleod23} and \citet{adams23}}\label{sec:m23}

We compare our sample to that of \citet{mcleod23} and \citet{adams23}, who have also selected $z \gtrsim$ 8.5 galaxies from the full CEERS dataset.  \citet{mcleod23} restrict their sample to galaxies with detection signal-to-noise greater than eight, thus their sample is smaller than ours, at 23 galaxies.  We find that 14 of these 23 galaxies are in our final sample.  We explored the properties of the remaining nine galaxies in our catalog.  We find that 2/9 objects have dominant $z >$ 9 solutions, but fail our $\Delta \chi^2$ criterion.  The remaining seven objects all have plausible high-redshift solutions (based on little-to-no detectable flux in F115W), but have dominant low-redshift solutions.  

While \citet{adams23} did not provide a catalog in their paper, we obtained the revised version of their sample via private communication.  Their final sample of $z >$ 8.5 galaxies consists of 55 objects, of which 25 are in our final $z >$ 8.5 galaxy sample.  All 30 not contained in our sample are present in our parent photometric catalog, all with a plausible redshift solution at $z >$ 7 (16/30 have $>$50\% of their integrated $\mathcal{P}(z)$ at $z >$ 7 in our catalog).  Two of these sources nominally pass all our selection cuts except their best-fitting photometric redshifts are $z \sim$ 7.5--8.2.  Of the remaining 28, 26 do not pass our $\Delta \chi^2$ cut, including 18 which do not pass our integrated $\mathcal{P}(z)$ cut.  This is likely driven in some cases by weakly positive ($\geq$1$\sigma$ significance) flux present in F606W and/or F814W in 14/28 objects, down-weighting high-redshift solutions.

%%%%% Figure sets for bio plots %%%%

\setcounter{figure}{6} 
%%%%%%%%%%%%%%%%%%%%%%%%%%%%%%%%%%%%

\subsubsection{Differences from \citet{finkelstein23}}\label{sec:f23}
In F23 we presented 26 $z >$ 8.5 candidates from the first epoch of CEERS, using 4 of the 10 CEERS NIRCam fields used here.  As our photometry and sample selection procedures here are slightly updated, we cross-checked our sample with the F23 sample.  We find that 20/26 galaxies presented in F23 are included in our sample here.  Of the six galaxies not included here, four of them originally satisfied our sample selection but were removed because they are now known to have spectroscopic redshifts $z_{spec} < 8.5$ (see \S 3.2.1).  These are ID$=$4774 (F23 ID 3908), 4777 (F23 ID 3910), 13256 (F23 ID 2159), and 23084 (F23 ID 1748).  The remaining two sources satisfied all sample selection criteria except the SNR below the Ly$\alpha$ break.  ID 2241 (F23 ID 1875) has a 1.9$\sigma$ detection in F606W and 1.98$\sigma$ detection in F814W, failing that criterion.  ID 8497 (F23 ID 7227) has a 2.6$\sigma$ detection in F115W, which fails this criterion for this object's $\mathcal{P}(z)$ distribution, which peaks at $z =$ 11.2.

In our present work over these four fields we identify 35 $z >$ 8.5 candidates, which includes 20 galaxies from F23 and 15 new sources.  We explored these new sources to see why they were not included in the F23 sample.  All 15 sources are present in the F23 photometric catalog.  Of these 15, 11 sources have best-fit photometric redshifts in the \citet{finkelstein23} catalog of $z >$ 8.5.  Comparing the photometric redshifts of these 11 sources in this previous catalog to our own, we find a median photometric redshift difference of zero (with a mean difference of 0.11, with the new values being slightly smaller).  The majority of these sources (8/11) were previously excluded as they failed the $\Delta \chi^2$ criterion (with values of $\sim$ 1–3).  The other three fell just on the other side of the F23 best-fit photo-$z$, detection threshold, or Ly$\alpha$ break non-detection thresholds.  Of the four objects fit at lower redshift, all three have significant high-redshift peaks, with $\int \mathcal{P}(z) >$ 7 of 0.73 (ID 10545), 0.43 (ID 17898) and 0.66 (ID 61620; with the remaining source, ID 20174, having a lower significance peak at $z \sim$ 9).

While this comparison makes it apparent that modest changes to photometry procedures can have not-insignificant effects on the composition of a high-redshift galaxy sample, the changes we implemented to our procedure here over that from F23 were done to increase photometric accuracy, primarily for the colors of faint galaxies.  The fact that all new sources which we select had significant high-redshift solutions in the F23 catalog adds confidence to our sample, though we fully acknowledge that spectroscopic confirmation of a \emph{majority} of this sample is needed for full confidence.  Fortunately, with the power of {\it JWST}, this is possible.

\section{Results}

Here we explore constraints on the abundance of galaxies at $z >$ 8.5 that we can place with our sample selected from the full CEERS survey.  In \S 4.2 we describe measurements of the cumulative surface density, while in \S 4.3 we describe the rest-frame UV luminosity function.  Both measurements require a correction for incompleteness, which we describe in \S 4.1.

\subsection{Completeness Simulations}\label{sec:completeness}

We quantify the completeness of our sample selection, broadly following F23, with an update here to implement an important size dependence.  Our completeness estimates come from complete end-to-end source injection simulations, injecting mock galaxies with a range of properties into our images, then performing photometry, photometric redshift measurements, and sample selection procedures identical to our that done on our real data.  In this way, we account for incompleteness due to both photometric effects, as well as sample selection effects.

For each of the 10 CEERS NIRCam pointings we run 50 simulation iterations.  Within each iteration we simulate 1000 sources over a uniform range of redshift from 8 $< z <$ 17.  The majority of the iterations had a log-normal distribution in F277W apparent magnitude peaking at $m \sim$ 29; we supplement these with additional simulations with a flat magnitude distribution to boost the number of brighter galaxies.  The result is a roughly flat distribution from m$_{F277W} =$ 22--26, with a larger, log-normal-shaped distribution from  m$_{F277W} =$ 26--30.

To simulate the fluxes in all observed {\it HST} and {\it JWST}/NIRCam filters, we use BC03 \citep{bruzual03} stellar population models with a distribution of stellar population age, dust attenuation and metallicity tuned to reproduce the expected (and now observed, e.g., \citealt{cullen23}) blue colors of very high-redshift galaxies (see \citealt{finkelstein15} for details on these models).  The result is a log-normal distribution of rest-UV colors, which peaks at F200W$-$F277W = $-$0.05, with a 68\% spread from F200W$-$F277W $= -$0.25 to $+$0.3, comparable to the measured colors of our observed objects (median of $-$0.1, 68\% spread from $-$0.3 -- 0.3).
%spectral slope $\beta$ which peaks at $\beta = -$2.3, with a shallow tail to $\beta = -$0.5, and a sharper drop to $\beta = -$3.2.  
The resulting model spectrum was then normalized to the F277W magnitude for a given object, with magnitudes in the remaining filters derived by integrating this spectrum through a given bandpass.

Source morphologies were created using \textsc{galfit} \citep{peng02} assuming a S\'{e}rsic profile with a log-normal distribution in the  S\'{e}rsic index $n$ (peaking at 1.2, with a minimum of $n=$1 and a tail to $n=$5), a log-normal distribution of the axis ratio with a peak at 0.8, and a uniform distribution of the position angle.  While in F23 we tuned the half-light radius ($r_h$) distribution such that the recovered $r_h$ distribution matched that observed in our sample, here we update our procedure to implicitly include the size of galaxies in our sample in the completeness calculation.  We thus use a uniform input distribution of $r_h$ from 1--8 pixels (0.03 -- 0.24\arcs).  Finally, the empirical PSF described in \S 2 is provided to \textsc{galfit} for a given band.

These simulated galaxy images are then created with \textsc{galfit} as 101$\times$101-pixel images, which we add at random positions to the real images (avoiding only image edges and regions of extremely high ERR-map values [$>$1000]).  Notably we do not avoid the positions of real objects, such that our simulations account for incompleteness due to sources along similar lines-of-sight to high-redshift galaxies.  We also create a full-frame image just of the simulated sources in F277W, on which we run \texttt{\textsc{SE}} to measure a ``noiseless'' version of the recovered \texttt{\textsc{SE}} $r_h$ in this filter for use with the completeness calculations (as we discuss below, the measured \texttt{\textsc{SE}} $r_h$ is biased low from the \textsc{galfit} input value, even in the noiseless image).

\begin{figure*}[!t]
\epsscale{1.15}
\plotone{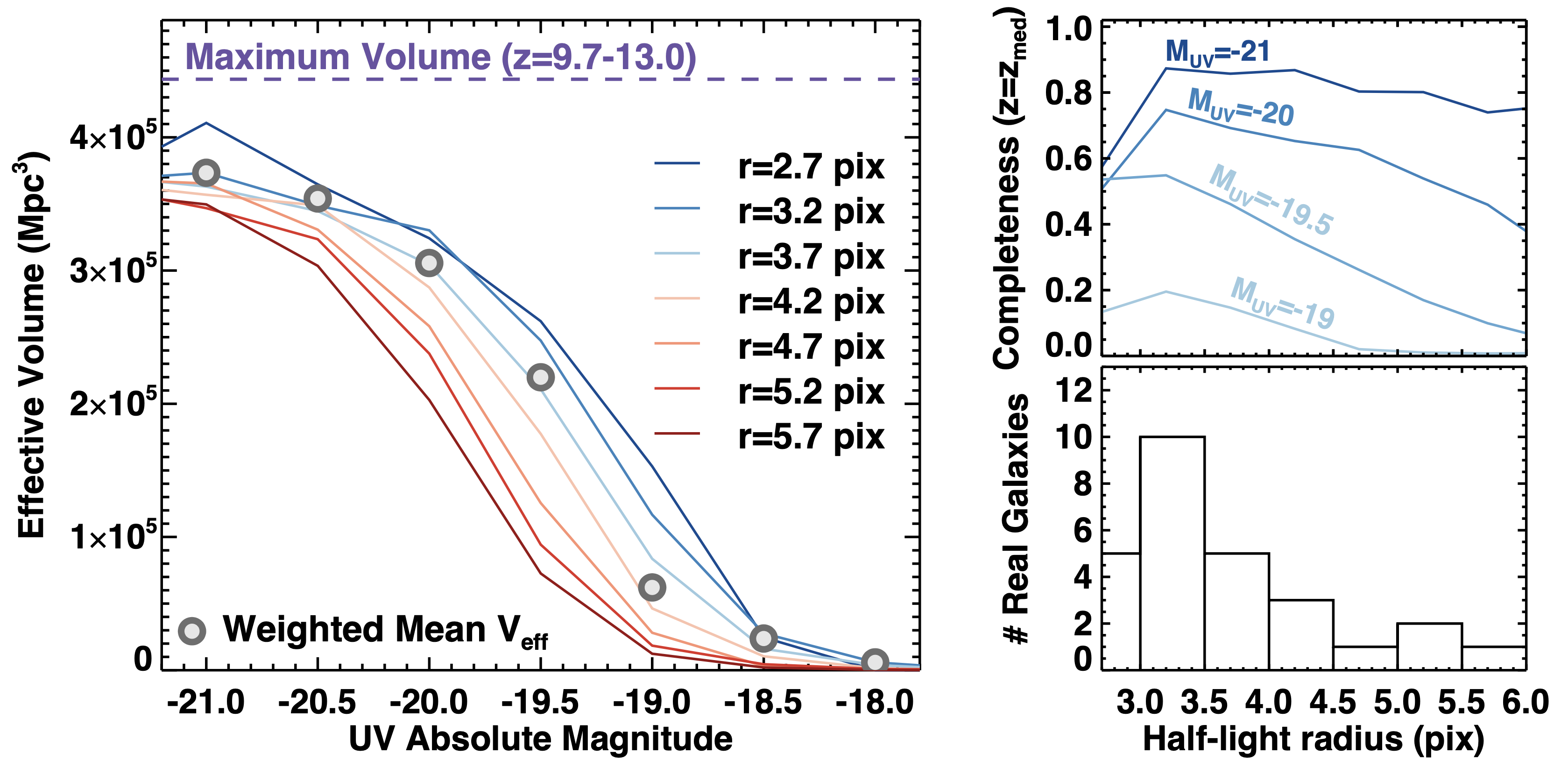}
\caption{The left panel shows the effective volume for our $z \sim$ 11 (9.7 $< z \leq$ 13) galaxy sample.  The purple dashed line shows the maximum volume one would obtain with a 100\% completeness over an ideal (though unrealistic) top-hat selection function.  The colored curves show our calculated effective volumes as a function of source half-light radii.  The circles show the volumes we use, calculated by weighting the volumes by the sizes of galaxies in a given magnitude bin.  The size distribution of our $z \sim$ 11 galaxy sample is given in the lower-right plot, while the completeness as a function of size at fixed input UV absolute magnitude is shown at the upper-right.  The completeness is very sensitive to the size, particularly at fainter magnitudes.}
\label{fig:veff}
\end{figure*}

The result of this process is a version of our data in all filters with these 1000 simulated sources included, as well as a catalog of their input properties (inclusive of the  \texttt{\textsc{SE}} $r_h$ measurement of the input image).  At this point, we run an identical process on these images as we did on our real data in \S 2 and \S 3, including creating the array of PSF-matched images, calculating photometry, aperture corrections, and PSF corrections.  Photometric catalogs are created, \textsc{EAZY} is run, and $M_{1500}$ is calculated.

The recovered catalogs are then matched against the input catalogs, resulting in a combined catalog of all recovered sources with their input and recovered properties.  As the imaging depths in the 10 CEERS fields are broadly similar (Table~\ref{tab:tab1}), we combine these catalogs from all 10 fields into one.  This final input catalog consists of 500,000 simulated sources, of which 343,196 sources are recovered by \texttt{\textsc{SE}} (the $\sim$70\% recovery fraction is expected given the larger number of very faint sources that were simulated).

While in F23 we calculated completeness only as a function of magnitude (with an assumed size distribution), here we explicitly add a size dependence, as many of the observed high-redshift galaxies are clearly resolved.  For a given bin in absolute magnitude, we calculate the completeness in bins of F277W half-light radius (with a bin width of 0.5 pixels, starting with a bin centered at 1.2 pixels).  For this process, we assume the intrinsic size as the half-light radius measured by \texttt{\textsc{SE}} on the noiseless images, as described above.  We show how the completeness depends on both UV absolute magnitude and half-light radius in Figure~\ref{fig:veff}.  While there is minimal dependence on size at the brightest magnitudes, as one goes fainter the completeness decreases more steeply for larger sources, as expected.

As we might expect the recovered sizes from both the simulated and real data to be biased, we compare these intrinsic sizes to those measured from the recovered sources. We find that the intrinsic sizes are $\sim$50\% larger.  To account for this bias when we calculate the completeness we multiply the recovered \texttt{\textsc{SE}} sizes of our real sources by this scale factor of 1.53 (this quantity has a small scatter of $\sigma=$0.06, and is independent of magnitude for $m < 29$).  We then set a minimum size for real sources of r$_h =$ 2.7 pixels, comparable to that expected for an unresolved source.  We also set a maximum size of r$_h =$ 6.0 pixels; one object has a larger (bias-corrected) \texttt{\textsc{SE}}-measured size, yet it is clear from the images these sizes are incorrectly measured due to the presence of a neighboring source.  For this object, we assume the median size from the sample.

\begin{figure*}[!t]
\epsscale{1.0}
\plotone{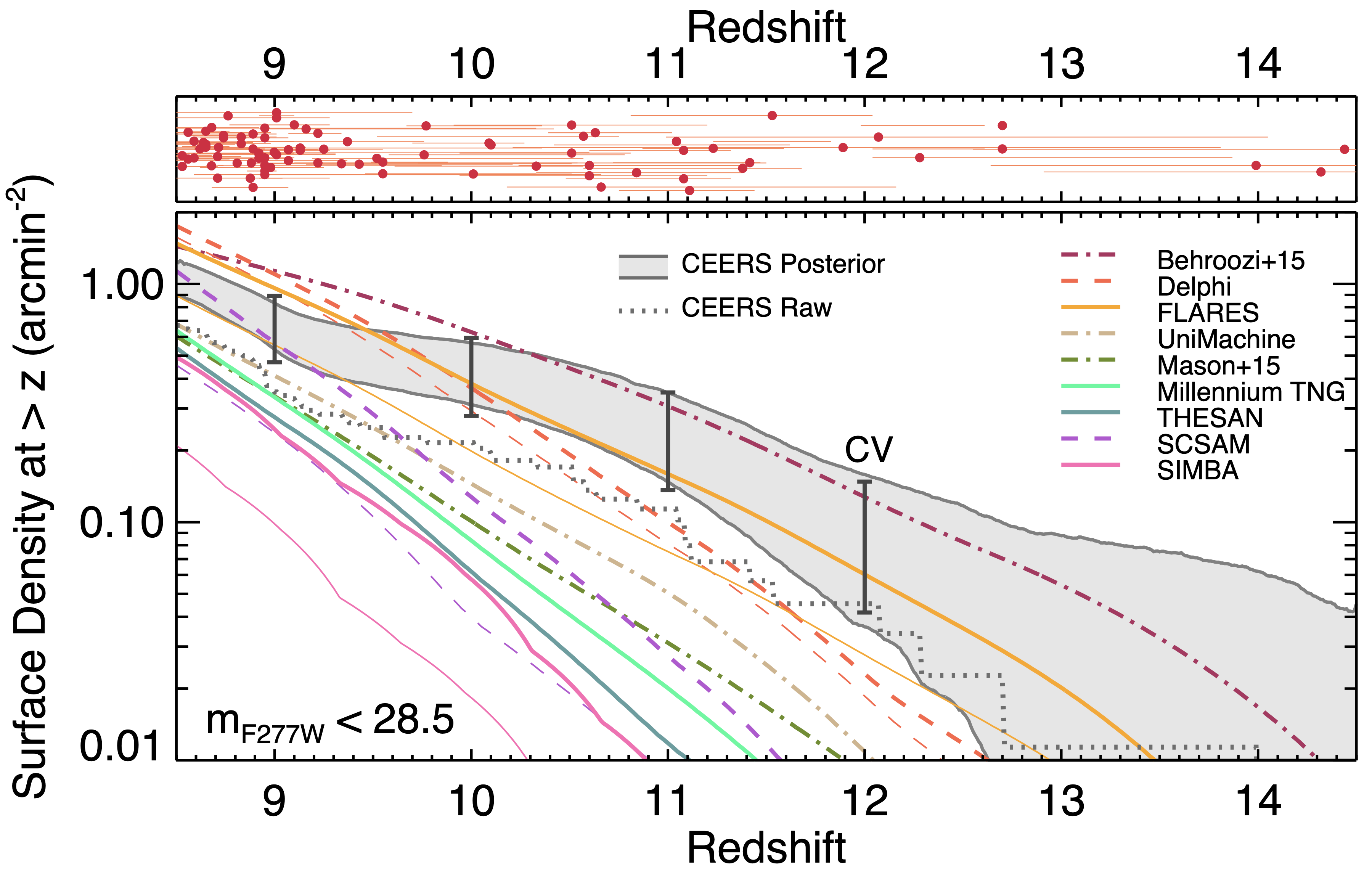}
\caption{The top panel shows the redshift distribution of our sample (scattered vertically for clarity), while the main panel shows the cumulative surface density of galaxies per unit surface area at redshift greater than $z$ for $m <$ 28.5.  The dotted line shows the raw counts from our CEERS $z >$ 8.5 galaxy sample, while the shaded region shows the 68\% confidence interval on the completeness corrected surface density (including only sources with completeness estimates $>$20\%), inclusive of photometric, photometric redshift, and Poisson uncertainties.  The vertical error bars show the estimated cosmic variance uncertainty.  The colored lines show predictions from a range of pre-launch simulations (including hydrodynamic, analytic, semi-analytic, and semi-empirical models), all also for $m <$ 28.5.  The observed surface density of galaxies lies above most predictions at $z >$ 10, and above all except the \citet{behroozi15} and FLARES models at $z >$ 11.  This confirms early results based on smaller samples that the observed abundance of $z \gtrsim$ 10 galaxies significantly exceeds most pre-launch physically-motivated expectations.}
\label{fig:sdens}
\end{figure*}

%We calculate the completeness in two ways.  
For our surface density analysis described in \S 4.2, we calculate a completeness per source, based on the fraction of recovered sources at the observed F277W magnitude, photometric redshift, and galaxy size.  For our rest-frame UV luminosity function analysis described in \S 4.3, we calculate the completeness in bins of absolute UV magnitude.  In each magnitude bin we calculate a completeness by weighting the recovery fractions in each size bin by the observed number of sources at each size.  We then integrate the comoving volume element across our redshift range with these completeness values, resulting in an effective volume for our sample in a given magnitude bin (a process similar to that done in \citealt{finkelstein15}).  These ``weighted mean" effective volumes are shown as the circles in Figure~\ref{fig:veff}.  They broadly trace the completeness curves for sources with r$_h \sim$ 3.5 pixels, comparable to the median size of objects in our sample.  One can see that assuming point sources in these simulations would lead to effective volume estimates modestly larger across $M_{UV} \sim -$20 to $-$19.  Our final effective volumes are listed in Table~\ref{tab:lftab}.

\subsection{The Cumulative Surface Density of Galaxies at $z >$ 8.5}

Here we explore the evolution of galaxies at $z >$ 8.5 via the cumulative surface density, updating the results from the first epoch of CEERS from F23.  Figure~\ref{fig:sdens} shows the surface density of objects with redshifts greater than $z$ for sources with $m_{F277W} <$ 28.5.  Following F23, we correct for incompleteness by counting each galaxy as one divided by the estimated completeness at the redshift and magnitude of a given galaxy, including here the size of the source (as well as the size bias correction discussed in \S 4.1).  To avoid significant uncertainty introduced by objects with very low completeness values, we further restrict this analysis to objects with completeness measurements greater than 20\% (this includes 58 objects of the 66 with F277W $<$ 28.5; this exclusion was not done in F23). We note that our magnitude limit is chosen as a compromise between maximizing the sample size while minimizing the completeness correction.   While 25\% of our sample is fainter than this limit, the majority of these fainter sources have completeness $<$20\%.  On the other hand, increasing the limit to F277W $<$ 28.0 would cut the sample in $\sim$half, reducing the constraining power of our observations.

The shaded region shows the 68\% confidence range on the cumulative surface density, derived via Monte Carlo simulations inclusive of photometric and photometric redshift uncertainties, as well as sampling the Poisson uncertainty.  We separately show estimated cosmic variance uncertainties at four different redshifts as the vertical error bars calculated based on the \textsc{bluetides} simulation \citep[][with caveats as discussed in F23]{bhowmick20}.  The dotted line shows the raw measurements with no correction.  While not shown for clarity, the values from \citet{finkelstein23} lie at the low end of our posterior distribution here (which is consistent with our finding of $\sim$3.4$\times$ more galaxies in $\sim$2.6$\times$ more area).

We compare here to the same nine {\it JWST} model-based pre-launch predictions as in F23, including semi-analytic models \citep[SAMs;][]{yung19a,yung20b,dayal17}, empirical models \citep{behroozi15,mason15, behroozi19}, and cosmological hydrodynamic simulations \citep{wilkins22a,kannan22,kannan22a,dave19}. As discussed extensively in F23, the compilation of model predictions here is inclusive of several different modeling approaches and assumptions (see \citet{somerville15a} for a thorough review).

Similar to F23, we find that at $z <$ 10, our observations are higher than most predictions, yet are consistent with the  \citet{behroozi15} empirical and \textsc{delphi} SAM models, as well as the FLARES hydro predictions with no dust attenuation.  At $z >$ 10, we find that our results lie significantly above all predictions with the exception of \citet{behroozi15}.  While this agreement is intriguing, this is not a physics-based model, but rather is based on the ansatz that the specific star formation rate in galaxies is proportional to the specific total mass accretion rate into halos. 
%While this model is intriguing it is not a physical model, but rather a fit to the data, extrapolated using the halo mass function (and in fact the UniverseMachine model is an updated prediction from a similar data-oriented analysis).  
We thus confirm the initial result from F23 that the observed abundance of $z >$ 10 galaxies discovered in CEERS lies above nearly all pre-launch predictions, where here the result is at higher confidence based on an updated sample $\sim$3$\times$ larger than the F23 sample. 
%though we do note that the scatter in model predictions is roughly the same scale as the difference between the median model and the data.

\subsection{The Evolution of the Rest-UV Luminosity Function at $z =$ 8.5--14.5}
\begin{figure*}[!t]
\epsscale{1.15}
\plotone{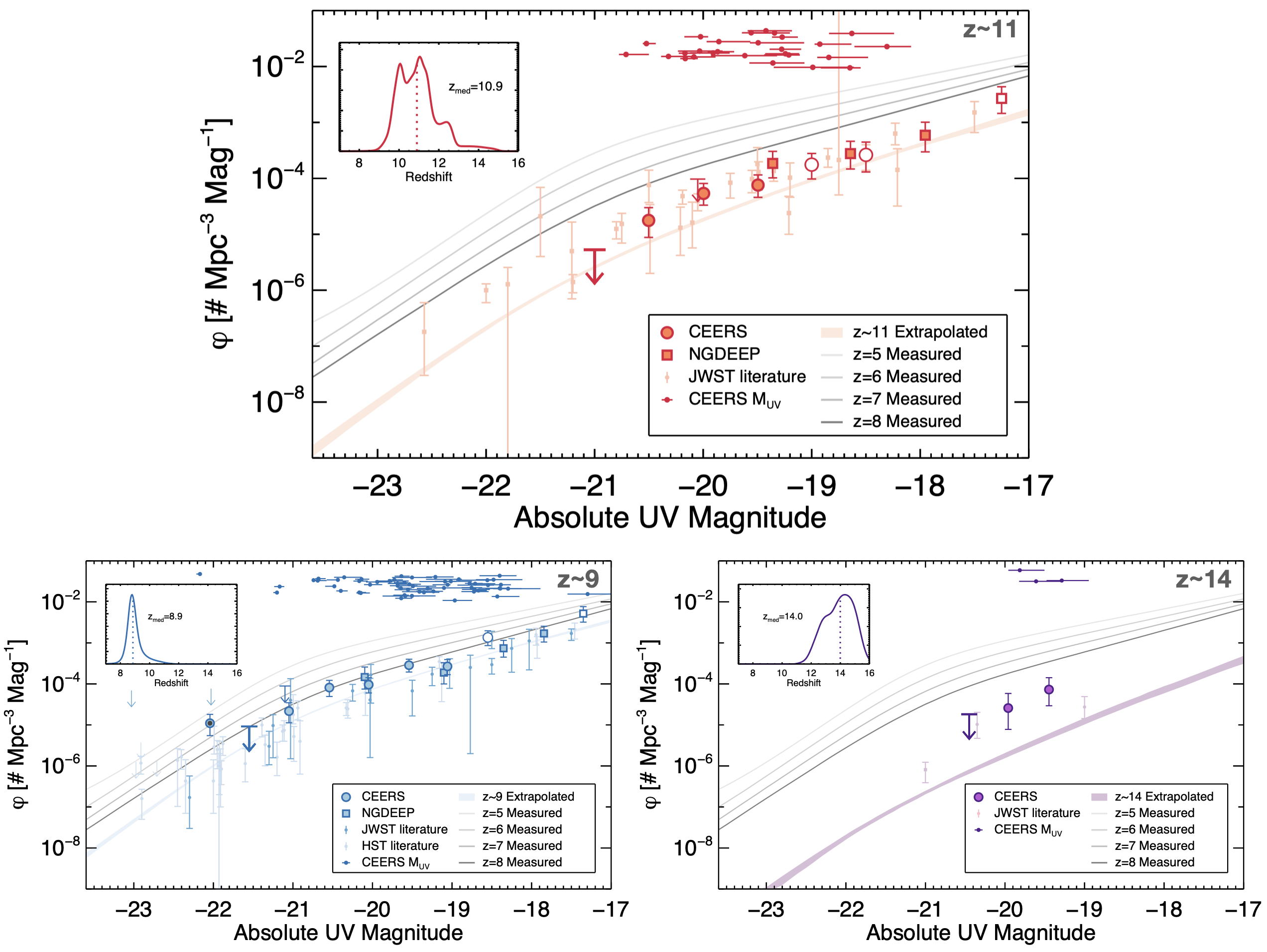}
\caption{The evolution of the rest-frame UV luminosity function, at $z \sim$ 11 (9.7 $< z_{best} \leq$ 13; top), $z \sim$ 9 (8.5 $< z_{best} \leq$ 9.7; bottom-left) and $z \sim$ 14 (13 $< z_{best} \leq$ 15; bottom-right).  The large circles show the calculated number densities from our sample (the small red dots denote the magnitudes of individual galaxies, offset vertically for clarity), while the squares show the results from NGDEEP \citep{leung23}.  The inset shows the stacked $\mathcal{P}(z)$ for each sample, with the dotted line denoting the median value of the $\mathcal{P}(z)$. Arrows show 1$\sigma$ upper limits in the first bin with no galaxies, while white-filled symbols denote bins which are $<$20\% complete.  The black dot in the brightest bin at $z \sim$ 9 indicates that this bin has only one object, the $z =$ 8.7 galaxy (which has AGN signatures) from \citep{larson22}.  Small symbols show literature results.  At $z \sim$ 9 the {\it HST} results are from \citet{bouwens19,bouwens21,bowler20,mcleod16,morishita18,stefanon19,rojasruiz20,bagley22,finkelstein22}, while the {\it JWST} results are from \citet{donnan23a,adams23,harikane23,perezgonzalez23,bouwens23b}.
The $z \sim$ 11 results shown are from \citet{donnan23a,donnan23b,mcleod23,adams23,perezgonzalez23,bouwens23b,harikane23,castellano23,franco23,casey23b}, while at $z \sim$ 14 we compare to \citet{donnan23a} and \citet{casey23b}.
The gray curves show the best-fitting double-power law (DPL) model from \citet{finkelstein22b} from $z =$ 5—8, while the light-shaded colored region shows this model empirically extrapolated to the median redshift for a given sample (where the width is the 68\% uncertainty on the luminosity function at this redshift from \citealt{finkelstein22b}).  Our CEERS results are generally consistent with previous luminosity function estimates, with smaller uncertainties reflecting our larger sample size.  We also note excellent agreement with the NGDEEP results where our samples overlap.  Our brighter CEERS results sit above the expected number densities for the empirically expected extrapolation from \citet{finkelstein22b}, with this offset increasing to higher redshift.} %The results at fainter luminosities (primarily from NGDEEP) are more broadly consistent with the extrapolations.}
\label{fig:lf}
\end{figure*}

\begin{figure}[!t]
\epsscale{1.2}
\plotone{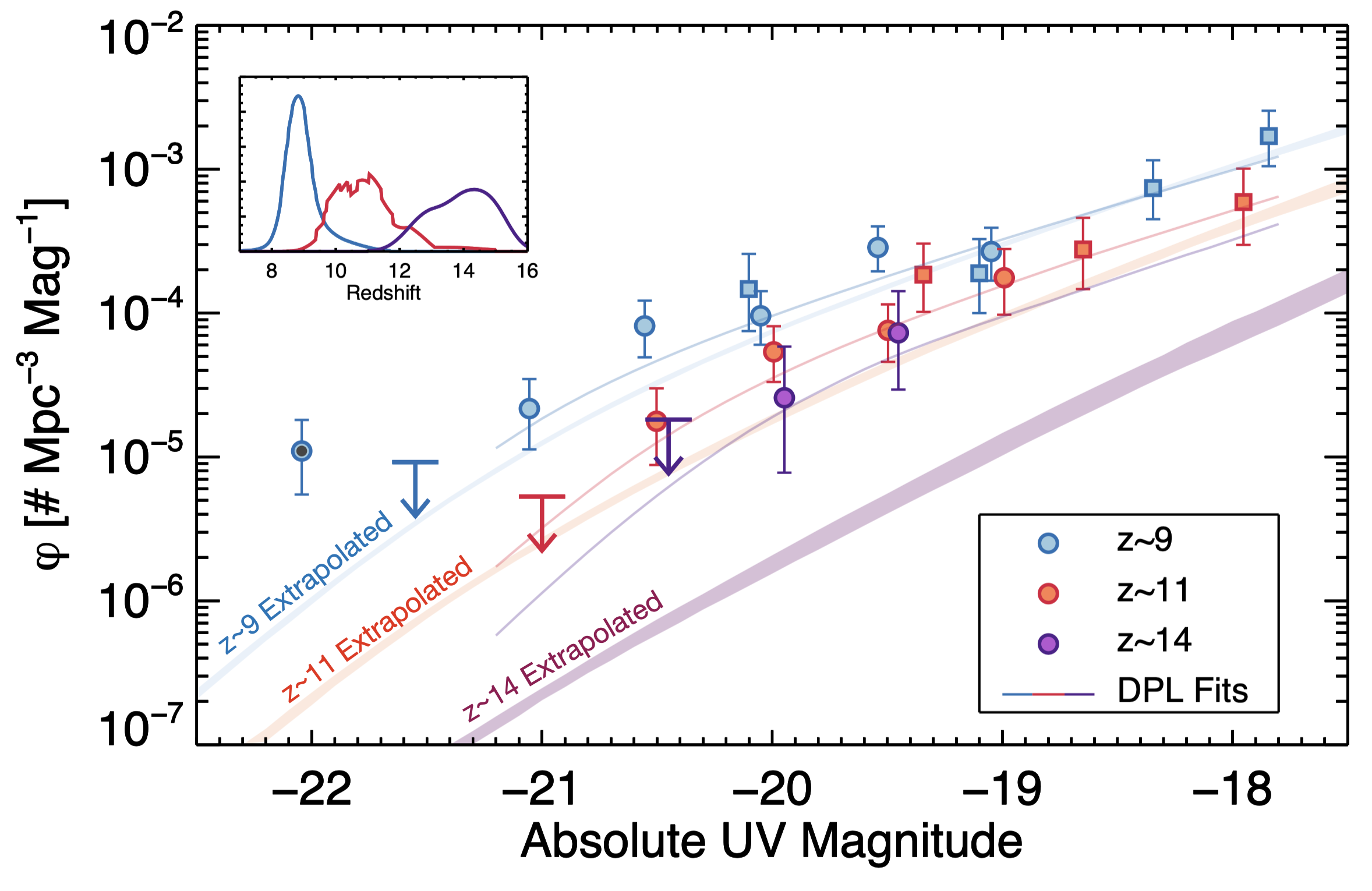}
\caption{Our measured UV luminosity functions at $z \sim$ 9, 11 and 14 are shown by the large symbols (circles for our CEERS results, and squares for NGDEEP from \citealt{leung23}).  The shaded regions are the same as in Figure~\ref{fig:lf}, showing the extrapolated UV luminosity functions from \citet{finkelstein22b}.  The inset panel likewise shows the same $\mathcal{P}(z)$ curves from Figure~\ref{fig:lf}, here plotted on the same scale.  The thin curves show the median DPL fit to the data at each redshift.  This figure highlights that brighter galaxies ($M_{UV} \lesssim -$20) have higher number densities than the extrapolated luminosity functions would predict.  While there is a known overdensity at $z \sim$ 8.7 \citep{larson22,whitler23} which could affect our lowest-redshift bin, there is no evidence for such overdensities at higher redshifts.}
\label{fig:lf_all}
\end{figure}

Here we present our measurements of the rest-frame UV luminosity function, measured separately for our $z \sim$ 9, $z \sim$ 11 and $z \sim$ 14 samples, where galaxies are again placed into a sample based on their best-fitting photometric redshift (or spectroscopic redshift when available).  Stacking the $\mathcal{P}(z)$'s of each sample (as shown in Figures~\ref{fig:lf} and \ref{fig:lf_all}), the median redshift in each bin is $z =$ 8.9, 10.9 and 14.0, respectively.  We note that our redshift bins become progressively wider for two reasons.  First, at these redshifts, unity redshift bins become small in terms of cosmic time spanned.  Second, our boundaries of $z =$ 9.7 and $z =$ 13 correspond to redshifts where the Ly$\alpha$ break falls directly between two NIRCam filters (F115W and F150W, and F150W and F200W, respectively), thus creating natural redshift demarcations given our available data.

We measure our luminosity function following the methodology of F23, here using a bin size of 0.5 magnitudes, with three key differences.  The first is that here we use three redshift bins, and in the bin in common ($z \sim$ 11) here our sample size is nearly $\sim$3$\times$ larger (27, compared to 10 in F23).  The second difference is our use of the size of the sources in the completeness correction, which as shown in \S 4.1 can affect the inferred effective volumes (Figure~\ref{fig:veff}).  When calculating the effective volume, we first calculate the effective volume as a function of absolute magnitude and size by integrating over the co-moving volume element as
\begin{equation}
V_{eff} (M_{UV}, r_h) = \int \frac{dV}{dz} C(z,M_{UV},r_h) dz, 
\end{equation}
where $C(z,M_{UV},r_h)$ is the completeness as calculated in \S 4.1.  In each magnitude bin, we then calculate a final effective volume as the weighted mean of the effective volumes in all size bins weighted by the number of sources in that magnitude bin at a given size (corrected for the $\sim$1.5$\times$ radius bias described in \S 4.1) as
\begin{equation}
    V_{final}(M_{UV}) = \frac{\sum_{i} V_{eff}(M_{UV},r_{h,i}) \times N(M_{UV},r_{h,i})}{\sum_{i} N(M_{UV},r_{h,i})},
\end{equation}
where $i$ is a radius index, and $N(M_{UV},r_{h,i})$ is the number of galaxies in a given UV magnitude and size bin.

The third difference comes when we estimate the number density in each bin via MCMC (see full details on this methodology in \citealt{finkelstein15}).  We do this to sample galaxy absolute magnitude posterior distributions such that galaxies can fractionally span multiple magnitude bins. In F23 we treated each magnitude bin separately.  Here, following \citet{leung23} each step in the MCMC chain samples the entire ensemble of galaxies for a given redshift bin, such that when a galaxy scatters out of one bin, it is accounted for in another bin (and thus is more realistic than what was done in F23).

Our measured number densities and final weighted-mean effective volumes are listed in Table~\ref{tab:lftab} for all three redshift bins.  For bins with no galaxies, we list the 84\% upper limit from the MCMC calculation.  We also demarcate fainter bins with completeness values $<$20\%; while we list these values, we caution that they are dominated by the completeness correction.

\begin{deluxetable}{cccc}
\vspace{2mm}
\tablecaption{CEERS Rest-UV Luminosity Functions}
\tablewidth{\textwidth}
\tablehead{\multicolumn{1}{c}{M$_{UV}$} & \multicolumn{1}{c}{Number$^{\dagger}$} & \multicolumn{1}{c}{Number Density} & \multicolumn{1}{c}{Effective}\\
\multicolumn{1}{c}{(mag)} & \multicolumn{1}{c}{$ $} & \multicolumn{1}{c}{(10$^{-5}$ mag$^{-1}$ Mpc$^{-3}$)} & \multicolumn{1}{c}{Volume (Mpc$^3$)}}
\startdata
\multicolumn{4}{c}{$z \sim$ 9}\\
\hline
$-$22.5&0&$<$0.9&187000\\
$-$22.0&1&1.1$_{-0.6}^{+0.7}$&187000\\
$-$21.5&0&$<$0.9&187000\\
$-$21.0&2&2.2$_{-1.0}^{+1.3}$&193000\\
$-$20.5&7&8.2$_{-3.2}^{+4.0}$&177000\\
$-$20.0&7&9.6$_{-3.6}^{+4.6}$&161000\\
$-$19.5&17&28.6$_{-9.1}^{+11.5}$&120000\\
$-$19.0&10&26.8$_{-10.0}^{+12.4}$&77900\\
\hline
$-$18.5$^{\ddagger}$&13&136.0$_{-49.9}^{+61.0}$&18600\\
\hline
\hline
\multicolumn{4}{c}{$z \sim$ 11}\\
\hline
$-$21.0&0&$<$0.5&373000\\
$-$20.5&3&1.8$_{-0.9}^{+1.2}$&354000\\
$-$20.0&8&5.4$_{-2.1}^{+2.7}$&306000\\
$-$19.5&8&7.6$_{-3.0}^{+3.9}$&220000\\
\hline
$-$19.0$^{\ddagger}$&5&17.6$_{-7.9}^{+10.3}$&62100\\
$-$18.5$^{\ddagger}$&3&26.3$_{-13.3}^{+18.2}$&23800\\
\hline
\hline
\multicolumn{4}{c}{$z \sim$ 14}\\
\hline
$-$20.5&0&$<$1.8&147000\\
$-$20.0&1&2.6$_{-1.8}^{+3.3}$&85400\\
$-$19.5&2&7.3$_{-4.4}^{+6.9}$&60600\\
\hline
\hline
\enddata
\tablecomments{$^{\dagger}$This column represents the nominal number of galaxies in a magnitude bin, though in the calculation of the luminosity function we account for galaxies moving between bins due to photometric and photometric redshift uncertainties.  $^{\ddagger}$The completeness in these bins is $<$20\%, thus these data points are not used to constrain the luminosity function evolution.}
\label{tab:lftab}
\vspace{-8mm}
\end{deluxetable}

In Figure~\ref{fig:lf} we plot these luminosity function measurements  compared to a wide range of literature results and extrapolations.  In this figure we also prominently show the results from NGDEEP at fainter luminosities from \citet{leung23} at $z \sim$ 9 and 11, as their photometry and sample selection process are nearly identical to our own (while we use the \citealt{leung23} measured number densities here, we verified that the results would be very similar had we calculated the number densities with their sample in our slightly modified redshift bins).  At $M_{UV} = -$19, where our results overlap, we find excellent agreement between CEERS and NGDEEP, with the NGDEEP results then continuing on faint-ward at a consistent slope at both $z \sim$ 9 and 11.  

The small faded symbols in Figure~\ref{fig:lf} show results from the literature.  At $z \sim$ 9 we find broadly good agreement with these literature results (which come from {\it HST}, {\it JWST} and ground-based studies).  We note that our brightest bin is higher than previous results - this bin contains a single object, the $z \sim$ 8.7 potential AGN from \citet{larson23a}.  Based on the confirmation of this object and another at a similar spectroscopic redshift, \citet{larson22} concluded that the EGS region is $\sim$4--10$\times$ denser than average at $z \sim$ 8.7 (see also \citealt{whitler23}).  While our unit-redshift bin used here mitigates this overdensity somewhat, at the bright end where numbers are small, our results are mildly higher than those from the literature, becoming broadly consistent at $M_{UV}$ $\gtrsim -$20.  We do note that when considering the full CANDELS area ($\sim$10$\times$ wider than we consider here), \citet{finkelstein22} found a volume density $\sim$5$\times$ lower in the magnitude bin including this object.

At $z \sim$ 11 we compare to a compilation of $z \sim$ 10--12 results from the literature \citep{donnan23a,donnan23b,mcleod23,bouwens23,harikane23,perezgonzalez23,castellano23,franco23,casey23b,adams23}.  While there is significant scatter, we find generally good agreement between our results and previously published values in the literature, where here our larger sample size (than most studies) results in smaller uncertainties.  We note in particular good agreement between our values and those from \citet{mcleod23} and \citet{adams23}, which are the only other previous studies to make use of the full CEERS area.

\begin{deluxetable*}{c|cc|cc|cc|cc|c}
\tablecaption{UV Luminosity Function Double Power-Law Parameters}\label{tab:dpl}
\tablewidth{\textwidth}
\tablehead{\multicolumn{1}{c}{$z$} & \multicolumn{2}{c}{log$_{\mathrm{10}}\phi^{\ast}$ (Mpc$^{-3}$)} & \multicolumn{2}{c}{$M^{\ast}$ (mag)} & \multicolumn{2}{c}{$\beta$} & \multicolumn{2}{c}{$\alpha$} & \multicolumn{1}{c}{log $\rho_{UV,<-17}$}\\
\multicolumn{1}{c}{$ $} & \multicolumn{1}{c}{Prior} & \multicolumn{1}{c}{Posterior} & \multicolumn{1}{c}{Prior} & \multicolumn{1}{c}{Posterior} & \multicolumn{1}{c}{Prior} & \multicolumn{1}{c}{Posterior} & \multicolumn{1}{c}{Prior} & \multicolumn{1}{c}{Posterior} & \multicolumn{1}{c}{(erg s$^{-1}$ Mpc$^{-3}$ Hz$^{-1}$)}}
\startdata
9&[$-$10,$-$1]&$-$4.4$_{-0.5}^{+0.6}$&[$-$22,$-$15]&$-$21.0$_{-0.5}^{+0.8}$&[$-$10,$-$3]&$-$4.9$_{-3.0}^{+1.5}$&[$-$5,$-$1]&$-$2.2$_{-0.3}^{+0.4}$&25.4$^{+0.1}_{-0.1}$\\
11&[$-$10,$-$1]&$-$4.4$_{-0.7}^{+0.8}$&[$-$22,$-$15]&$-$20.4$_{-0.7}^{+1.0}$&[$-$10,$-$3]&$-$5.1$_{-2.9}^{+1.6}$&[$-$5,$-$1]&$-$2.2$_{-0.4}^{+0.6}$&25.0$^{+0.2}_{-0.2}$\\
14&[$-$10,$-$1]&$-$4.8$_{-0.8}^{+0.3}$&$M^{\ast}_{z=11}$($\pm$0.1)&$-$20.4$_{-0.1}^{+0.1}$&$\beta_{z=11}$($\pm$0.1)&$-$5.1$_{-0.1}^{+0.1}$&[$-$5,$-$1]&$-$2.55$_{-1.40}^{+1.05}$&24.9$^{+1.1}_{-0.7}$\\
\enddata
\tablecomments{Constraints on the rest-UV luminosity function assuming a double power law form.  We place priors on $M^{\ast}$ and $\beta$ at $z \sim$ 14 to match the values measured at $z \sim$ 11.  The final column lists the specific luminosity density, obtained by integrating the luminosity functions at magnitudes brighter than $-$17.}
\end{deluxetable*}

Each panel of Figure~\ref{fig:lf} also shows an empirical extrapolation of the UV luminosity function from \citet{finkelstein22b}.  In this study, they fit an evolving double-power-law model to all available UV luminosity function data at $z =$ 3--9, assuming that the double power-law (DPL) parameters $\phi^{\ast}$, $M^{\ast}$, $\beta$ and $\alpha$ vary smoothly with 1$+z$ (they also simultaneously fit the evolution of the AGN UV luminosity function with a separate DPL, though at the magnitudes we consider here star-forming galaxies were found to dominate).  In each panel we show the measured DPL fits at $z =$ 5--8 as the gray lines, then showing the extrapolation to a given redshift range as the light shaded region (where for the extrapolations, we used the MCMC chains from \citealt{finkelstein22b} to generate samples of the DPL parameters for each redshift).  The upper and lower bounds show the median DPL at the upper/lower bound of the FWHM from the stacked $\mathcal{P}(z)$, as shown in the inset panels.  

We show this more clearly in Figure~\ref{fig:lf_all}, where we overplot both our measured number densities and these extrapolated luminosity functions for all three of our redshift samples.  Based on pre-launch expectations of either a smoothly or rapidly declining luminosity function at $z >$ 8, we would expect to see our results fall either on or below these extrapolations.  While the results at $z \sim$ 9 are consistent with this extrapolation (with the exception of the brightest points, which are affected by the known overdensity), we see clearly and significantly that our $z \sim$ 11 observed number densities lie modestly above this extrapolation, while the $z \sim$ 14 observed number densities lie even higher above their extrapolated region.  

\subsection{Double Power Law Fits}
By combining with NGDEEP, we are able to sample 3--4 magnitudes of dynamic range in $M_{UV}$ at $z \sim$ 9 and $z \sim$ 11.  We therefore fit a double-power law function to our observations, following evidence that this functional form better represents the UV luminosity function at high redshifts than a Schechter function \citep[e.g.][]{bowler15,bowler20,finkelstein22b}.
We fit a DPL to each of our three redshift bins independently via a MCMC algorithm (following the methodology of \citealt{leung23}).  We assign fairly uninformative priors on all parameters at $z \sim$ 9 and 11; at $z =$ 14 due to our poor observational constraints we fix $M^{\ast}$ and $\beta$ to the $z \sim$ 11 values (within a small tolerance).  We list the priors and posterior results in Table~\ref{tab:dpl}.  The median DPL fit is shown as the thin line in Figure~\ref{fig:lf_all}.  While this does a reasonable job of representing the data, the uncertainties, particularly on $\beta$ and $M^{\ast}$ at all redshifts, and on the faint-end slope $\alpha$ at $z \sim$ 14 are presently quite large.  We do note that our measured faint-end slope of $-$2.2 at $z =$ 11 is consistent with the value from \citet{leung23}, though they imposed more restrictive priors on $\beta$ and $M^{\ast}$, thus our uncertainty is higher.

\subsection{Evolution with Redshift}
\begin{figure*}[!t]
\epsscale{1.15}
\plotone{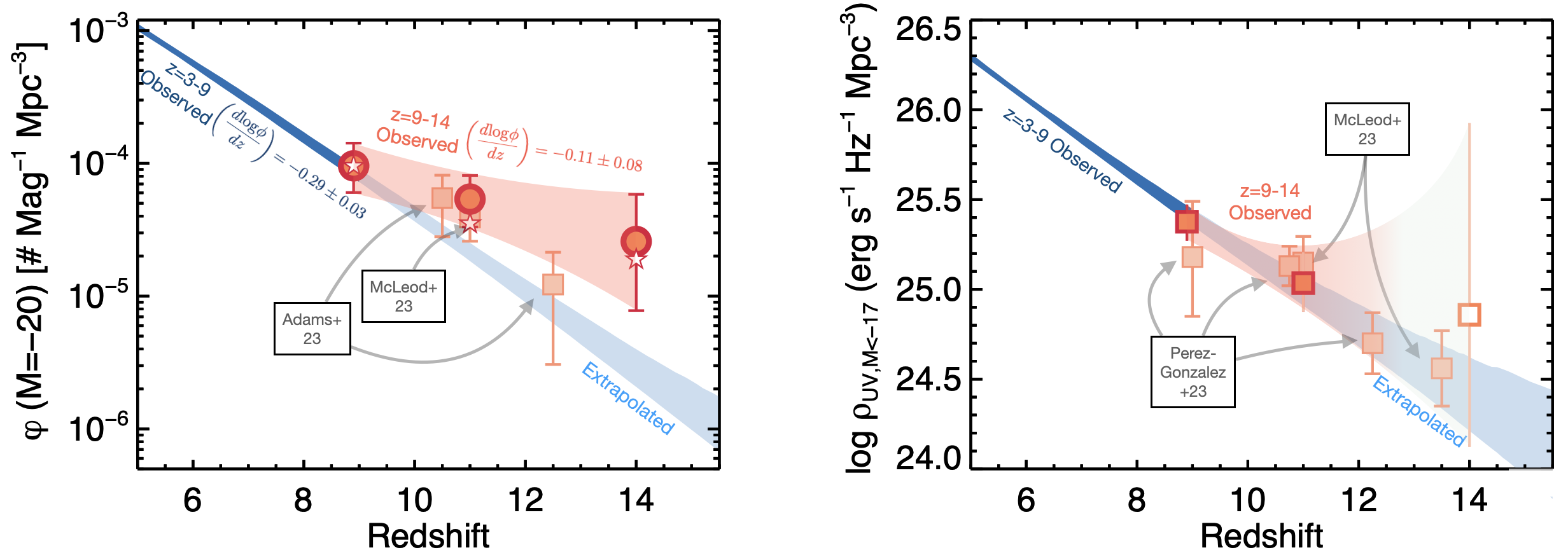}
\caption{Left) The evolution of the observed number density at $M_{UV} = -$20.  The red circles show the observed number density at this absolute magnitude from CEERS (connected by the light red shaded region; the small stars show the DPL fit values at this magnitude). The dark blue region shows the measured value from \citet{finkelstein22b}, and the lighter shaded region shows the extrapolation of the \citet{finkelstein22b} results to higher redshift.  While pre-launch expectations were that the number densities at $z >$9 would either continue the observed trend at $z =$ 3--9, or evolve more rapidly downward with increasing redshift, we find that the number density of bright galaxies surprisingly flattens at $z >$ 9, where we measure a change in slope $d\rm log \phi$/$dz$ between $z <$ 9 and $z >$ 9 at 2.1$\sigma$ significance.  Right) The evolution of the integrated specific UV luminosity density, obtained by integrating double-power law fits to our observed luminosity functions to $M_{UV} = -$17.  The evolution here is less clear, with increased uncertainties (particularly at $z \sim$ 14, which is shown faded to represent its large uncertainty) making it less clear whether the this quantity also has a flatter evolution at higher redshift.}
%Contrary to the results in the left panel for bright galaxies, this quantity, which is dominated by fainter galaxies, follows the extrapolated values out to at least $z \sim$ 11, and possibly $z \sim$ 14 (though the lack of constraining faint data thus far at $z \sim$ 14 leaves us unable to constrain that redshift; this point is shown faded to represent its large uncertainty).  These results together imply that whatever changing physical processes at $z >$ 9 one invokes to explain our observations must primarily affect bright galaxies only.}
\label{fig:evolution}
\end{figure*}

To explore the evolution in the UV luminosity function in more detail, in the left-hand panel of Figure~\ref{fig:evolution} we show our observed number densities for the CEERS bin closest to $M_{UV} = -$20 at each of our three redshift bins, on top of the expectations for this number density from the \citet{finkelstein22b} extrapolation. This figure clearly shows that the observed number densities diverge from the observed evolution at $z =$ 3 to 9, flattening at higher redshifts.  We quantify this by measuring the slope ($d \rm log \phi$/$dz$) both for previous observations at $z =$ 3--9, and our results here at $z =$ 9--14.  We find that this slope changes from $d \rm log \phi$/$dz$ $= -$0.29 $\pm$ 0.03 at $z =$ 3--9 to $-$0.11 $\pm$ 0.08 at $z =$ 9--14.  Thus, while the abundance of bright ($M_{UV} = -$20) galaxies evolves somewhat steeply at a constant slope at $z =$ 3--9, this evolution is \emph{flatter} towards higher redshifts at the 2.1$\sigma$ significance level. 

For this calculation we have used as our fiducial values the actual measured number densities at $M_{UV} = -$20. In the left panel of Figure~\ref{fig:evolution} we show as small stars the values of $\phi(M=-$20) from the median DPL model at each redshift.  The differences from the observed values are negligible at $z =$ 9 and 14, while at $z =$ 11 this parameterized value is slightly below the observed value at $z =$ 11.  Using these DPL values, the observed evolutionary slope steepens to $d \rm log \phi$/$dz = -$0.18 $\pm$ 0.07 (reducing the significance of the flattening to 1.5$\sigma$).

To consider whether the evolution in the abundance of faint galaxies changes at $z >$ 9, in the right-hand panel of Figure~\ref{fig:evolution} we show the specific UV luminosity density ($\rho_{UV}$), obtained by integrating the UV luminosity function from our DPL fits to $M_{UV} < -$17 (integrating each step of the chain to calculate the median and 68\% confidence range); this quantity is dominated by the abundance of faint galaxies given the steep faint-end slopes (we note that we plot this quantity rather than the number densities at fainter magnitudes due to the lack of constraints at $z \sim$ 14, visible with the very large error at this highest redshift in the plotted integrated quantity).  Interestingly, our measured specific UV luminosity densities at $z =$ 9 and $z =$ 11 are fully consistent with the extrapolated values (at $z \sim$ 14 the poor constraints on the faint-end slope leave the uncertainty is too large to reach definitive conclusions, thus we show this as a faded data point).  This can also be observed in Figure~\ref{fig:lf_all}, as the faintest data points at $z \sim$ 9 and 11 are consistent with the extrapolated luminosity function.  A similar specific UV luminosity density is found by \citet{perezgonzalez23}, while \citet{mcleod23} finds a slightly elevated value, though consistent with the empirical extrapolation within the uncertainties.  

Taking both panels of Figure~\ref{fig:evolution} together, we find clear evidence that the evolution of the number density of bright galaxies is observed to flatten at $z >$ 9, while the evolution of the integrated UV luminosity density, which is dominated by the abundance of fainter galaxies, is less clear, and may possibly follow the lower-redshift evolutionary trend extrapolated to higher redshift.  We discuss potential physical explanations for this intriguing result in the following section.

\section{Discussion} \label{sec:discussion}

In \S 4 we presented strong evidence that \emph{(i)} the abundance of galaxies at $z >$ 9 is in excess of nearly all pre-{\it JWST} launch simulation predictions as well as above extrapolations from lower-redshift observations, and \emph{(ii)} the evolution of the abundance of bright ($M_{UV} = -$20) galaxies is flatter at $z =$ 9--14 than at $z =$ 3--9.  Here we explore several potential explanations for these observations.  Potential explanations could be due to  galaxies being brighter than predicted or more numerous (e.g., horizontal evolution in the luminosity function rather than vertical).  However, while the former is relatively easily achievable via a variety of reasonable physical modifications (as we discuss below) the latter would require major revisions to modern cosmology (e.g., more dark matter halos than expected), which we consider less likely.

\subsection{Redshift Accuracy}
One valid concern with early {\it JWST} studies is that selection techniques which worked well at lower redshift would begin to fail.  In particular, while the physics behind Ly$\alpha$-break-based selection should persist at these high redshifts, it is possible that heretofore unknown populations of contaminants could have adverse affects.  While one could model this contamination based on simulations, it relies on said simulations correctly modeling the colors of \emph{all} potentially contaminating populations \citep[e.g.,][]{larson23a,harikane23}, which is unlikely, particularly prior to \emph{JWST} observations.

Spectroscopic validation of photometric redshifts is thus required.  Unlike the past decade, when only the brightest {\it HST} $z >$ 6 galaxies could have redshifts validated via either weak Ly$\alpha$ emission \citep[e.g.,][]{finkelstein13,zitrin15,oesch15,hoag19,jung19a,jung20,larson22} or Ly$\alpha$ breaks for the brightest sources \citep[e.g.][]{oesch16}, {\it JWST}'s spectroscopic capabilities allow easy rest-optical-based spectroscopic redshifts out to $z \approx$ 9.5 (beyond which [\ion{O}{3}] redshifts out of the NIRSpec window) and Ly$\alpha$ continuum-based redshifts (with the NIRSpec prism mode) to arbitrarily higher redshifts \citep[e.g.][]{curtislake23,fujimoto23,arrabalharo23a,arrabalharo23b,hainline23,fujimoto23b}.

As the CEERS spectroscopic component was observed in the second epoch \emph{and} a DDT NIRSpec followup program was performed, a significant number of CEERS high-redshift candidates were spectroscopically observed in Cycle 1.  Of our original sample of 93 candidate galaxies, 17 had NIRSpec spectroscopic observations, with redshifts originally presented in \citet{fujimoto23,arrabalharo23a,arrabalharo23b,harikane23,larson23b}.  As discussed in \S 3.2.1, only one of these 17 had a ``catastrophically" (defined as $|z_{spec} - z_{phot}|/(1+z_{spec}) > 0.3$) incorrect redshift (similar success was seen in the UNCOVER survey by \citealt{fujimoto23b}).  Of the remaining 16, all had $z_{spec} >$ 7.8.  We also note that four of the galaxies in our sample at $z_{phot} >$ 9.7 were spectroscopically observed by CEERS with no spectroscopic detection.  For these sources, the absence of strong emission lines is plausibly consistent with $z_{spec} >$ 9.6.  These results imply that it is unlikely that significant contamination from low-redshift galaxies is affecting our results (with the caveat that the sample of galaxies confirmed is as-yet small and fairly biased towards brighter sources).  

We next consider whether any smaller, yet non-negligible, systematic offsets in redshift could play a role.  A trend for the photometric redshifts to be overestimated at $z \gtrsim$ 8 has been reported, with results from CEERS \citep{arrabalharo23b,fujimoto23}, JADES \citep{hainline23} and UNCOVER \citep{fujimoto23b} showing the photometric redshifts to be over-estimated by $<\Delta z> =$ 0.45 ($\pm$ 0.11), 0.26 ($\pm$ 0.04) and 0.28 ($\pm$ 0.33), respectively.  As discussed in these studies, suchaed offsets indicate a mismatch between the galaxy spectra and the adopted photometric redshift templates due to a variety of physical effects, including an increasing neutral fraction and/or enhanced DLA absorption \citep[e.g.,][]{umeda23,heintz23}.  

For our specific sample of galaxies in this work (13 galaxies with $z_{spec} >$ 8.5), we find a median (mean) offset of $z_{phot} - z_{spec} =$ 0.1 (0.3) with a standard deviation of 0.2. As discussed in \S 3.2.1 our sample excludes three galaxies selected with $z_{phot} >$ 8.5 with $z_{spec}$ $\sim$ 8; including these objects does not change the median offset, but it does increase the mean offset (to 0.5) and the standard deviation (to 0.3).  These three objects in particular highlight the difficulty of working at $z \lesssim$ 9 within the CEERS dataset due to the bluest {\it JWST} filter being F115W.  Upcoming F090W imaging from PID 2234 (PI Ba\~{n}ados) will improve this situation, probing below the Ly$\alpha$ break at $z \sim$ 8.  At present these larger $\Delta z \sim$ 1--2 offsets affect only a small fraction (3/16) of the spectroscopic sample.   Should these larger offsets exist in the rest of the non-confirmed sample, the instances of these objects we do see imply it would primarily affect the $z \sim$ 9 results (though we note that ID$=$4777 [$z_{spec}=$7.993] is in the $z \sim$ 11 galaxy sample, though it has a very broad $\mathcal{P}[z]$).

%We consider two potential impacts these offsets could have on our interpretation of the results --  lowering the median redshift of the sample (meaning we should compare to the extrapolated results at slightly lower redshift), and the reduced distance modulus will resulting in slightly fainter absolute magnitudes.  Of these two, the former is the larger affect, as the difference in the distance modulus for $\Delta z =$0.3 (1) at $z =$ 11 is only $\sim$0.04 (0.13) mag.

We simulate the potential impact of these systematic redshifts offsets by re-measuring our observed luminosity function values in the same MCMC manner as above, where here in each step of the MCMC chain we assign simulated spectroscopic redshifts for each object. 
For objects which already have spectroscopic redshifts, we keep those values.  For the remaining objects, we draw randomly from the observed $\Delta z$ ($=z_{phot}-z_{spec}$) distribution, adding $\Delta z$ to the photometric redshift to simulate a (potentially biased) spectroscopic redshift.  For each new ``$z_{spec}$", we re-measure a new value of M$_{UV}$.  We perform this redshift assignment and subsequent SED-fitting-based $M_{UV}$ measurement prior to running the MCMC chain, pre-computing 100 random draws of $\Delta z$ and the corresponding $M_{UV}$, then drawing from these pre-computed values randomly at each step of the chain.  Through this process we simulate the effect of the potential redshift bias both on the specific galaxy samples (as objects can move between redshift bins, potentially lowering the median redshift), as well as the impact on the absolute magnitudes (due to changes in the distance modulus).  Both effects can combine to affect the number density, though the former is the larger affect as the difference in the distance modulus for $\Delta z =$0.3 (1) at $z =$ 11 is only $\sim$0.04 (0.13) mag.

First we examine the impact on the median redshift in each bin, which we find to be fairly minimal: $z_{med} =$ 8.9, 10.9 and 14.2 (unchanged for the $z \sim$ 9 and $z \sim$ 11 bins, and 0.2 higher in the $z \sim$ 14 bin).  The corresponding impact on the number densities at $M_{UV} = -$20 are also modest, with these values being 12\%, 14\% and 36\% lower at $z \sim$ 9, 11 and 14.  Measuring $d\rm log \phi$/$dz$ using both these new median redshift values and the corresponding simulated number densities, we find $d\rm log \phi$/$dz = -$0.14 $\pm$ 0.09 over $z =$ 9--14, not significantly different from our fiducial value of $-$0.11 $\pm$ 0.08.  We note that while for this exercise we restricted the $\Delta z$ sample to $z _{spec} >$ 8.5, we found that including the three $z_{spec} \sim$ 8 galaxies did not change the results.

We thus conclude that the measurable redshift bias from the available spectroscopic confirmations is unlikely to be the primary cause of the observed change in slope at $z >$ 9 in the evolution of the abundance of bright galaxies.  We acknowledge again that the existing number of spectroscopic redshifts is small, and biased towards primarily bright sources.  In addition, some of these redshifts come from the Ly$\alpha$ break only, and these values have been measured to be up to $\Delta z \sim$0.2 different from more secure emission-line-based redshifts \citep{fujimoto23b}.  Larger samples of spectroscopic confirmations of galaxies in this epoch are needed to increase confidence that any redshift bias does not affect the measured number densities.

%The first is that, considering Figure~\ref{fig:specz}, the majority of galaxies have rather small offsets, with 8/13 $z_{spec} >$ 8.5 galaxies having $\lvert\Delta$$z\rvert$ $<$ 0.2.  
%The second is that even in the worst-case scenario where we consider that the full sample could, on average, have the photometric redshift over-estimated by $\Delta$$z =$ 0.5, it would not explain the observed offset in the UV luminosity function, as the difference in distance modulus ($\mu$) at such high redshifts is minimal (e.g., $\mu$($z=$10.5) $-$ $\mu$($z=$10.0) $=$ 0.07 mag, while the CEERS results are offset by $\Delta$$M \sim$ 0.5 (Figure~\ref{fig:evolution}).  These results strongly suggest that a significant number of catastrophic redshift outliers is highly unlikely to affect our stample, thus in the following we consider more physical processes to explain our observations.
%Add here from evolution plot: z=11 point would need to be dz=1.5 to match expections, z=14 would need dz=3
%Finish by saying its unlikely to be biased by catastrophic z failures or systematic offsets.

\subsection{AGN Contribution}

Accreting supermassive black holes, particularly when they are not obscured from view, can emit quite strongly in the rest-UV \citep[e.g.,][]{stevans14}.  While the contribution of AGN light to the rest-UV emission from high-redshift galaxies has been fairly unconstrained, some notable examples do exist at $z \sim$ 7 \citep[e.g.][]{fujimoto22b,endsley23b}.  However, the evolution in the AGN UV luminosity function suggests that AGNs do not dominate the rest-UV emission in galaxies (e.g, non-quasars) at high redshift \citep[e.g.,][]{finkelstein22b}.  However, with {\it JWST}'s spectroscopic abilities, it is worth revisiting whether AGN could be contributing significantly to the UV emission from galaxies.

\begin{figure*}[!t]
\epsscale{1.0}
\plotone{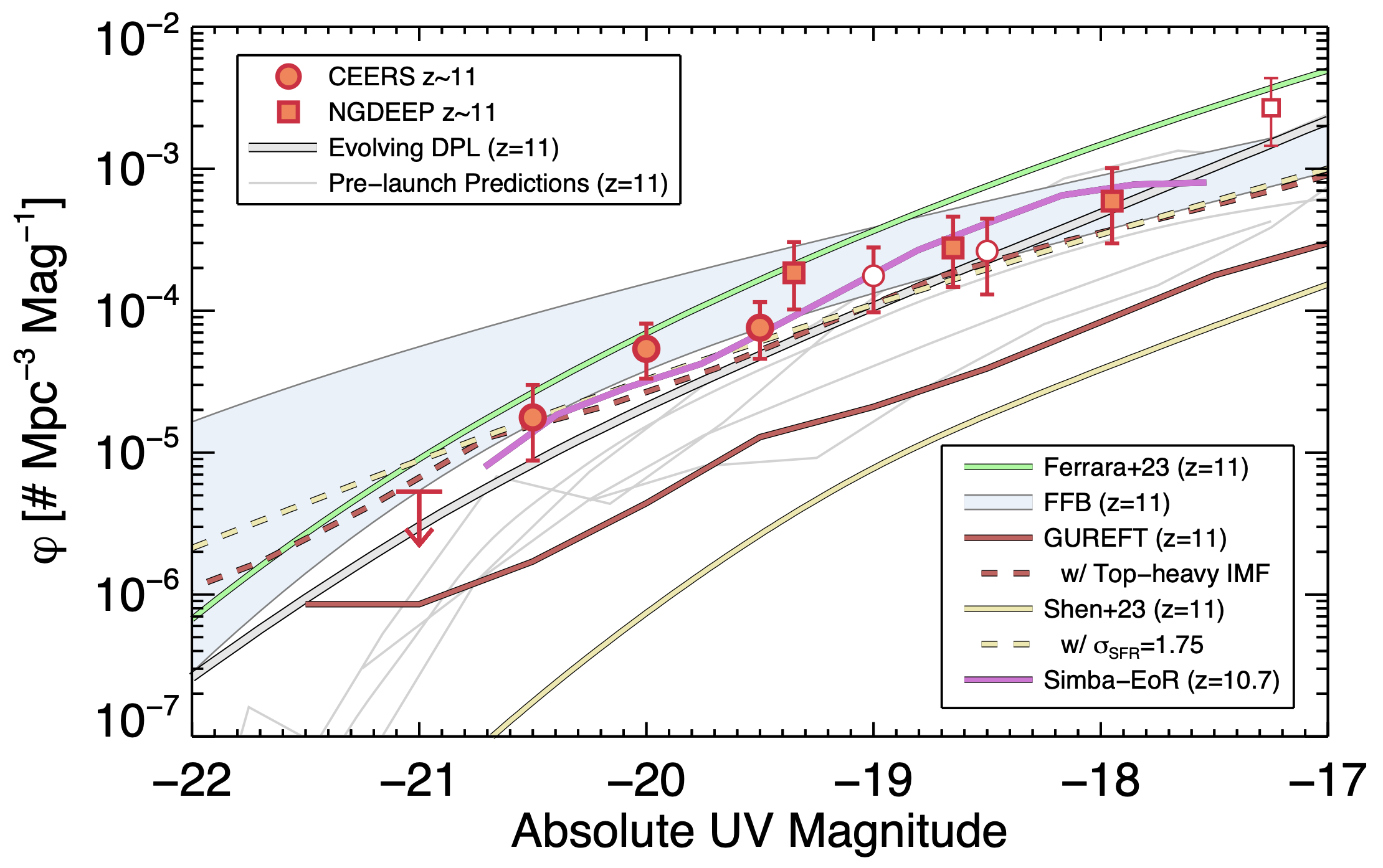}
\caption{A comparison of the observed $z \sim$ 11 UV luminosity function from CEERS and NGDEEP (symbols are the same as in the top panel of Figure~\ref{fig:lf}) to model predictions.  Pre-launch predictions from FLARES, DELPHI, UniverseMachine, THESAN and BlueTides are shown as the thin gray lines, while the colored lines show more recent predictions from \citet[][green]{ferrara23}, \citet[][the blue shaded region shows a range of maximum star-formation efficiency from 0.2--1]{dekel23}, \citet[][yellow; the dashed line includes a strong stochastic star-formation component]{shen23}, \citet[][red; the dashed line indicates a top-heavy IMF UV luminosity enhancement of 3$\times$]{yung23} and Jones et al.\ (in prep, purple).  The thick gray line shows the empirical DPL luminosity function from \citet{finkelstein22b} extrapolated to $z =$ 11.  These predictions show that a variety of potential physical solutions can predict a $z \sim$ 11 luminosity function in agreement with observations.}
\label{fig:lf_sims}
\end{figure*}

Early observations hint that growing super-massive black holes are indeed somewhat common at $z \sim$ 5--9, with signs of potential AGN activity found in dozens of galaxies \citep[e.g.,][]{kocevski23,larson23a,harikane23b,leung23,labbe23,bogdan23}, with several sources containing spectroscopically confirmed broad-line AGN \citep[e.g.,][]{kocevski23,harikane23b,matthee23,larson23a,maiolino23b,kokorev23,furtak23,greene23}.  While this may suggest AGN could contribute to the UV luminosity, the many of these sources have unique two-component SEDs with fairly flat UV spectral slopes, with a change to a steeply rising red slope in the rest-UV optical \citep[e.g.,][]{kocevski23,barro23,matthee23,labbe23}.   While an AGN jet could trigger enhanced star-formation \citep[e.g.][]{duncan23}, here we aim to assess whether the UV emission we observe is dominated by emission from an AGN accretion disk.  For these red AGN in particular, the point-source morphology in the longest wavelength bands strongly suggests that obscured AGN light is dominating the rest-optical emission, while advanced SED modeling is needed to robustly constrain the amount of AGN contribution to the rest-frame UV.  Such a contribution remains possible as scattered UV light from a partially obscured AGN could in theory contribute to the observed UV emission \citep[e.g.,][]{kocevski23,barro23,labbe23,greene23}, though the resolved nature of the rest-UV emission in these galaxies indicates stellar emission may dominate.

AGN have also recently been identified in objects with more typical galaxy-like morphologies and SED slopes.  %\citet{kocevski23} discussed the case of CEERS 3210, the first confirmed (via broad H$\alpha$) {\it JWST} red AGN, noting that the rest-UV spectral slope is fairly blue, and could be due to either scattered AGN light or stellar emission (or some combination).  
\citet{larson23a} inferred the presence of an AGN in a $z =$ 8.7 galaxy (via a 2.5$\sigma$ significant broad-H$\beta$ line), and noted that the SED has a flat slope through the rest-near-infrared (aided by MIRI observations, \citealt{papovich23}) suggesting stellar light is dominating at all observed wavelengths.  \citet{maiolino23a} inferred via extremely high gas densities that the nucleus of the well-known galaxy GN-z11 (at $z =$ 10.6) likely hosts an AGN; analysis of this object by \citet{tacchella23b} shows that 2/3 of the rest-UV continuum emission emanates from the nucleus, hinting that this object could be AGN dominated in the UV.  \citet{harikane23b} discuss 11 confirmed broad-lined AGN in the CEERS survey, and found that the majority of them showed extended morphologies in the rest-UV (with most of the rest being extremely UV-faint reddened AGNs), suggesting that much of the rest-UV emission is stellar in origin.  Although some of these observations indicate that the AGN contribution to the total UV luminosity is negligible, this might not always the case, depending on the phase of the AGN duty cycle (which affects the contrast between the AGN and the host galaxy). A possible high AGN fraction has been argued in the recent NIRSpec follow-up for lensed galaxies at $z =$ 8.5--13.2 \citep{fujimoto23b}. 
 
These early observations do not yet collectively paint a clear picture of the contribution of AGN to the rest-frame UV emission from early galaxies.  It is clear that AGNs exist in these epochs, though many discovered so far appear to be primarily obscured.  Deciphering the relative contribution from stars and AGNs to the emergent UV emission, including constraining the extent to which scattered UV light from obscured AGNs plays a role, will require a combination of more advanced SED modeling techniques along with deep $\sim$1--2$\mu$m spectroscopy.  Until such analyses can be conclusively done, AGNs remain a possible scenario to explain the high abundance of bright galaxies at early times.

\subsection{Change in Physical Processes}

The remaining explanations for the observed UV luminosity enhancement involve changes in the physical processes regulating the ratio of observed UV light to the halo masses of these galaxies.  We further explore this here, aided by a variety of recent theoretical results motivated by early {\it JWST} observations of the $z >$ 9 universe, summarized in Figure~\ref{fig:lf_sims}.  This figure compares our observed UV luminosity function (combined with that of NGDEEP) to recent predictions from \citet{ferrara22}, \citet{yung23}, \citet{dekel23}, \citet{shen23}, and Jones et al.\ (in prep).  We also compare to pre-launch predictions from FLARES \citep{lovell21,vijayan20,wilkins22b}, DELPHI \citep{dayal17}, UniverseMachine \citep{behroozi20}, THESAN \citep{kannan22} and BlueTides \citep{feng16,wilkins17}. 
In this figure we compare our observations to these model predictions made at $z =$ 11 when possible (when not we interpolate the number densities in log space), as this is the median of the stacked redshift probability distribution for the galaxies which make up our luminosity function sample (see inset panel of Figure~\ref{fig:lf}).  We note that differences between models can be due both to the underlying methodology and/or sub-grid physics, as well as the procedures used to generate observed luminosities.

\subsubsection{Significant Evolution in Attenuation}

\citet{ferrara23} have developed a physical model which successfully reproduces the observed $z =$ 7 UV luminosity function via a dust implementation which is designed to match both the shape of the UV luminosity function and the $z \sim$ 7 obscured SFR results from the ALMA REBELS survey \citep{bouwens22a}.  Evolving their model to higher redshifts, they naturally predict a slowing in the evolution of the UV luminosity function from $z \sim$ 9 to 11 due to significantly reduced attenuation in galaxies.  This arises in their model due to outflows driven by very high (``super-Eddington'') specific SFRs, which can efficiently drive gas (and dust) out from these galaxies \citep{ferrara23b}.  They discuss that this is supported by the data as early results on the colors of $z >$ 8 galaxies showed that they were fairly blue \citep[e.g.,][]{finkelstein22c,cullen23,topping22,papovich23} and thus likely contained little dust.

As shown in Figure~\ref{fig:lf_sims}, this model does significantly better than pre-launch simulations, in fact slightly over-predicting the observed UV luminosity function.  We thus explore whether significant color evolution is observed, particularly in modestly bright galaxies, which could further support this model.  To study potential evolution with redshift, we select a sample of galaxies from the CEERS catalog at $z \sim$ 6--8.  The sample selection was identical to that done here for $z >$ 8.5, simply evolving the redshift cuts to lower redshift (primarily requiring $\int\mathcal{P}$($z$ $>$ 4) $\geq$ 0.7, $z_{best}$ $>$ 5.0, and $\mathcal{S}_z =$ 6, 7 or 8).  The full sample of $>$ 2000 sources was visually inspected, resulting in a final sample of 1018, 574, and 224 galaxies at $z \sim$ 6, 7 and 8, respectively.

\begin{figure}[!t]
\epsscale{1.15}
\plotone{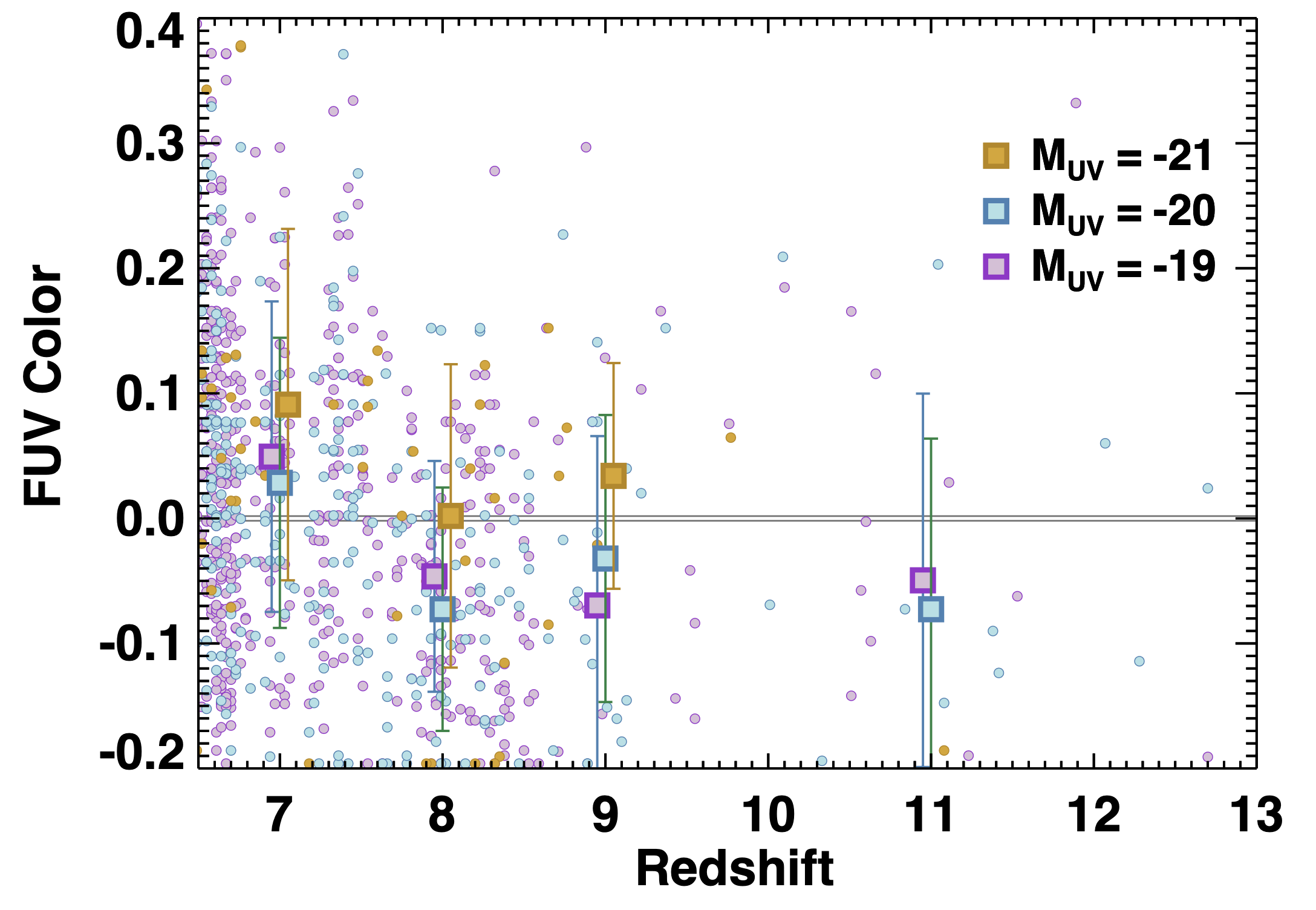}
\caption{The evolution of the FUV color $\mathcal{C}_{FUV}$ with redshift, color-coded by M$_{UV}$.  The large squares show median values in bins of redshift and magnitude (error bars show the 1$\sigma$ spread), for bins with $\geq$ five sources.  While reddened galaxies exist at $z \sim$ 7--8, the median colors are still fairly blue.  Notably, the median FUV color for $M_{UV} = -$20 galaxies is similar at $z \sim$ 8 to $z \sim$ 11, suggesting a significant drop in dust attenuation is unlikely to explain the high abundance of bright $z \sim$ 11 galaxies.}
\label{fig:colors}
\end{figure}

As discussed in \S 3.3, we defined a rest-far-UV color $\mathcal{C}_{FUV}$, and calculated it for galaxies in both our $z >$ 8.5 and $z =$ 6--8 galaxy samples.  We note that this specific FUV color is very sensitive to dust attenuation due to a narrow color baseline, in that $\Delta \mathcal{C}_{FUV} =$ 0.05 is equivalent to $\Delta A_{V} =$ 0.1.  In Figure~\ref{fig:colors} we plot $\mathcal{C}_{FUV}$ versus redshift, color-coded by M$_{UV}$.  We plot the median in bins of redshift ($z =$ 6.5--7.5, 7-5--8.5, 8.5--9.7, and 9.7--13) and while there is significant scatter in color at all redshifts, the median color is fairly blue.  Focusing on the middle magnitude bin (M$_{UV} = -$20) where our results have the strongest constraining power (too few brighter galaxies exist at $z >$ 9 to make conclusions), we see that the median FUV color does not significantly evolve from $z \sim$ 8 to $z \sim$ 11.  

This initial exploration into the evolution of UV colors does not support rapid changes in the evolution of the attenuation between $z \sim$ 7 to 11, as would be implied by the \citet{ferrara23} model.  Rather, the majority of these $z >$ 6 galaxies appear roughly dust free, implying that the ALMA REBELS/detected galaxies used to calibrate the \citet{ferrara23} model are not indicative of the bulk of the $z \sim$ 7 galaxy population.  In particular, \citet{papovich23} showed that with the inclusion of MIRI photometry, typical galaxies at $z >$ 7 are fairly blue.  We do note that finding so little dust is in itself a puzzle, as these galaxies should have made significant amounts of dust in the process of building their observed stellar masses. Though UV-faint dusty sources may indeed exist in this epoch \citep{rodighiero23}, in our UV-selected sources the dust must get destroyed (via, e.g., reverse shocks in supernovae), or be removed via winds even down to $z \sim$ 7 in these sources.  We acknowledge however that the scatter in $C_{FUV}$ is large and that more advanced UV spectral slope modeling with larger samples covering a wider dynamic range in stellar mass and $M_{UV}$ can provide further insight.

\subsubsection{Change in Conversion from Mass to UV Luminosity}

If reduced attenuation is unlikely to be the dominant effect explaining the discrepancy between models andobservations, then it is prudent to consider changes in the processes of star formation which could result in enhanced UV luminosities.  A very straightforward potential explanation could be a change in the initial mass function (IMF).  As discussed by \citet{finkelstein23} and \citet{harikane23}, a change in the characteristic stellar mass from $\sim$1 M\sol\ to $\sim$10 M\sol\ is expected when the cosmic microwave background temperature is higher and the gas metallicities are lower \citep[e.g.,][]{larson98,Bromm2002,tumlinson06,steinhardt23}.  Such a change would decrease the mass-to-UV light ratio by factors of up to several \citep[e.g.,][]{raiter10,zackrisson11}.

As one example of this, we show the new semi-analytic model predictions based on merger trees from  the  \textsc{gureft}  simulation suite \citet{yung23, yung23b}.  This is an update to the ``Santa Cruz SAM" predictions \citep{somerville15b,yung19a}, now using $N$-body based merger trees constructed with finely spaced snapshots at very high-redshift to better capture halo growth at early times.  Their fiducial model still under-predicts the observations, implying that the Extended Press-Schechter based merger trees used in the previously published models were not the sole reason for the discrepancy. However, as they discuss in their paper, the shape of their UV luminosity function appears consistent with the data.  They show that if they decrease the mass-to-light ratio by a factor of $\sim$3 (dashed line), as would be plausible for a top-heavy IMF, their model becomes consistent with the observations.

A smaller UV luminosity enhancement of only $\sim$1.5$\times$ would be needed to bring the empirically extrapolated DPL UV luminosity function from \citet{finkelstein22b} at $z =$ 11 (thick gray line) into agreement with our observations.  However the discrepancy between this empirical extrapolation is greater for brighter galaxies than for fainter galaxies, such that a flat UV luminosity boost at all magnitudes would lead to an over-prediction of the faint-end due to the empirically predicted faint-end slope being steeper ($\alpha = -$2.5) than that observed from NGDEEP ($\alpha = -$2.2).  While our results cannot constrain the IMF presently, deep rest-UV spectroscopic observations could be capable of detecting high-ionization lines such as \ion{He}{2} or [\ion{Ne}{5}], which could indicate the presence of very massive stars in these galaxies \citep[e.g.,][]{tumlinson00,bromm01,schaerer03,olivier22,cleri23}.

One caveat to an increased light-to-mass ratio due to a top-heavy IMF would be that it would be accompanied by an increased feedback-to-mass ratio, because the massive stars that produce UV light are also those that produce energetic radiation, strong stellar winds and Type II supernovae. Changing the IMF would also change the integrated chemical yields, which could impact cooing. The models above have yet to self-consistently model the increased feedback and modified chemical yields of a top-heavy IMF for high-redshift galaxies.  However, other studies \citep[e.g.,][]{Fontanot14} have found that evolving top-heavy IMFs (e.g., top-heavy IMFs for galaxies with high SFRs) have tended to decrease the mass of stars formed relative to models with universal IMFs.

\subsection{High Star Formation Efficiency}

Even without a change in the IMF, the UV luminosities of very high-redshift galaxies could be enhanced if the rates of gas conversion into stars was increased.  As discussed in \citet{finkelstein23}, most models adopt fairly long gas-depletion timescales, based on observations of nearby galaxies.  While perhaps a coincidence, it is interesting that among the pre-{\it JWST} predictions only the \citet{behroozi15} model, which assumes a negligible gas depletion timescale, can match the observations in Figure~\ref{fig:sdens}.  The physical processes which may lead to a decrease in this timescale are not currently obvious, though as the dependence of star-formation with gas density \citep{schmidt59,kennicutt98b} is super-linear \citep[e.g.][]{vallini23}, the very high gas densities present in early-universe halos likely play a role. Very high-efficiency star formation has been observed in super-star clusters in local galaxies \citep[e.g.][]{turner15,smith2020,costa2021}. The high cloud surface densities in the environments of these superstar clusters, which are rare in the nearby Universe, could be common in galaxies at $z>9$. Theoretical work on molecular cloud scales has also shown that both stellar winds and radiation feedback may become ineffective at very high cloud surface densities, leading to higher cloud-scale star formation efficiencies \citep{grudic20,lancaster21,menon23}. 
%This could be a common scenario at $z >$ 9, where cold accretion from the cosmic web is thought to be very common and effective in powering star formation, especially toward the higher-end of the mass spectrum.

In this context, we explore two additional predictions.  The first is an updated version of the \textsc{Simba} model \citep{dave19} known as \textsc{Simba-EoR} (Jones et al., in prep.).  This simulation includes a new subgrid ISM model that co-evolves dust and $H_2$ by explicitly tracking all formation and destruction mechanisms, which turns out to yield more $H_2$ at low metallicities relative to \textsc{Simba}.  This leads to earlier star-formation in halos, increasing the luminosities of early galaxies compared to \textsc{Simba} (which, as shown in Figure~\ref{fig:sdens} significantly under-predicts $z >$ 9 observations).  The \textsc{Simba-EoR} predictions are well matched to the CEERS observations, despite no specific tuning to EoR data, though the faint-end slope appears higher than that observed.  It is worth noting that the galaxies are not predicted to be dust-free; without accounting for extinction, the Simba-EoR predictions would be clearly above the observations.

We also compare to the Feedback-Free Starburst (FFB) physical model introduced by \citet{dekel23}, using the luminosity function, including a dust attenuation prescription, presented in Li et al.\ (in preparation). This model predicts that in massive galaxies at $z \sim 10$, where the gas density is above a threshold of $\sim 3\times 10^3$ cm$^{-3}$ and the gas-phase metallicity is below $\sim 0.2$ Z\sol, star formation in thousands of globular-cluster-like clouds is expected to proceed on a free-fall timescale shorter than the $\sim 2$ Myr interval between a starburst and the onset of effective stellar and supernova feedback, thus allowing high star-formation efficiency free of suppression by feedback\footnote{See also \citealt{renzini23}, who propose a related picture in which stars above 20 $M_\odot$ collapse directly to black holes, thereby failing to deposit energy in the ISM.}. 
In Figure~\ref{fig:lf_sims} we show as the blue shaded region the predictions of this model with a maximum SFE ranging from 20\% (lower bound) to 100\% (upper bound).
These predictions match our observations reasonably well over the range where we detect galaxies. %though before correction for dust they may over-predict the abundance of brighter galaxies. 
This model predicts an enhanced bright end due to the predicted high SFE preferentially at high halo masses, aided by the fairly low levels of dust attenuation and the high level of star-formation stochasticity predicted in the FFB phase.
%These predictions reasonably well-match our observations over the range where we detect galaxies, though they tend to over-predict the abundance of brighter galaxies.  Interestingly, this model predicts a shallower faint-end slope than other models, while showing an enhanced bright end.  This is due both to its prediction of a high SFE preferably at high halo masses, fairly low levels of dust attenuation, and a high level of star-formation stochasticity.

\subsubsection{Stochastic Star Formation}

In \S 4 we presented Figure~\ref{fig:evolution}, which showed that the evolution of the abundance of bright galaxies appears to flatten at $z >$ 9, while the specific UV luminosity density (integrated to $M_{UV} = -$17) is not inconsistent with evolution with a consistent slope from lower redshift.
 It is important to note that presently these integrals are uncertain -- the observations are also consistent with an elevated value at the faint end within the uncertainties.  Future work combining the available deep fields (NGDEEP, MIDIS, JADES, etc.) will soon improve these constraints. Should these future studies conclude that any flattening evolution is observed primarily at brighter luminosities, we comment here on possible physical drivers of such a differential evolution.  First, should any of the processes discussed above be at play primarily in more massive halos, it could lead to a preferentially enhanced abundance of bright galaxies.  However, it seems unlikely that there would be dramatic differences in the star-formation efficiency or IMF across a fairly limited dynamic range of UV absolute magnitudes (and thus presumably halo mass).

One plausible explanation for the observed behavior could be a significant increase in the stochasticity of star formation \citep[e.g.,][]{shen23,sun23,mirocha23}.  As shown by \citet{shen23}, introducing a variability in the conversion from halo mass to UV luminosity (which could encompass both star-formation stochasticity, as well as variations of dust attenuation, metallicity, etc.), leads to an ``upscattering" in the UV luminosity function.  Due to the steep faint-end slope, more galaxies will scatter from faint-to-bright luminosities than from bright-to-faint luminosities, which can lead to a shallower bright end (in an effect similar to Eddington bias).  They explore a range of scatter values (encompassed by the variable $\sigma_{UV}$ which describes the Gaussian width of the kernel scattering the UV luminosities).  As we show in Figure~\ref{fig:lf_sims}, $\sigma_{UV} =$ 1.75 yields a predicted UV luminosity function in good agreement with our observations, though it does overpredict abundances at $M_{UV} \leq -$21.  \citet{pallottini23} explored the level of stochasticity present within their high-resolution hydrodynamic simulations, and found that while stochasticity was present, it was at a lower level, equivalent to $\sigma_{UV} \sim$ 0.6.  This is just one simulation, so it remains to be seen if higher levels of stochasticity are plausible.  We also note that when stochasticity is not included, the fiducial UV luminosity function of the empirical \citet{shen23} model is quite low (solid gold line in Figure~\ref{fig:lf_sims}).  Taking, for example, the UV luminosity function from the physics-based Santa Cruz {\sc GUREFT} SAMs \citep{yung23}, lower levels of UV scatter would be needed to match the observations, plausible more consistent with simulation results.

Observational evidence is emerging that high-redshift galaxies have significant variability in their star-formation rates.  \citet{looser23a} discovered a surprisingly quiescent galaxy at $z =$ 7.3, which shows evidence for a $\sim$10-20 year lull in star-formation activity after a recent burst.  They extended this study in \citet{looser23b}, finding with a spectroscopic analysis that lower-mass galaxies at high redshift have particularly bursty star-formation histories.  \citet{endsley23} came to a similar conclusion after analyzing the photometry of a much larger sample of galaxies, finding that fainter galaxies at higher redshifts show lower [\ion{O}{3}] equivalent widths than their brighter counterparts, which they interpret as evidence that the brightest galaxies are frequently experiencing a recent upturn in star-formation activity (see also \citealt{tacchella23}).

While evidence thus far indicates that galaxies at higher redshifts and lower masses may host more variable star-formation histories, we cannot yet conclude whether this is the major physical driver of the slower-than-expected evolution of the UV luminosity function at brighter luminosities.  Deep {\it JWST}/PRISM spectroscopy could yield a dataset capable of constraining star-formation histories via SED modeling.  Another interesting test to constrain this possibility was proposed by \citet{munoz23}, who noted that if variability in the ratio of UV luminosity to halo mass was significant, many UV-bright galaxies would exist in lower-mass halos, which could be distinguishable via a weaker clustering strength than if there was a more direct correlation between UV luminosity and halo mass.  Such a measurement would require a wide and deep photometric survey capable of identifying a sufficient source density of $z >$ 10 galaxies.  This may be possible with COSMOS-Web (PID 1727, PIs Kartaltepe \& Casey; \citealt{casey23a}), and will certainly be possible with deep field observations with the \emph{Nancy Grace Roman Space Telescope} (in particular with the added $K_s$ filter).

\section{Conclusions}

We have presented the results of a comprehensive search for $z >$ 8.5 galaxies in the full NIRCam dataset from the Cosmic Evolution Early Release Science survey.  We created a new photometric catalog aimed at measuring accurate colors for faint galaxies, as well as robust estimates of the total flux, implementing multiple key improvements over our previous work \citep{finkelstein23}.  We identify a sample of 88 candidate $z >$ 8.5 galaxies, with 55 galaxies in our $z \sim$ 9 sample (selected over 8.5 $< z <$ 9.7), 27 galaxies in our $z \sim$ 11 sample (selected over 9.7 $< z <$ 13), and three galaxies in our $z \sim$ 14 sample (selected over 13 $< z < $ 15).  Notably, 13 of our galaxies are spectroscopically confirmed, eight in the $z \sim$ 9 sample ($z_{spec} =$ 8.63--9.00), and five in the $z \sim$ 11 sample ($z_{spec} =$ 9.77 -- 11.42).

We perform advanced source-injection simulations to assess our source completeness, accounting for both photometric and photometric redshift recovery, explicitly accounting for the sizes of our sources.  While the impact of this update is minimal for bright ($M_{UV} \sim -$ 21) galaxies, it does result in modestly (10--30\%) lower effective volumes than the assumption of a point source at fainter ($M_{UV} \sim -$ 19.5) luminosities.  

We use these completeness estimates to first compare the cumulative surface density of galaxies in our sample to a variety of pre-{\it JWST}-launch simulation predictions, finding that the observed abundance of galaxies is higher than any physical model prediction at $z >$ 10, with this tension increasing with increasing redshift.  Our results are in the least tension with the empirical model of \citet{behroozi15}, which posits that the specific star-formation rate tracks the specific halo accretion rate; at these high redshifts this would imply very short timescales for star-formation.  While it remains to be seen if this is physical, it would be consistent with the observed abundance of UV-bright galaxies.

We calculate the rest-UV luminosity function in our three redshift bins.  Comparing to previous results, we find general agreement, though our uncertainties are typically smaller due to our larger sample sizes.  Notably, we see evidence of the known $z \sim$ 8.7 overdensity in the EGS field \citep{finkelstein22,larson22,whitler23} in the brighter bins of our $z \sim$ 9 UV luminosity function.

We analyze the evolution of the UV luminosity function from $z \sim$ 9 to $z \sim$ 14 in two ways.  First, we examine the evolution of galaxies at $M_{UV} = -$ 20.  While the abundance of galaxies at this fixed UV luminosity has been conclusively measured to rise smoothly from $z =$ 9 to 3, we find evidence for a significant flattening.  Extrapolating the observed evolution from $z =$ 9 to 3 to higher redshifts, one would expect the abundance of galaxies to rise by a factor of $\gtrsim$ 20 from $z \sim$ 14 to $z \sim$ 9.  Conversely, we measure a rise of 4.3 ($\pm$ 3.7) over this epoch.  Phrasing this another way,
$d \rm log \phi$/$dz$ $= -$0.29 $\pm$ 0.03 at $z =$ 3--9, while we find $-$0.11 $\pm$ 0.08 at $z =$ 9--14.  We also explore the total integrated specific UV luminosity density, fitting double-power law models to our observed luminosity functions, and integrating to $M_{UV} = -$17.  Interestingly, we find that this quantity \emph{follows} the observed extrapolation at $z >$ 9.  This hints that whatever new physical processes in play at these epochs may primarily affect bright galaxies, though the uncertainty in the integrated specific UV luminosity density at $z \gtrsim$ 11 is presently high.

We discuss a variety of potential physical causes for the observed results.  The high yield of spectroscopic confirmations implies that significant sample contamination is unlikely, though confirmation of a larger fraction of our galaxy sample would increase confidence in this claim.  We find, based on blue colors for not only our galaxies, but galaxies at similar UV luminosities at $z =$ 6--9, that a significant drop in dust attenuation at earlier times is unlikely to be the dominant explanation.  Rather, we find that models which implement a combination of increased star-formation efficiency and/or an increased degree of bursty, stochastic star formation at these redshifts are more consistent with our observations.  A change in the underlying IMF may also play a role.% cannot rule out a change in the IMF, though any change would primarily need to be in brighter/more massive galaxies, which seems less likely.

We find slight evidence that the physics at play may be more important in bright galaxies than faint galaxies.  Should this represent a dependence of star formation efficiency on halo mass, it would be expected to imprint signatures into a broad range of other cosmological probes. A key example is the topology of neutral hydrogen 21cm intensity maps, where such a differential star-formation efficiency effect would introduce an additional non-linear bias in the power-spectrum analysis. This signature may be detectable with ongoing and upcoming experiments, such as  the Hydrogen Epoch of Reionization Array (HERA; \citealt{liu2016}) and the Square Kilometer Array (SKA; \citealt{Thelie2023}), thus providing independent cross-checks on our results.

Our results show that the abundance of bright galaxies at $z >$ 9 robustly exceeds expectations based on pre-launch observations.
%, which prior to {\it JWST}'s launch were of, at best, a continued smooth decline in the abundance of such galaxies.  
While many possibilities exist to explain these observations, each of them are directly empirically testable with a modest investment in further {\it JWST} spectroscopy.  Deep NIRSpec followup can not only confirm the redshifts of all galaxies in this sample, but it can also probe diagnostic emission lines for either AGN activity or the presence of very massive stars.  Such observations could empirically measure the star-formation histories, testing models of stochasticity.  %Direct metallicity measurements would also constrain models which implement high star-formation efficiencies, which require less-than-Solar metallicities \citep[e.g.][]{dekel23}.  
As we are still very early in the {\it JWST} mission, it is highly likely such observations will become available in the near future, answering these key questions about star and galaxy formation at early times.

\facility{HST (ACS, WFC3)}
\facility{JWST (NIRCam)}

\begin{acknowledgements}
We thank Andrea Ferrara and Nathan Adams for sharing their data.  We also thank Andrea Ferrara, Pawan Kumar, Om Gupta, and Julian Mu\~{n}oz for helpful conversations.  We acknowledge that the location where this work took place, the University of Texas at Austin, that sits on indigenous land. The Tonkawa lived in central Texas and the Comanche and Apache moved through this area. We pay our respects to all the American Indian and Indigenous Peoples and communities who have been or have become a part of these lands and territories in Texas, on this piece of Turtle Island. We acknowledge support from NASA through STScI ERS award JWST-ERS-1345.  PGP-G acknowledges support from grants PGC2018-093499-B-I00 and PID2022-139567NB-I00 funded by Spanish Ministerio de Ciencia e Innovaci\'on MCIN/AEI/10.13039/501100011033, FEDER, UE. RA acknowledges support from ANID Fondecyt 1202007. 

\end{acknowledgements}

\clearpage

\appendix

\section{Table of 8.5 $\leq z \leq$ 10 Sources}

Here we include a table of galaxies in our sample at 8.5 $< z <$ 9.7, split into Table~\ref{tab:lesshighz} and Table~\ref{tab:lesshighz2}.

\begin{deluxetable*}{cccccccccc}
\vspace{2mm}
%\tabletypesize{\small}
\tablecaption{Summary of 8.5 $\lesssim z \lesssim$ 9.5 Candidate Galaxies}
\tablewidth{\textwidth}
\tablehead{\multicolumn{1}{c}{ID} & \multicolumn{1}{c}{RA} & \multicolumn{1}{c}{Dec} & \multicolumn{1}{c}{m$_{F277W}$} & \multicolumn{1}{c}{M$_{1500}$} & \multicolumn{1}{c}{$\mathcal{C}_{FUV}$} & \multicolumn{1}{c}{$\int_7^{20} \mathcal{P}(z)$} & \multicolumn{1}{c}{$\Delta \chi^2$} & \multicolumn{1}{c}{Photometric} & \multicolumn{1}{c}{Spectroscopic}\\
\multicolumn{1}{c}{$ $} & \multicolumn{1}{c}{(J2000)} & \multicolumn{1}{c}{(J2000)} & \multicolumn{1}{c}{(mag)} & \multicolumn{1}{c}{(mag)} & \multicolumn{1}{c}{(mag)} & \multicolumn{1}{c}{$ $} & \multicolumn{1}{c}{$ $} & \multicolumn{1}{c}{Redshift} & \multicolumn{1}{c}{Redshift}}
\startdata
CEERS-13452&214.861602&52.904604&28.2&$-$19.4$^{+0.1}_{-0.2}$&$-$0.16$^{+0.11}_{-0.01}$&1.00&12.1&9.55$^{+0.78}_{-0.09}$&Nz\\
CEERS-76575&215.015299&52.913706&28.5&$-$19.1$^{+0.0}_{-0.2}$&$-$0.08$^{+0.09}_{-0.03}$&1.00&30.3&9.55$^{+0.33}_{-0.06}$&---\\
CEERS-100239&214.800532&52.725500&29.0&$-$18.5$^{+0.1}_{-0.4}$&$-$0.04$^{+0.14}_{-0.07}$&0.97&5.6&9.52$^{+1.05}_{-0.33}$&---\\
CEERS-26109&214.809675&52.858695&28.5&$-$19.2$^{+0.1}_{-0.2}$&$-$0.14$^{+0.15}_{-0.00}$&0.98&6.8&9.43$^{+0.78}_{-0.27}$&---\\
CEERS-91724&214.902804&52.794311&27.1&$-$20.4$^{+0.2}_{-0.1}$&0.15$^{+0.09}_{-0.19}$&0.92&5.1&9.37$^{+0.39}_{-0.51}$&---\\
CEERS-99689&214.802248&52.730517&28.4&$-$19.1$^{+0.1}_{-0.3}$&0.17$^{+0.09}_{-0.17}$&0.91&4.6&9.34$^{+1.17}_{-0.39}$&---\\
CEERS-90326&214.889422&52.793006&29.1&$-$18.4$^{+0.1}_{-0.3}$&$-$0.10$^{+0.05}_{-0.06}$&0.99&8.8&9.25$^{+0.30}_{-0.27}$&---\\
CEERS-45970&214.977355&52.926497&27.0&$-$20.4$^{+0.1}_{-0.1}$&0.02$^{+0.12}_{-0.07}$&1.00&27.0&9.22$^{+0.12}_{-0.15}$&---\\
CEERS-17198&214.858821&52.881221&28.5&$-$18.9$^{+0.0}_{-0.4}$&0.10$^{+0.00}_{-0.22}$&0.94&4.6&9.22$^{+1.32}_{-0.12}$&---\\
CEERS-88331&214.954423&52.852402&28.4&$-$18.4$^{+0.1}_{-0.4}$&0.05$^{+0.29}_{-0.04}$&0.99&5.4&9.16$^{+1.26}_{-0.21}$&---\\
CEERS-74442&214.958350&52.872526&28.3&$-$19.5$^{+0.3}_{-0.2}$&0.04$^{+0.09}_{-0.13}$&0.96&5.7&9.13$^{+1.23}_{-0.42}$&---\\
CEERS-42447&214.795552&52.767286&28.3&$-$19.7$^{+0.1}_{-0.1}$&$-$0.15$^{+0.03}_{-0.04}$&0.99&8.1&9.13$^{+0.15}_{-0.21}$&Nz\\
CEERS-58138&214.876471&52.844055&28.2&$-$19.6$^{+0.1}_{-0.1}$&$-$0.18$^{+0.11}_{-0.00}$&0.99&7.0&9.10$^{+0.18}_{-0.33}$&---\\
CEERS-17898&214.873638&52.887711&29.1&$-$18.4$^{+0.3}_{-0.1}$&$-$0.04$^{+0.14}_{-0.10}$&0.90&4.5&9.07$^{+0.15}_{-2.73}$&---\\
CEERS-39128&214.746942&52.747625&27.9&$-$19.6$^{+0.3}_{-0.2}$&$-$0.16$^{+0.23}_{-0.01}$&0.86&4.7&9.07$^{+0.30}_{-1.26}$&---\\
CEERS-64676&215.125148&52.986537&28.9&$-$18.3$^{+0.4}_{-0.2}$&$-$0.02$^{+0.32}_{-0.00}$&0.96&6.7&9.01$^{+0.69}_{-0.72}$&---\\
CEERS-56878&214.888127&52.858988&27.7&$-$19.9$^{+0.2}_{-0.1}$&$-$0.15$^{+0.12}_{-0.02}$&0.98&7.4&9.01$^{+0.30}_{-0.30}$&Nz\\
CEERS-1398&214.937205&52.965351&29.2&$-$17.3$^{+1.4}_{-0.3}$&0.49$^{+0.76}_{-0.05}$&0.96&4.3&9.01$^{+0.27}_{-1.53}$&---\\
CEERS-61419&214.897231&52.843854&28.1&$-$19.3$^{+0.1}_{-0.1}$&0.13$^{+0.04}_{-0.07}$&1.00&15.3&8.95$^{+1.65}_{-0.06}$&8.998 $_{-0.001}^{+0.001}$\\
CEERS-5007&214.966722&52.968284&28.5&$-$19.2$^{+0.2}_{-0.1}$&$-$0.16$^{+0.11}_{-0.04}$&0.96&6.7&8.98$^{+0.27}_{-0.36}$&---\\
CEERS-47105&214.919100&52.877077&28.8&$-$18.4$^{+0.2}_{-0.3}$&0.19$^{+0.04}_{-0.22}$&1.00&16.1&8.95$^{+0.90}_{-0.24}$&---\\
CEERS-64887&215.059748&52.939334&27.3&$-$20.1$^{+0.1}_{-0.1}$&$-$0.01$^{+0.08}_{-0.05}$&0.99&7.8&8.95$^{+0.12}_{-0.18}$&---\\
CEERS-44441&214.968699&52.929650&26.9&$-$20.7$^{+0.1}_{-0.0}$&$-$0.02$^{+0.04}_{-0.04}$&1.00&52.6&8.95$^{+0.06}_{-0.09}$&---\\
CEERS-13388&214.846175&52.894002&28.4&$-$19.5$^{+0.3}_{-0.3}$&0.08$^{+0.10}_{-0.13}$&0.94&6.0&8.95$^{+1.38}_{-0.66}$&---\\
CEERS-19548&214.876146&52.880826&27.4&$-$19.6$^{+0.1}_{-0.1}$&0.14$^{+0.04}_{-0.06}$&1.00&64.5&8.95$^{+0.15}_{-0.24}$&---\\
CEERS-30173&214.781164&52.817432&28.0&$-$19.6$^{+0.2}_{-0.3}$&$-$0.02$^{+0.14}_{-0.01}$&1.00&16.7&8.95$^{+1.71}_{-0.36}$&---\\
CEERS-48308&214.989485&52.919175&28.2&$-$19.5$^{+0.1}_{-0.1}$&$-$0.12$^{+0.16}_{-0.02}$&0.95&6.6&8.92$^{+0.15}_{-0.36}$&---\\
CEERS-65503&215.079067&52.949637&28.6&$-$19.0$^{+0.2}_{-0.2}$&0.08$^{+0.07}_{-0.10}$&1.00&36.6&8.92$^{+0.39}_{-0.33}$&---\\
CEERS-78591&214.994190&52.876475&27.6&$-$19.8$^{+0.2}_{-0.2}$&0.08$^{+0.07}_{-0.10}$&1.00&16.2&8.92$^{+0.18}_{-0.36}$&---\\
\enddata
\tablecomments{A summary of the key properties for the first half of the 55 galaxies in our sample with 8.5 $\leq z \leq$ 9.7. Spectroscopic redshifts come from \citet{arrabalharo23a}, \citet{arrabalharo23b}, \citet{fujimoto23}, \citet{larson23a}, and \cite{tang23}.}
%\vspace{-8mm}
\end{deluxetable*} \label{tab:lesshighz}

\begin{deluxetable*}{cccccccccc}
\vspace{2mm}
%\tabletypesize{\small}
\tablecaption{Summary of 8.5 $\lesssim z \lesssim$ 9.5 Candidate Galaxies}
\tablewidth{\textwidth}
\tablehead{\multicolumn{1}{c}{ID} & \multicolumn{1}{c}{RA} & \multicolumn{1}{c}{Dec} & \multicolumn{1}{c}{m$_{F277W}$} & \multicolumn{1}{c}{M$_{1500}$} & \multicolumn{1}{c}{$\mathcal{C}_{FUV}$} & \multicolumn{1}{c}{$\int_7^{20} \mathcal{P}(z)$} & \multicolumn{1}{c}{$\Delta \chi^2$} & \multicolumn{1}{c}{Photometric} & \multicolumn{1}{c}{Spectroscopic}\\
\multicolumn{1}{c}{$ $} & \multicolumn{1}{c}{(J2000)} & \multicolumn{1}{c}{(J2000)} & \multicolumn{1}{c}{(mag)} & \multicolumn{1}{c}{(mag)} & \multicolumn{1}{c}{(mag)} & \multicolumn{1}{c}{$ $} & \multicolumn{1}{c}{$ $} & \multicolumn{1}{c}{Redshift} & \multicolumn{1}{c}{Redshift}}
\startdata
CEERS-6184&214.950081&52.949266&28.0&$-$19.5$^{+0.3}_{-0.0}$&0.08$^{+0.07}_{-0.04}$&1.00&14.3&8.92$^{+0.03}_{-0.57}$&---\\
CEERS-89895&214.885377&52.792716&27.9&$-$20.0$^{+0.2}_{-0.1}$&$-$0.20$^{+0.04}_{-0.00}$&0.97&5.6&8.89$^{+0.18}_{-0.21}$&---\\
CEERS-11960&214.907630&52.944612&29.1&$-$18.7$^{+0.3}_{-0.0}$&$-$0.07$^{+0.07}_{-0.04}$&1.00&31.9&8.89$^{+0.12}_{-0.48}$&---\\
CEERS-10545&214.997038&52.960082&28.2&$-$19.6$^{+0.1}_{-0.0}$&$-$0.20$^{+0.07}_{-0.00}$&0.99&6.6&8.89$^{+0.06}_{-0.30}$&---\\
CEERS-61381&214.901253&52.846996&28.5&$-$18.7$^{+0.2}_{-0.2}$&0.30$^{+0.04}_{-0.21}$&1.00&11.0&11.29$^{+0.21}_{-1.56}$&8.881 $_{-0.001}^{+0.001}$\\
CEERS-7078&215.011708&52.988303&27.1&$-$20.4$^{+0.0}_{-0.1}$&$-$0.10$^{+0.09}_{-0.00}$&1.00&75.9&8.98$^{+0.06}_{-0.06}$&8.876 $_{-0.002}^{+0.002}$\\
CEERS-96512&214.800253&52.749561&28.6&$-$18.7$^{+0.2}_{-0.2}$&$-$0.07$^{+0.17}_{-0.03}$&0.87&4.5&8.83$^{+0.15}_{-0.75}$&---\\
CEERS-82881&214.996399&52.854157&27.9&$-$19.7$^{+0.1}_{-0.1}$&$-$0.06$^{+0.09}_{-0.06}$&1.00&19.9&8.83$^{+0.18}_{-0.24}$&---\\
CEERS-4702&214.994404&52.989378&27.5&$-$20.2$^{+0.1}_{-0.0}$&$-$0.07$^{+0.07}_{-0.03}$&1.00&22.0&8.98$^{+0.12}_{-0.12}$&8.809 $_{-0.003}^{+0.003}$\\
CEERS-43833&214.938642&52.911749&26.8&$-$20.7$^{+0.1}_{-0.0}$&0.07$^{+0.04}_{-0.06}$&1.00&80.3&9.01$^{+0.09}_{-0.09}$&8.763 $_{-0.001}^{+0.001}$\\
CEERS-83492&215.079631&52.909565&27.8&$-$19.6$^{+0.0}_{-0.1}$&$-$0.02$^{+0.09}_{-0.08}$&1.00&22.6&8.74$^{+0.15}_{-0.18}$&---\\
CEERS-13544&214.844768&52.892103&26.6&$-$20.5$^{+0.0}_{-0.1}$&0.23$^{+0.00}_{-0.04}$&1.00&81.5&8.74$^{+0.12}_{-0.12}$&---\\
CEERS-43725&214.967532&52.932953&26.3&$-$21.2$^{+0.0}_{-0.0}$&0.03$^{+0.00}_{-0.04}$&1.00&69.2&8.68$^{+0.06}_{-0.09}$&8.715 $_{-0.001}^{+0.001}$\\
CEERS-65379&215.137032&52.991109&28.6&$-$19.0$^{+0.3}_{-0.2}$&0.06$^{+0.12}_{-0.13}$&0.90&4.3&8.71$^{+0.57}_{-0.60}$&---\\
CEERS-89295&214.883206&52.794364&28.7&$-$19.4$^{+0.2}_{-0.1}$&$-$0.19$^{+0.07}_{-0.00}$&0.99&9.1&8.71$^{+0.12}_{-0.33}$&---\\
CEERS-37891&214.745310&52.753653&28.0&$-$19.7$^{+0.2}_{-0.1}$&$-$0.19$^{+0.14}_{-0.00}$&0.89&4.0&8.68$^{+0.24}_{-0.63}$&---\\
CEERS-81061&215.035392&52.890667&25.0&$-$22.2$^{+0.0}_{-0.0}$&0.19$^{+0.00}_{-0.04}$&1.00&104.0&8.68$^{+0.06}_{-0.03}$&8.679 $_{-0.001}^{+0.001}$\\
CEERS-66635&215.120033&52.972564&27.3&$-$20.2$^{+0.1}_{-0.1}$&$-$0.10$^{+0.13}_{-0.03}$&1.00&28.8&8.65$^{+0.15}_{-0.18}$&---\\
CEERS-79589&214.989581&52.866557&26.5&$-$20.7$^{+0.1}_{-0.0}$&0.15$^{+0.04}_{-0.00}$&1.00&60.1&8.65$^{+0.12}_{-0.09}$&---\\
CEERS-81784&215.008674&52.868309&26.3&$-$21.2$^{+0.1}_{-0.1}$&$-$0.08$^{+0.10}_{-0.02}$&1.00&11.3&8.65$^{+0.12}_{-0.12}$&---\\
CEERS-25535&214.838706&52.882221&29.1&$-$18.4$^{+0.3}_{-0.2}$&0.00$^{+0.13}_{-0.09}$&0.98&8.2&8.65$^{+0.51}_{-0.39}$&---\\
CEERS-90671&214.961276&52.842364&28.1&$-$18.8$^{+0.2}_{-0.1}$&0.15$^{+0.09}_{-0.02}$&1.00&18.8&8.68$^{+0.21}_{-0.27}$&8.638 $_{-0.001}^{+0.001}$\\
CEERS-12240&214.902237&52.939370&28.4&$-$18.6$^{+0.1}_{-0.4}$&0.51$^{+0.00}_{-0.43}$&1.00&23.4&8.62$^{+0.30}_{-0.21}$&---\\
CEERS-88342&214.961212&52.857134&28.7&$-$18.5$^{+0.4}_{-0.2}$&0.26$^{+0.11}_{-0.15}$&0.99&8.3&8.59$^{+0.51}_{-0.45}$&---\\
CEERS-61620&214.904392&52.848203&29.2&$-$18.7$^{+0.1}_{-0.1}$&$-$0.20$^{+0.00}_{-0.00}$&0.97&4.7&8.59$^{+0.15}_{-0.27}$&---\\
CEERS-88518&214.942159&52.842306&28.2&$-$19.1$^{+0.4}_{-0.2}$&0.08$^{+0.15}_{-0.13}$&0.97&7.0&8.56$^{+0.42}_{-0.57}$&---\\
CEERS-20174&214.885146&52.883650&28.5&$-$18.8$^{+0.1}_{-0.1}$&$-$0.08$^{+0.12}_{-0.04}$&0.89&4.8&8.56$^{+0.21}_{-0.51}$&---\\
CEERS-88437&214.943826&52.844229&27.5&$-$20.1$^{+0.1}_{-0.1}$&0.04$^{+0.00}_{-0.13}$&1.00&11.6&8.53$^{+0.18}_{-0.12}$&---\\
CEERS-78984&214.984209&52.866090&27.8&$-$19.6$^{+0.1}_{-0.1}$&0.02$^{+0.00}_{-0.05}$&1.00&30.4&8.53$^{+0.15}_{-0.15}$&---\\
\enddata
\tablecomments{A summary of the key properties for the second half of the sample of galaxies in our sample with 8.5 $\leq z \leq$ 9.7. Spectroscopic redshifts come from \citet{arrabalharo23a}, \citet{arrabalharo23b}, \citet{fujimoto23}, \citet{larson23a}, and \cite{tang23}.}
%\vspace{-8mm}
\end{deluxetable*} \label{tab:lesshighz2}

\section{Removed Sources}

\begin{figure*}[!t]
\epsscale{1.2}
\plotone{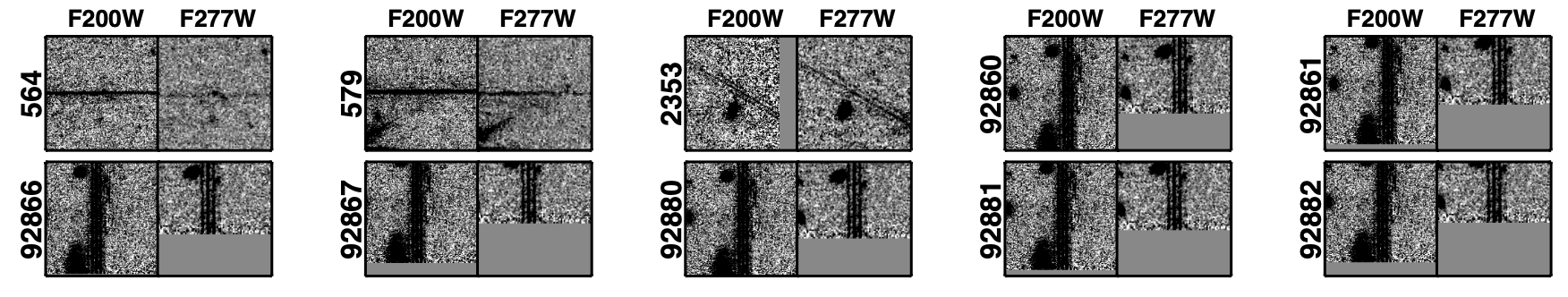}
\caption{Cutout images, 5\arcs\ on a side, of objects originally selected, but identified via visual inspection as being diffraction spikes.}
\label{fig:bad_diffspikes}
\end{figure*}

\begin{figure*}[!t]
\epsscale{1.2}
\plotone{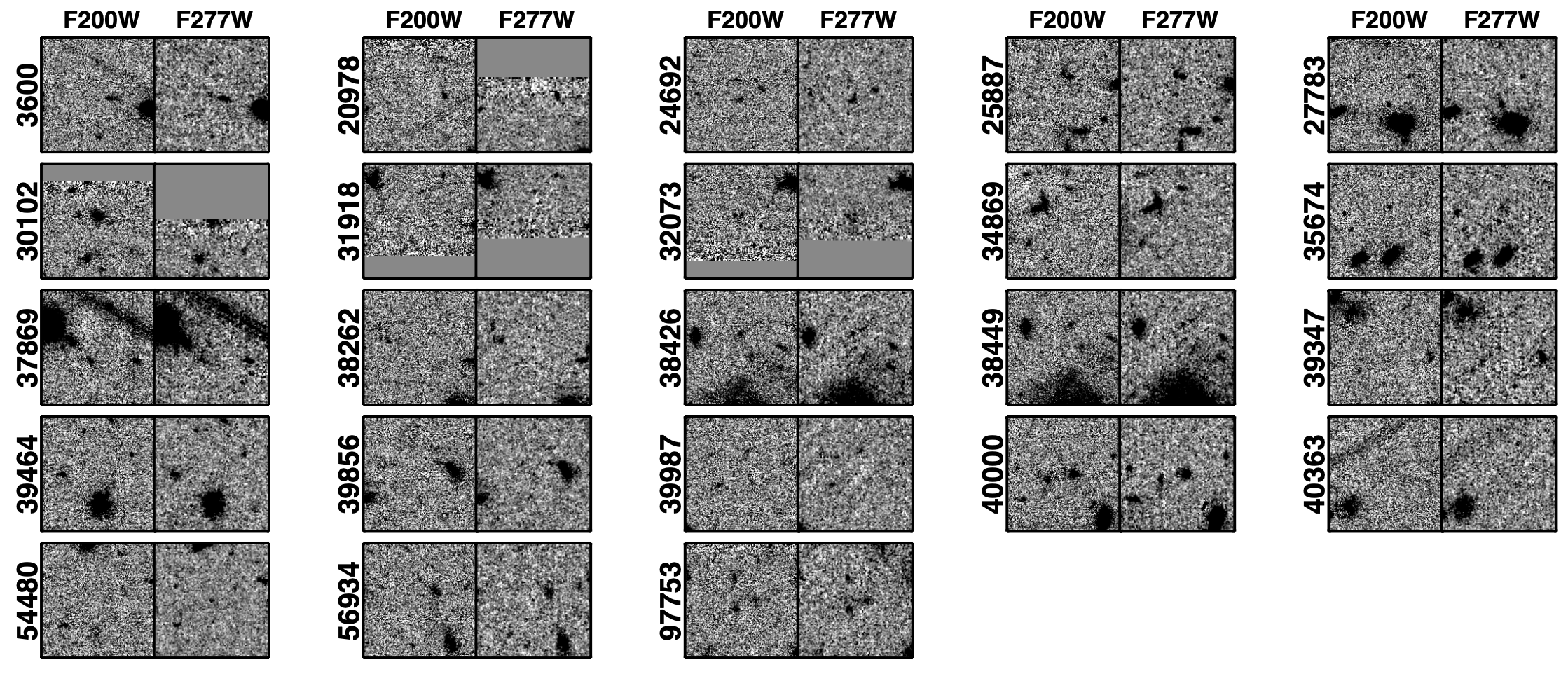}
\caption{Same as Figure~\ref{fig:bad_diffspikes}, for objects identified as bad pixels.}
\label{fig:bad_badpix}
\end{figure*}

\begin{figure*}[!t]
\epsscale{1.2}
\plotone{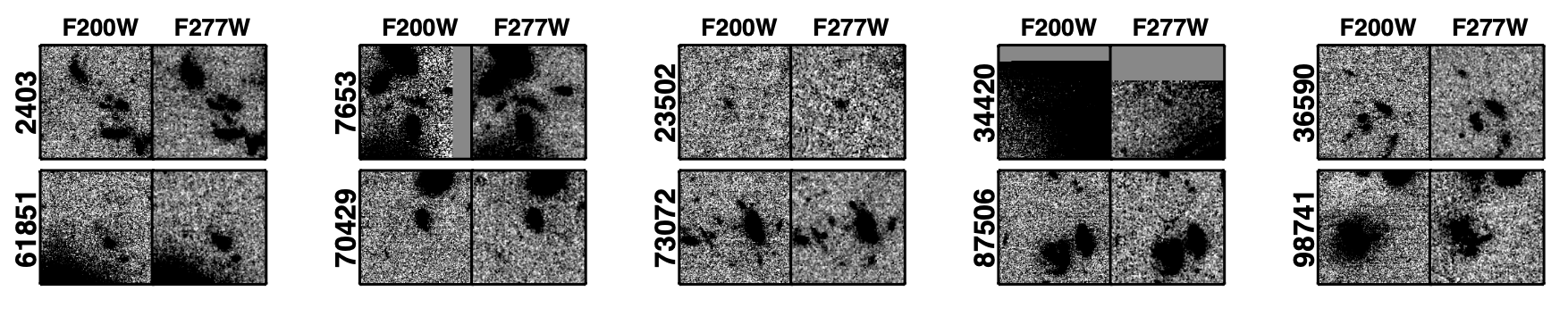}
\caption{Same as Figure~\ref{fig:bad_diffspikes}, for objects identified as oversplit portions of nearby brighter galaxies.}
\label{fig:bad_oversplit}
\end{figure*}

\begin{figure*}[!t]
\epsscale{1.2}
\plotone{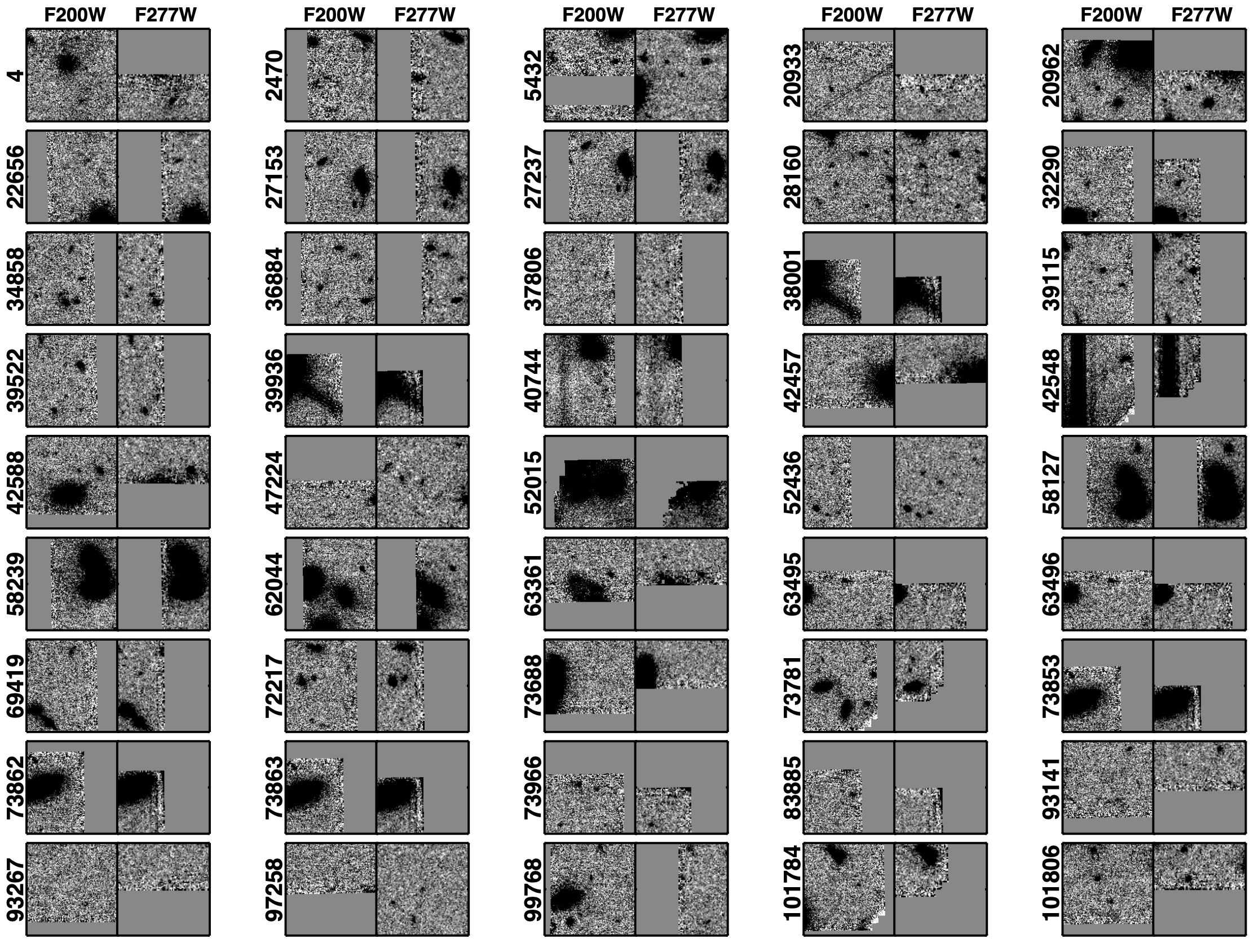}
\caption{Same as Figure~\ref{fig:bad_diffspikes}, for objects identified as being associated with chip edges.}
\label{fig:bad_edges}
\end{figure*}

\begin{figure*}[!t]
\epsscale{1.2}
\plotone{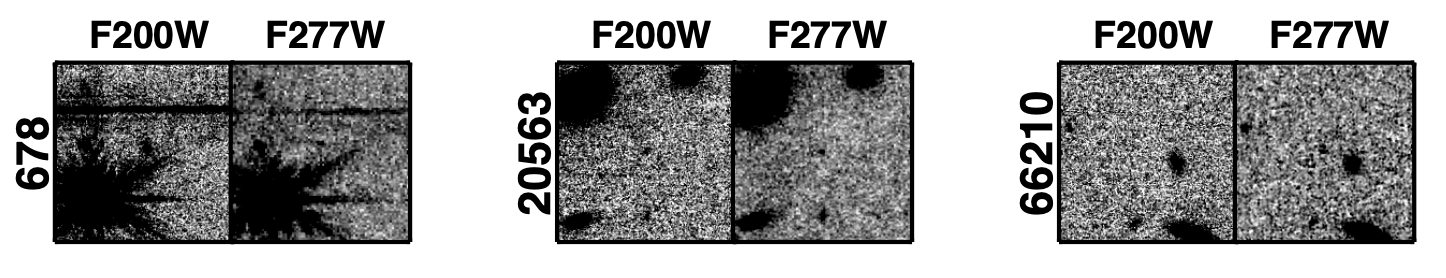}
\caption{Same as Figure~\ref{fig:bad_diffspikes}, for objects identified as being affected by bad photometry.}
\label{fig:bad_badphotom}
\end{figure*}

In Section~\ref{sec:selectioncriteria} we described our visual inspection process to ensure a clean, robust sample of high-redshift galaxies.  As discussed, 91 sources were removed.  Here we present tables (Tables~\ref{tab:appendix-spurious1} and \ref{tab:appendix-spurious2}) of these removed sources along with 5\arcs\ cutout images in the F200W and F277W filters (Figure~\ref{fig:bad_diffspikes}, \ref{fig:bad_badpix}, \ref{fig:bad_edges}, \ref{fig:bad_oversplit} and \ref{fig:bad_badphotom}).  We also include a table of the five sources removed in \S 3.2.1; four due to having $z_{spec} <$ 8.5, and one with similar colors as the $z \sim$ 16 candidate confirmed at $z_{spec} =$ 4.9 (Table~\ref{tab:appendix-spectab}).

\begin{deluxetable*}{cccc}
\vspace{2mm}
%\tabletypesize{\small}
\tablecaption{Objects Removed from the Sample During Visual Inspection}
\tablewidth{\textwidth}
\tablehead{\multicolumn{1}{c}{ID} & \multicolumn{1}{c}{RA} & \multicolumn{1}{c}{Dec} & \multicolumn{1}{c}{Notes}\\
\multicolumn{1}{c}{$ $} & \multicolumn{1}{c}{(J2000)} & \multicolumn{1}{c}{(J2000)} & \multicolumn{1}{c}{$ $}}
\startdata
4&214.947775&52.980444&Edge\\
564&215.005249&53.017773&Diffraction Spike\\
579&215.006814&53.018834&Diffraction Spike\\
678&215.007828&53.019061&Bad Photometry (affected by diffraction spike)\\
2353&214.945032&52.966213&Diffraction Spike\\
2403&215.012421&53.014295&Oversplit (possibly unique source)\\
2470&214.964578&52.979483&Edge\\
3600&214.943365&52.959006&Bad Pixels(s)\\
5432&215.000777&52.989639&Edge\\
7653&215.021371&52.990815&Oversplit\\
20563&214.958980&52.933775&Bad Photometry (flux visible in F115W)\\
20933&214.767704&52.854815&Edge\\
20962&214.735949&52.832000&Edge\\
20978&214.768561&52.855113&Bad Pixels(s)\\
22656&214.780702&52.854494&Edge\\
23502&214.766563&52.840230&Oversplit (spurious signal near bright object)\\
24692&214.755849&52.827409&Bad Pixels(s)\\
25887&214.775760&52.838390&Bad Pixels(s)\\
27153&214.798863&52.844780&Edge\\
27237&214.799254&52.844685&Edge\\
27783&214.779457&52.827965&Bad Pixels(s)\\
28160&214.780257&52.826895&Edge\\
30102&214.749567&52.841891&Bad Pixels(s)\\
31918&214.811074&52.829806&Bad Pixels(s)\\
32073&214.834670&52.845385&Bad Pixels(s)\\
32290&214.743795&52.784616&Edge\\
34420&214.772999&52.805015&Oversplit\\
34858&214.755731&52.778280&Edge\\
34869&214.782948&52.797710&Bad Pixels(s)\\
35674&214.766152&52.781819&Bad Pixels(s) (plausibly real source)\\
36590&214.723366&52.746324&Oversplit\\
36884&214.802322&52.801134&Edge\\
37806&214.769166&52.771157&Edge\\
37869&214.783779&52.781243&Bad Pixels(s)\\
38001&214.690385&52.748166&Edge\\
38262&214.799921&52.790796&Bad Pixels(s)\\
38426&214.792791&52.784917&Bad Pixels(s)\\
38449&214.793033&52.784921&Bad Pixels(s)\\
39115&214.774698&52.768211&Edge\\
39347&214.792824&52.779987&Bad Pixels(s)\\
39464&214.808944&52.790945&Bad Pixels(s)\\
39522&214.776287&52.767365&Edge\\
39856&214.811602&52.791096&Bad Pixels(s)\\
39936&214.690515&52.748090&Edge\\
39987&214.807915&52.787863&Bad Pixels(s)\\
\enddata
\tablecomments{Properties of objects removed from the sample during visual inspection.  The notes column gives the primary reason, as well as indicating which objects plausible still could be real candidates, but were conservatively removed due to the stated concerns.}
%\vspace{-8mm}
\end{deluxetable*} \label{tab:appendix-spurious1}

\begin{deluxetable*}{cccc}
\vspace{2mm}
%\tabletypesize{\small}
\tablecaption{Objects Removed from the Sample During Visual Inspection}
\tablewidth{\textwidth}
\tablehead{\multicolumn{1}{c}{ID} & \multicolumn{1}{c}{RA} & \multicolumn{1}{c}{Dec} & \multicolumn{1}{c}{Notes}\\
\multicolumn{1}{c}{$ $} & \multicolumn{1}{c}{(J2000)} & \multicolumn{1}{c}{(J2000)} & \multicolumn{1}{c}{$ $}}
\startdata
40000&214.800141&52.786022&Bad Pixels(s)  (plausibly real source)\\
40363&214.796309&52.777817&Bad Pixels(s)\\
40744&214.780832&52.764953&Edge\\
42457&214.762724&52.743756&Edge\\
42548&214.788142&52.761069&Edge\\
42588&214.824026&52.786252&Edge\\
47224&214.982491&52.920795&Edge\\
52015&214.878570&52.875991&Edge\\
52436&214.861696&52.861844&Edge (plausibly real source)\\
54480&214.872931&52.859585&Bad Pixels(s)\\
56934&214.819248&52.809749&Bad Pixels(s)\\
58127&214.851212&52.826107&Edge\\
58239&214.851631&52.825884&Edge\\
61851&214.840216&52.801356&Oversplit\\
62044&214.865103&52.818660&Edge\\
63361&214.917040&52.847539&Edge\\
63495&215.092356&52.969912&Edge\\
63496&215.092471&52.969992&Edge\\
66210&215.112944&52.969707&Bad Photometry (only significant flux visible in one filter)\\
69419&215.118348&52.955698&Edge\\
70429&215.152240&52.974036&Oversplit (plausibly real source)\\
72217&215.129726&52.949675&Edge\\
73072&215.108332&52.929579&Oversplit (plausibly real source)\\
73688&215.111283&52.928601&Edge\\
73781&215.136874&52.945927&Edge\\
73853&214.946038&52.869056&Edge\\
73862&214.946343&52.868885&Edge\\
73863&214.946212&52.868955&Edge\\
73966&214.999418&52.905737&Edge\\
83885&214.852858&52.806349&Edge\\
87506&214.932964&52.841802&Oversplit (plausibly real source)\\
92860&214.965779&52.829727&Diffraction Spike\\
92861&214.965915&52.829654&Diffraction Spike\\
92866&214.965694&52.829574&Diffraction Spike\\
92867&214.965862&52.829489&Diffraction Spike\\
92880&214.965712&52.829665&Diffraction Spike\\
92881&214.965845&52.829598&Diffraction Spike\\
92882&214.965961&52.829537&Diffraction Spike\\
93141&214.915928&52.795313&Edge\\
93267&214.956713&52.822962&Edge\\
97258&214.787857&52.736119&Edge\\
97753&214.842961&52.771072&Bad Pixels(s)\\
98741&214.853712&52.773300&Oversplit\\
99768&214.833470&52.752889&Edge\\
101784&214.858663&52.756439&Edge\\
101806&214.864006&52.760037&Edge\\
\enddata
\tablecomments{A continuation of the previous table.}
%\vspace{-8mm}
\end{deluxetable*} \label{tab:appendix-spurious2}

\begin{deluxetable*}{ccccccc}
\vspace{2mm}
%\tabletypesize{\small}
\tablecaption{Objects Removed due to NIRSpec Information}
\tablewidth{\textwidth}
\tablehead{\multicolumn{1}{c}{ID} & \multicolumn{1}{c}{RA} & \multicolumn{1}{c}{Dec} & \multicolumn{1}{c}{m$_{F277W}$} & \multicolumn{1}{c}{$\int_7^{20} \mathcal{P}(z)$} & \multicolumn{1}{c}{Photometric} & \multicolumn{1}{c}{Spectroscopic}\\
\multicolumn{1}{c}{$ $} & \multicolumn{1}{c}{(J2000)} & \multicolumn{1}{c}{(J2000)} & \multicolumn{1}{c}{(mag)} & \multicolumn{1}{c}{$ $} & \multicolumn{1}{c}{Redshift} & \multicolumn{1}{c}{Redshift}}
\startdata
CEERS-4774&215.005185&52.996577&27.0&0.92&8.92$^{1.35}_{0.66}$&8.005$\pm$0.001\\
CEERS-4777&215.005366&52.996697&28.0&0.99&10.12$^{0.93}_{0.69}$&7.993$\pm$0.001\\
CEERS-13256&214.914550&52.943023&26.5&1.00&16.45$^{0.18}_{0.45}$&4.912$\pm$0.001\\
CEERS-23084&214.830688&52.887770&28.2&0.98&8.77$^{0.45}_{0.69}$&7.769$\pm$0.003\\
CEERS-43382&214.951146&52.923539&28.8&0.94&16.84$^{2.34}_{0.93}$&---\\
\enddata
\tablecomments{Properties of objects removed from the sample.  The first four have spectroscopic redshifts of $z <$ 8.5.  The final object was not spectroscopically confirmed, but has a photometric redshift of $z >$ 16, and exhibits an observed SED extremely similar to CEERS-13256, which is confirmed by \citet{arrabalharo23a} to be at $z =$ 4.912.}
%\vspace{-8mm}
\end{deluxetable*} \label{tab:appendix-spectab}

\end{document}